\documentstyle[prd,aps,floats]{revtex}
\input{psfig.tex}
\tighten
\begin{document} 
\draft
%\twocolumn[\hsize\textwidth\columnwidth\hsize\csname@twocolumnfalse\endcsname
%DONT FORGET ``]'' below when de-commenting the above line

\preprint{Imperial/TP/97-98/9, DAMTP/R-97/53}
\title{A detailed study of defect models for
cosmic structure formation} 
\author{Andreas Albrecht{$^{1}$}, Richard A. Battye{$^{1,2}$} and James Robinson{$^{1,3}$}}
\address{${}^1$ Theoretical Physics Group, Blackett Laboratory, Imperial
College, Prince Consort Road, 
 London SW7 2BZ,  U.K.\\
${}^2$ Department of Applied Mathematics and Theoretical Physics, University of Cambridge, \\ Silver Street, Cambridge CB3 9EW, U.K. \\
${}^3$ Department of Astronomy, Campbell Hall, University of California, Berkeley CA 94720, U.S.A.
}
\maketitle
\begin{abstract}
%\narrowtext
We calculate predictions from wide class of `active' models of
cosmic structure formation which allows us to scan the space of
possible defect models.  We calculate the linear cold dark matter power
spectrum  and Cosmic Microwave Background (CMB) anisotropies over all
observable scales using a full linear Einstein-Boltzmann code.
Our main result, which has
already been reported, points to a serious problem reconciling the
observed amplitude of the large-scale galaxy distribution with the COBE 
normalization. Here, we describe our methods and results in detail. The problem is present 
for a wide range of defect parameters, which can be used to represent
potential differences among defect models, as well as possible
systematic numerical errors. We explicitly examine the
impact of varying the defect model parameters and we show how the
results substantiate these conclusions. 
The standard scaling defect models
are in serious conflict with the current data, and we show how attempts to
resolve the problem by considering non-scaling defects or modified stress-energy components would require
radical departures from what has become the standard picture. 
\end{abstract}

\date{\today}

%]

\renewcommand{\thefootnote}{\arabic{footnote}}
\setcounter{footnote}{0}

\section{Introduction}

Topological defects are an almost generic phenomena in nature and have
been already detected in a number of laboratory systems
(see, for example,  ref.\cite{condensed}), where
symmetry breaking phase transitions take place. Probably the most
exciting possibility, however, is that they are formed during  
spontaneous symmetry breaking at a phase transition in the early
universe \cite{kib1,VS,HK}, since they could act as the primordial seeds for
galaxy formation; 
the most plausible models being the so called cosmic string
\cite{zeld,vil} and cosmic texture \cite{TurSperg} theories. 

Theories for galaxy formation can be described as either `passive' or
`active' \cite{ACFM}. In passive theories, such as those predicted by
the inflationary paradigm, all the perturbations are set up
effectively as (super-horizon) initial conditions at very early times, which then evolve
under a deterministic linear evolution, until very late when
non-linear processes take over on the very smallest scales. By
contrast, perturbations are created on all scales at all times in
active models making predictions much more difficult to calculate. Typically, one has to deal with the fundamental non-linearity of
the source over a large dynamic range --- approximately 25 orders of
magnitude --- from the defect formation to the present. 

The last few months have seen dramatic progress in pinning down
the predictions from defect models of cosmic structure formation in
what we shall describe as the standard scenario, that is defect
motivated stress-energy components with an assumption of perfect
scaling from formation to the present day. Three
groups\cite{PSelTa,ABRa,ACDKSS} have performed calculations which 
integrate the linear Einstein-Boltzmann equations using the latest
technology for different models of the defect stress-energy two-point
functions to produce predictions of power spectra for the cold dark matter (CDM)
density field and the Cosmic  Microwave Background (CMB)
anisotropies. This article gives a detailed presentation of the
methods and results of ref. \cite{ABRa}. 

There are two traditional approaches to the study of defect
dynamics. Some authors have used large-scale simulations to provide
the sources for their CMB and structure formation calculations
\cite{PST,CT,DGS,ACSSV}, while others have developed analytic models which
attempt to describe the statistical properties of the defects
\cite{kib2,ACK,MSa}. Even using the latest technology, modelling
the source using simulations is severely constrained by dynamic
range. In this work, we will use a model based approach to calculate
the two-point correlation functions, which act as sources for a
state-of-the-art linear Einstein-Boltzmann solver. This has a number
of 
advantages and also disadvantages when compared with the complementary
simulation based approach; the real strength being that one is not
constrained to a particular defect based scenario, allowing one to
explore all possible scenarios and understand the robustness of any
claims that one might make. The downside  is that one must take care
to construct the source stress-energy, which has many possible degrees
of freedom, in a way which is at least physically plausible. This is
generally done by comparison to some kind of simulation.  

The recent work by Pen {\em et al} \cite{PSelTa} and our collaboration 
\cite{ABRa}, has exposed an apparently deep conflict between COBE
normalized defect models  and the observed galaxy correlations on
large scales.
The conflict was quantified in
ref.\cite{ABRa} in terms 
of $b_{100}$, the bias between the dark matter and baryon
distributions on scales of $100h^{-1}$Mpc required to match the COBE
normalized defect models with the galaxy data. 
Although some evidence for such a problem has existed for some time,
uncertainties about whether the computations had sufficient dynamic
range meant the precise quantitative details of the problem were not
fully understood. Most previous work on this problem relied on separate
calculations for the large-angle CMB, which is normalized to COBE, and
the linear matter power spectrum, using analytic expressions to
relate the two, via, for example, the mass per unit length in the case
of cosmic strings. Our calculations (as well as those of Pen {\em et al}
\cite{PSelTa}) do not use any such
extrapolation, with the perturbations in the matter (dark and baryonic)
and the photon (plus neutrinos) distributions being calculated in a
self consistent way across all observable scales. Hence, a single
normalization to COBE gives the normalized linear matter power
spectrum. 

We should note that the scale of $100h^{-1}$Mpc was chosen for three
reasons. Firstly, $100h^{-1}$Mpc is sufficiently large that
non-linear effects are not expected to affect the
power spectrum. Secondly, the discrepancy in the power
spectra at this scale is at its most extreme. And finally, it is
unlikely that the power spectrum on these scales can be affected by
changes in the cosmogony, for example, the introduction of hot dark
matter (HDM). 

The price of our solution to the dynamic range problem is that (1) we
can only calculate the power spectrum of the matter and CMB since we
have only included the two-point functions of the source \cite{AS}
--- no non-Gaussian effects are included --- and (2) the 
results depend on the validity of the simple scaling picture over many 
orders of magnitude of cosmic expansion.  Even though there is
substantial support for this assumption, both from numerical simulations and
analytic modelling\cite{AT,BB,AShe,kib2,ACK,MSa}, there are also reasons to think that it may not be
perfect.  One of the important components of our work is an
investigation of possible deviations from scaling.     
Our approach to modelling the defect two-point functions has also allowed
us to explore other variations, besides deviations from scaling.
These variations represent possible differences between defect models
and possible systematic errors in numerical defect simulations.
We have found that
solving the $b_{100}$ problem requires extreme departures from the 
standard picture. (Interestingly, we are learning that defect networks
in $\Omega_{matter} <1 $ cosmologies may exhibit interesting levels
non-scaling behaviour\cite{us,paul}.) 

In the next section, we discuss some aspects of the linearized
Einstein-Boltzmann solver, in particular the inclusion of source
stress-energy, the Einstein-Boltzmann equations for vector perturbations and ways of calculating the
ensemble average for incoherent perturbations. Section III presents in 
detail our modelling of the source two-point functions, by reference
to a specific string motivated model. The results for the standard
scaling model are presented along with quantification of the
$b_{100}$ problem. We also show that simple modifications to the model and cosmological parameters, such as the Hubble constant and baryon density, have little impact on the $b_{100}$ problem. In Section IV we discuss possible deviations from
the standard scaling assumption and section V explores
further modifications to the model, which might lead to an improvement in the
comparison with the data. We find that it is possible to get better
agreement, although
only the most extreme modifications come close on $100h^{-1}$Mpc
scales. The aim of this  paper is to present a pedagogical exposition
of our work, so that the expert can reproduce and interpret our
results. In the final section, we discuss its relation  to that of
others and point to directions of future investigation.

\section{Linear Einstein-Boltzmann solver}

\subsection{CMBFAST}

In order to calculate CMB anisotropies, one must solve the linearized
Einstein-Boltzmann equations. Recent years have seen techniques
developed to solve these equations to very high precision $(\sim 1\%)$
for the standard adiabatic models based on inflation \cite{various},
culminating in the public release of CMBFAST \cite{cmbfast} which can
perform this task in under a minute on a modern workstation. The
standard Boltzmann method involves evolving over 3000 highly oscillatory
linear ODEs from some time deep in the radiation era to the present
day, which can take many hours. The line of sight method \cite{lineofsight}
used in this code reduces the time drastically by splitting the
prohibitively oscillatory geometric effects from the dynamical processes
due to, for example, the Doppler effect and potential evolution. This
reduces the number of ODEs down to about 30, but adds an integration
along the line of sight.  

It is usual to express temperature anisotropies in terms of a
decomposition into spherical harmonics, 
\begin{equation}
{\Delta T\over T}(\theta,\phi)=\sum_{l=0}^{\infty}\sum_{m=-l}^{l} a_{lm}Y_{lm}(\theta,\phi)\,.
\end{equation}
For a Gaussian random field, such as those generated by inflation, the
statistics of anisotropies are entirely specified by the angular power
spectrum $C_l=\langle |a^2_{lm}| \rangle$ 
where the angled brackets denote an ensemble average.

For the moment, let us assume that we only require the anisotropies for
a simple inflationary model which creates no appreciable vector and tensor
fluctuations. In this case, the angular power spectrum is given by 
\begin{equation}
C_l={2\over\pi}\int k^2dk \langle \Delta^S_l(k,\tau_0)\Delta^{S*}_l(k,\tau_0) \rangle\,,
\label{clscalar}
\end{equation}
where the photon distribution function $\Delta^S_l(k,\tau_0)$ at the
present day $\tau=\tau_0$ is given by the line of sight integration, 
\begin{equation}
\Delta^S_l(k,\tau_0)=\int_0^{\tau_0}d\tau\left(S^0_T(k,\tau)j^{00}_l\left[k(\tau_0-\tau)\right]+S^1_T(k,\tau)j^{10}_l\left[k(\tau_0-\tau)\right]+S^2_T(k,\tau)j^{20}_l\left[k(\tau_0-\tau)\right]\right)\,,
\label{scalarsource}
\end{equation}
and $S^0_T(k,\tau)$, $S^1_T(k,\tau)$, $S^2_T(k,\tau)$ can be deduced
from ref.\cite{HW}, with  
\begin{equation}
j^{00}_l(x)=j_l(x)\,,\quad j^{10}(x)=j^{\prime}_l(x)\,,\quad j^{20}_l(x)={1\over 2}\left(3j_l^{\prime}(x)+j_l(x)\right)\,. 
\end{equation}
For the coherent limit, implicit in phase focused \cite{ACFM} inflationary models, one
can perform the ensemble average by simply replacing $\langle
\Delta^S_l(k,\tau_0)\Delta^{S*}_l(k,\tau_0) \rangle =
|\Delta^S_l(k,\tau_0)|^2$. 

In the rest of this section, we  will  describe how this approach can
be modified to include active sources such as cosmic defects. Firstly,
we show how simple coherent scalar sources may be added. Then we
discuss the inclusion of the vector and tensor sources, almost generic
in any active model. Finally, we show how one may perform the ensemble
average when the source is not coherent and discuss the various implications of decoherence. We have already noted that for non-Gaussian
sources, such as the topological defects under consideration here, the
angular power spectrum does not entirely specify the nature of the
of anisotropies. However, most realistic models are thought to lead to
only mildly non-Gaussian anisotropies through the Central Limit Theorem
for the superposition of non-Gaussian probability
distributions. Hence, it should  be a useful discriminant between
different models for structure formation.  We will discuss the efficacy of using  power spectra to distinguish between different models of structure formation in the conclusions.

\subsection{Coherent active scalar sources}
\label{cohere}
As a first step, therefore, let us introduce an independent covariantly conserved
component of stress-energy $\Theta_{\mu\nu}$ into the Einstein
equations, 
\begin{equation}
G_{\mu\nu}=8\pi\left[T_{\mu\nu}+\Theta_{\mu\nu}\right]\,,
\end{equation}
where $T_{\mu\nu}$ is the stress-energy of CDM, baryons, photons and
neutrinos present in a particular cosmogony. This extra component,
usually assumed to be `stiff', that is, unaffected by gravity at first order, adds a forcing term to the Einstein-Boltzmann equations which represents the active sources. 

For the moment let us assume that the Fourier transform of the
stress-energy can be decomposed as $\Theta_{0i}=\Theta_D\hat k_i$ and
\begin{equation}
\Theta_{ij}={1\over 3}\delta_{ij}\Theta+\left(\hat k_i \hat
k_j-{1\over 3}\delta_{ij} \right)\Theta^S\,,
\label{scalarsplit}
\end{equation}
where $\hat k_i$ is a unit vector in Fourier space, $\Theta_D$ is the velocity
field, $\Theta$ is the isotropic pressure, or three times the pressure, and $\Theta^S$ is the
anisotropic stress. The conservation equations for this decomposition
are 
\begin{eqnarray}
\dot\Theta_{00}+{\dot a \over a}\left(\Theta_{00}+\Theta\right)-\Theta_D=0
\,, \cr 
\dot\Theta_D+{2\dot a \over a}\Theta_D+{1\over
3}k^2\left(\Theta+2\Theta^S\right)=0\,. 
\label{stressen}
\end{eqnarray}
Hence, in order to incorporate this source stress-energy, one must add
(\ref{stressen}) to the ODE solver, modify the linearized Einstein
equations to include the forcing terms and specify two quantities
from $\Theta_{00}$, $\Theta_D$, $\Theta$ and $\Theta^S$. 

The initial conditions for the Einstein-Boltzmann equations must also
be modified. The idea is that one sets up initial conditions for a
pure growing mode deep in the radiation era. In order to enforce
causality, one requires that components of the pseudo-stress-energy
tensor $\tau_{\mu\nu}$ be zero\cite{PST}, which creates a balance
between the initial metric perturbations, the defect stress-energy and
the matter perturbations. However, there is a residual degree of
freedom which allows us to just simply set everything to
zero. Physically, this implies that the initial conditions are
unimportant relative to the actual sources themselves, which is
implicit in the distinction between passive and active sources.  

The most common defect based models are thought to scale for most of
the history of the universe and so for the moment at least, we
specialize our discussion to scaling sources. This requires that
$\tau^{1/2}\Theta_{00}(k,\tau)=F_1(k\tau)$,
$\tau^{1/2}\Theta(k,\tau)=F_2(k\tau)$,
$\tau^{1/2}\Theta_D(k,\tau)=F_3(k\tau)$ and
$\tau^{1/2}\Theta^S(k,\tau)=F_4(k\tau)$, with the functions $F_i(x)$
having well defined power series expansions around $x=0$. Moreover,
further constraints can be placed on the leading  
order behaviour of these functions by causality and analyticity
\cite{PSelTc}. In particular, this implies that,   
\begin{eqnarray}
&\langle \Theta_{00}\Theta^*_{00} \rangle = Zk^0 +{\cal O}(k^2)\,,\quad 
\langle \Theta^{S}\Theta^{S*} \rangle = Y k^0 +{\cal O}(k^2)\,,\quad
\langle \Theta_{00}\Theta^{S*} \rangle = X k^2+{\cal O}(k^4)\,,&\cr
&\langle \Theta\Theta^*\rangle = W k^0+{\cal O}(k^2)\,,\quad \langle\Theta_D\Theta^*_D\rangle = V k^4+{\cal O}(k^6)\,,&
\label{correl}
\end{eqnarray}
with  all the other correlators and cross-correlators being deduced in a similar way or by assuming stress-energy conservation. In the coherent limit $\langle\Theta_{00}\Theta^{S*}\rangle=\langle\Theta_{00}\Theta^*_{00}\rangle^{1/2}
\langle\Theta^{S}\Theta^{S*}\rangle^{1/2}$, which using (\ref{correl}) implies that $Y=0$ and in fact $\langle\Theta^{S}\Theta^{S*}\rangle\sim k^4$. This very specialized limit leads to some slight subtleties which will not in general be present for active
 sources. This will be discussed in the section on incoherent sources. 

\begin{figure}
\hskip 1.75in\psfig{file=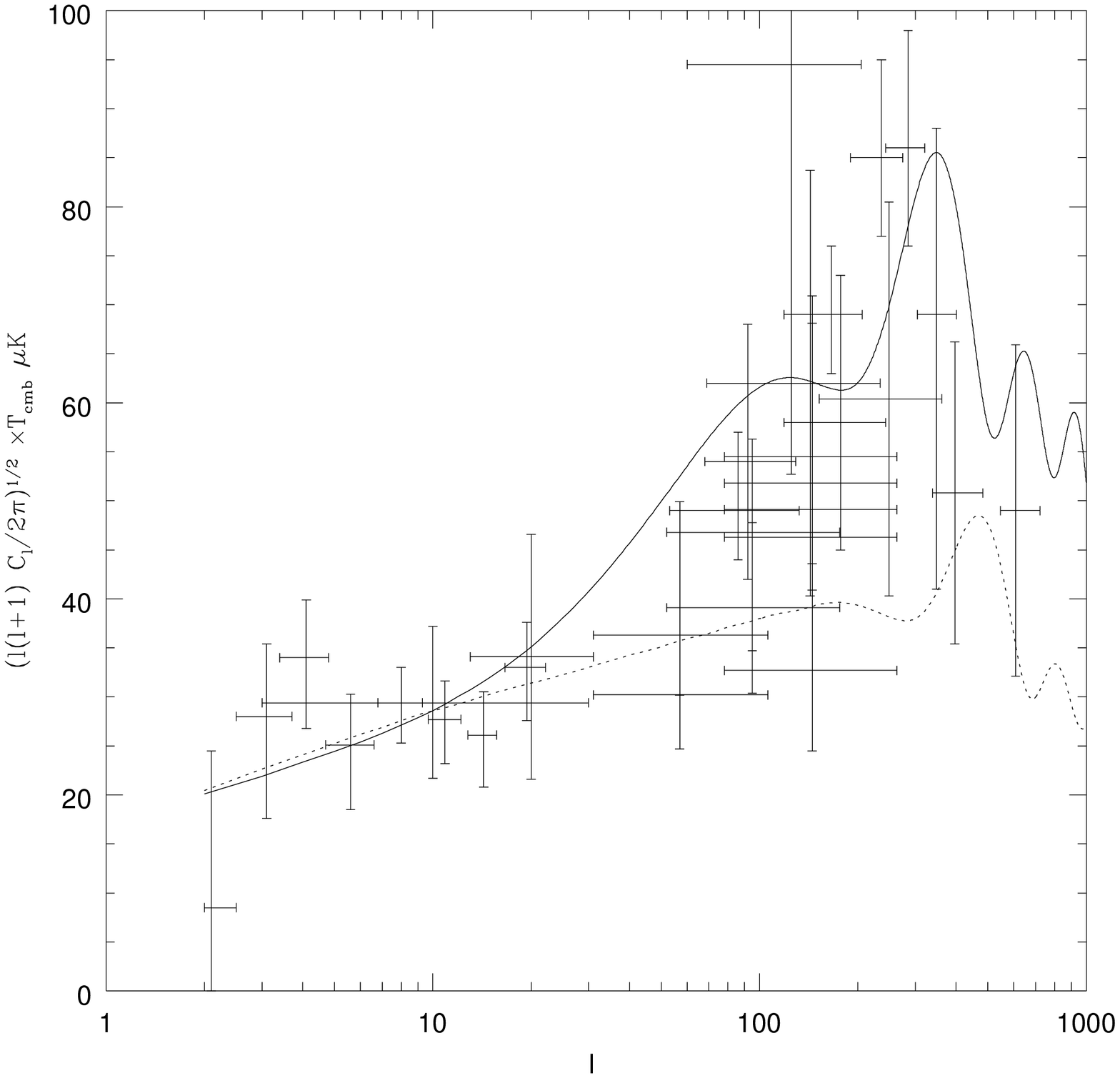,width=3.4in}
%\vskip 3.4in
\caption{The angular power spectrum for the simple coherent sources
(a,solid line) and (b,dotted line) normalized to COBE. Note no vector
or tensor components are calculated here.} 
\label{fig-press}
\end{figure}

One simple choice which is consistent with the coherent limit is to define \cite{HSW}
\begin{equation}
4\pi\Theta={3\alpha\over\tau^{1/2}}{\sin Ak\tau\over Ak\tau}\,,\quad
4\pi\Theta^S={\beta\over\tau^{1/2}}{6\over B^2-C^2}\left({\sin Bk\tau\over
Bk\tau} 
-{\sin Ck\tau\over Ck\tau}\right)\,,
\end{equation}
where $A$, $B$, $C$, $\alpha$ and $\beta$ are constants. The angular power spectra for this
source are presented in Fig.~\ref{fig-press} for (a) $\alpha=1$, $\beta=0$ and $A=1$
(b) $\alpha=1$, $\beta=1$, $A=1$, $B=1$ and $C=0.5$. Taking into account the arbitrary normalization used in ref.\cite{HSW}, the results seem to be
identical.

At a very technical level, various modifications to internal systems
parameters of CMBFAST were also required. The source function for
$\Delta^S_l(k,\tau)$ is dependent on three variables and the code
sets up a discrete three-dimensional array containing
these values with spacings $\Delta l$, $\Delta k$ and $\Delta\tau$,
before performing the line of sight integration
over $l$ values. Obviously, the most
accurate results are obtained by using the smallest possible values
of $\Delta l$, $\Delta k$ and $\Delta\tau$; the cost being an
increase in the CPU time. We systematically reduced all three spacings
from the values in the original version
and found that the results converged when $\Delta k$ was reduced by about a factor of two, with
$\Delta l$ and $\Delta\tau$ being left the same. We also varied the
number of equations solved by the ODE integrator and found that there
was no discernible improvement from any increase.  We concluded that the line of sight integration method is also useful to calculate the predictions of active models for structure formation.

\subsection{Generalization to include vector and tensor modes}

The split of the energy momentum tensor (\ref{scalarsplit}) is not the most general since it includes only scalar sources. A more general split is 
\begin{equation}
\Theta_{0i}=\Theta_D\hat{k}_i+\Theta_{0i}^{V} \,,\quad
\Theta_{ij}={1\over 3}\delta_{ij}\Theta+\left(
\hat{k}_i\hat{k}_j-{1\over 3}\delta_{ij}\right)\Theta^{S}+\left(
\hat{k}_i\Theta^{V}_j+\hat{k}_j\Theta^{V}_i\right)+\Theta_{ij}^{T}\,,
\end{equation}
which includes vector $(\Theta_{0i}^{V},\Theta^V_i)$ and tensor $(\Theta_{ij}^{T})$ sources such that 
\begin{equation}
\hat{k}_i\Theta_{0i}^{V}=\hat{k}_i\Theta^{V}_{i}=\hat{k}_i\Theta^{T}_{ij}=
\hat{k}_j\Theta^T_{ij}=\Theta^T_{ii}=0\,.
\end{equation}
Without loss of generality, one can fix the direction in Fourier space and  choosing $\hat{k}=(0,0,1)$ gives 
\begin{equation}
\Theta_{\mu\nu}=
\left( \begin{array}{cccc}
\Theta_{00} & \Theta_{01}^{V} & \Theta_{02}^{V} & \Theta_D \\
\Theta_{01}^V &{1\over 3}\Theta-{1\over 3}\Theta^{S}+\Theta_{11}^T & \Theta_{21}^T & \Theta_1^V \\
\Theta_{02}^V & \Theta_{21}^T & {1\over 3}\Theta-{1\over 3}\Theta^S-\Theta_{11}^T & \Theta_2^V\\
\Theta_D & \Theta_1^V & \Theta_2^V & {1\over 3}\Theta+{2\over 3}\Theta^{S}
\end{array} \right)\,,
\end{equation}
with $\Theta_{03}^V=\Theta_{3}^V=\Theta_{3i}^T=\Theta_{i3}^T=0$ for $i=1,2,3$ and $\Theta^T_{11}+\Theta^T_{22}=0$.
There appear to be four independent vector components 
$(\Theta_1^V,\Theta_2^V,\Theta_{01}^V,\Theta_{02}^V)$ and two tensor
components $(\Theta_{11}^T,\Theta_{21}^T)$. However, two of the vector
components are related to the other two by stress-energy conservation,
and hence there are also just two independent components.  

These vector and tensor sources can be identified with their respective contributions to the angular power spectrum $C_l=C_l^S+C_l^V+C_l^T$ which are defined in analogy to (\ref{clscalar}) by
\begin{equation}
C_l^{I}={2\over\pi}\int k^2dk \langle \Delta^{I}_l(k,\tau_0)\Delta^{I*}_l(k,\tau_0) \rangle\,,
\label{clall}
\end{equation}
where the index $I$ runs over $I=S,V,T$ corresponding to the scalar, vector, tensor contributions. The scalar photon distribution function is given in (\ref{scalarsource}), with the vector and tensor versions defined by
\begin{eqnarray}
&\Delta^{V} _l(k,\tau_0)=\displaystyle{\int_0^{\tau_0}d\tau\left(V^{1}_T(k,\tau)j^{11}_l\left[k(\tau_0-\tau)\right]
+V^{2}_T(k,\tau)j^{21}_l\left[k(\tau_0-\tau)\right]\right)}\,, &\cr &
\Delta^{T}_l(k,\tau_0)=\displaystyle{\int_0^{\tau_0}d\tau\left(T^{2}_T(k,\tau)j^{22}\left[k(\tau_0-\tau)\right]\right)}\,, &
\label{allsource}
\end{eqnarray}
where $V^{1}_T$, $V^{2}_T$, $T^{2}_T$ are the source functions for the vector and tensor perturbations and
\begin{equation}
j_l^{11}(x)=\sqrt{l(l+1)\over 2}{j_l(x)\over x}\,,\quad j_l^{21}(x)=\sqrt{3l(l+1)\over 2}\left({j_l(x)\over x}\right)^{\prime}\,,\quad j_l^{22}(x)=\sqrt{{3\over 8}{(l+2)!\over (l-2)!}}{j_l(x)\over x^2}\,.
\end{equation}

The Boltzmann equations for the scalar and tensor perturbations are
well studied both analytically and numerically, but the same is not
true for vector perturbations. Here, we include the Boltzmann
equations for the vector perturbations, along with a discussion of
their salient features. For a more in depth study of scalar, vector
and tensor perturbations, the reader is referred to ref.\cite{HW}
which also includes all the other relevant Boltzmann equations. 

We split up the photon distribution function for the vector sources into its angular multipole moments 
\begin{equation}
\Delta^{V}(k,\tau_0,\mu)=\sum_{l=1}^{\infty}(-i)^l(2l+1)P_l(\mu)\Delta^V_l(k,\tau_0)\,,
\end{equation}
where $\mu=\cos\theta$ is the angular variable. With this decomposition the Boltzmann equations become
\begin{equation}
\dot\Delta_{l}^V=k\left({\sqrt{l^2-1}\over 2l+1}\Delta^V_{l-1}-{\sqrt{l^2+2l}\over 2l+1}\Delta^V_{l+1}\right)-\dot\kappa\Delta^V_l+S_l
\end{equation}
for $l>0$, where $\dot\kappa$ is the differential cross-section due to Thomson scattering,
\begin{equation}
S_1={1\over 3}\dot\kappa\left(V_{B}+\dot V\right)\,,\quad S_2 ={1\over 5}P^V= {1\over 10}\dot\kappa\left(\Delta^V_2-\sqrt{6}E^V_2\right)\,,
\end{equation}
and $S_l=0$ for $l>2$. In the above expression, $E^V_2$ is the quadrapole of the electric component of the photon polarization distribution and $V$ is the vector metric perturbation which satisfies
\begin{equation}
\dot V + {2\dot a\over a}V = -{64\pi G\over 25\sqrt{3}k}\left(\rho_{\gamma}\Delta^V_2+\rho_{\nu}N^V_2\right)-{8\pi G\over k}\Theta^V\,,
\end{equation}
where $\Theta^V$ is the vector source, either $\Theta^V_1$ or $\Theta^V_2$, $\rho_{\gamma}$ is the average photon density, $\rho_{\nu}$ is equivalent quantity for neutrinos and 
$N^V_2$ is the quadrupole component of the neutrino distribution. $V_{B}$ is the vector (vortical) component  of the baryon velocity which satisfies 
\begin{equation}
\dot V_{B}=\dot V -{\dot a\over a}\left(V_{B}-V\right) + {\dot\kappa\over R}\left(3\Delta^V_1-V_B\right)\,,
\end{equation}
where $R=3\rho_{b}/4\rho_{\gamma}$ and $\rho_{b}$ is the average baryon density.
As already discussed the standard approach is to solve these linear
differential equations equations for $l<l_{\rm max}$, plus the equivalent equations for
the polarization distribution functions, which we have not included
here for brevity.  
The line of sight integration method which we use only
requires $l_{\rm max}\approx 7$, hence reducing the amount of CPU time
required. However, we need one final ingredient, the source
functions for vector perturbations, which are given by 
\begin{eqnarray}
V^1_T=\dot\kappa e^{-\kappa}\left( V_{B} - V\right)\,,\cr
V^2_T=\dot\kappa e^{-\kappa} P^V + {1\over\sqrt{3}}e^{-\kappa}kV\,,
\end{eqnarray}
where 
\begin{equation}
\kappa(\tau)=\int_{\tau}^{\tau_0}\dot\kappa(\tau^{\prime}) d\tau^{\prime}\,,
\end{equation}
is the optical depth due to Thomson scattering. We should note that
exactly analogous expressions exist for both the electric and magnetic
components of the polarization. These are also presented in
ref.\cite{HW}. 

The important thing to notice is that vector perturbations are not
created at any significant level in the absence of a source. This can
be seen from the Einstein equation: if $\Theta_V=0$ then $V\propto
a^{-2}$ since the effects of $\Delta^V_2$ and $N^V_2$ are
negligible. Hence, they are not present in inflationary models, but
are generic in any active model. 

\subsection{Incoherent sources}

In the previous few sections we have studied how the linearized
Einstein-Boltzmann equations can be modified to include active sources
which create scalar, vector and tensor anisotropies and we have shown
how it is possible to relate to the measured quantities, such as  the
angular power spectrum, when the sources are coherent. Such sources
are, however, very unnatural, generically relying on assumptions such
as pure spherical symmetry to maintain coherence. Therefore, we are
forced to consider methods for calculating the ensemble average for
theories which include incoherent sources. 

This subject was studied in a series of works which first discussed
the equally unnatural, but probably more realistic, totally incoherent
limit \cite{ACFM,MACFa} and then multi-parameter models which allowed
for a gradual relaxation from total coherence to total decoherence
\cite{MACFb}. One key advantage that the current work has over, for
example, ref.\cite{MACFb} 
is that once a given source model is chosen there is no uncertainty in
relative normalization of the anisotropy created by acoustic 
oscillations at the time of last scattering and that created much
later by the ISW effect.  The methods used in the earlier work
suffered from large uncertainties in this aspect of the
calculation. We shall see that decoherence is actually prevalent in
most active models and the contribution to the anisotropy from the surface
of last scattering does not, in general, give rise to noticeable
acoustic type signatures in the angular power spectrum. We also find
a suppression of the anisotropy created at last scattering
with respect to the ISW effect, not anticipated in
refs.\cite{ACFM,MACFa,MACFb}. 

The main problem in dealing with decoherence is that the unequal time
correlators (UETC) do not factorize, that is,  
\begin{equation}
\label{incohfactorize}
\langle P(k,\tau_1)P^*(k,\tau_2)\rangle\ne \langle P(k,\tau_1)P^*(k,\tau_1) \rangle^{1/2}
\langle P(k,\tau_2)P^*(k,\tau_2) \rangle^{1/2}\,,
\end{equation}
for some arbitrary quantity $P(k,\tau)$ and $\tau_1\ne\tau_2$. There
are two methods which one can use to overcome this problem. The first
\cite{Ta,Tb}, used in ref.\cite{PSelTa}, is to treat the UETC,
evaluated at the discrete times used in the linear Einstein-Boltzmann
solver, as a matrix which is symmetric and hence diagonalizable. This
diagonalization yields a change of basis and the source can then be
written as a sum of coherent sources, 
\begin{equation} 
\langle P(k,\tau_1)P^*(k,\tau_2)\rangle=\sum_i\lambda_i v_i(k,\tau_1)v^*_i(k,\tau_2)\,,
\end{equation}
where the $\lambda_i$ are the eigenvectors, ordered such that 
$\lambda_1>\lambda_2>..>\lambda_n$, and $v_i(k,\tau)$ are the
orthogonal, coherent basis functions. Since everything is linear, one
can use each of these basis functions, as a source in the modified
version of CMBFAST and the sum of the individual angular power spectra
yields the total. Although each of these coherent sources is
has a degree of acausality, it appears to give good
convergence from using only the ten largest eigenvalues \cite{PSelTa},
which indicates that the full calculation is well behaved.

The second method \cite{ACFM}, which is used in ref.\cite{ABRa} and
the current work, is to not work directly with the UETC. Instead, we
create an ensemble of source histories which has the same two-point
correlation statistics as the required UETC. The exact process for
doing this is discussed in the next section, but once it is done, the
ensemble average can be estimated by averaging the angular power
spectra from many individual source histories. Since our
ensemble is finite, one can also calculate its standard deviation to
give an idea of how accurate the calculated average is. The results
presented in this work used either 100 or 400 source histories to give
small statistical errors. However, it was possible to gain a
qualitative feel from as little as 40 source histories. 

This method is clearly more computationally intensive than the
eigenvalue decomposition requiring CMBFAST to be run over 100 times as
opposed to about 10. We improved the turnaround speed by  doing the
calculations in parallel and 100 realizations took about an hour on 16$\times$R10000 processors of the new SGI  Origin
2000 owned the UK Computational Cosmology Consortium. Setting aside
the apparent computational inefficiency\footnote{Obviously there will
be some overhead in doing the eigenvalue decomposition for any
particular UETC and also in creating the UETC from 
simulations. Therefore, we are not quite comparing like with like.}, we
believe that this method is more physically transparent since each of
the source histories will be causal and at the very 
least provides a useful check on the diagonalization method.

For these incoherent source histories the nature of the anisotropic
stress sources will be seen to be important on super-horizon scale,
since it is implicitly
linked to the vector and tensor sources. As
mentioned earlier, in the special case of the coherent limit $\langle
\Theta^S\Theta^{S*}\rangle \sim k^4$, which is not true in
general. Using similar arguments, it was shown in ref.\cite{PSelTc},
that  
\begin{equation}
\label{superhor}
\langle|\Theta^S|^2\rangle : \langle|\Theta^V_i|^2\rangle : \langle |\Theta^T_{ij}|^2\rangle = 3:2:4\,,
\end{equation}
where all indices are summed. The reason for this is that each of the
components is linked via the anisotropic part of the space-space
component of the source stress-energy tensor. Using a simple model, it
was deduced that  
\begin{equation}
C_l^{S} : C_l^V : C_l^T = 1.46 : 1 : 0.29\,,
\end{equation}
at around $l=10$. While it is true that, the vector and tensor
contributions to the angular power spectrum are likely to be similar
to the scalar contributions 
on large angular scales, there is no general formula or constraint for
this ratio and indeed, the model we present in the subsequent sections
will be seen to have a larger scalar component than that of
refs.\cite{PSelTa,PSelTc}. 
We should note that the incoherent case, where $\langle
\Theta^S\Theta^S\rangle \sim k^0$, the formation of anisotropies along
the line of sight, that is, the ISW effect, will be influenced by super-horizon correlations, in a way not
possible in a coherent model.  

\section{The standard scaling model}

\subsection{Modeling the source histories}
\label{scalingmodel}
Here, we present the model for defect two-point functions based on
a description of scaling cosmic strings. 
First, we set out the motivation for the model, then
we give the mathematical details, explaining
each of the parameters. Finally, we show some two-point
functions calculated within the framework of the model, demonstrating
an acceptable level of agreement between these two-point functions and those
measured in simulations. 

The starting point for our model 
is ref.\cite{VHS}.
In this work, measurements  of the string two-point correlation
functions in Minkowski space simulations of network evolution were made and 
a strikingly simple analytic model was put forward, capable
of reproducing the important features of these two-point functions
with good accuracy. The basic assumption of this model is that
a string network
can be 
represented as a collection of randomly oriented straight segments,
each of length $\xi t$, where $t$ is the physical time and $\xi$ is a
constant parameterizing the coherence length of the string.
To model the motion of the strings, 
each of these segments is assigned a randomly oriented velocity whose 
magnitude is chosen from a Gaussian distribution with zero mean and
standard deviation $v$. 

Under these assumptions, 
an analytic expression can be derived for the string energy two-point
correlation function as follows: the Fourier transformed stress
energy tensor of a string in Minkowski space is given by
\begin{equation}
\Theta_{\mu\nu}({\bf k},t)=\mu\int d \sigma
\left( \dot{\bf X}_\mu\dot{\bf X}_\nu -  
{\bf X}'_\mu {\bf X}'_\nu \right) e^{i {\bf k}.{\bf X}(\sigma,t)}\,,
\end{equation}
and hence the energy two-point correlator is given by 
\begin{equation}
\langle\Theta_{00}(k,t_1)\Theta^*_{00}(k,t_2)\rangle
=
\langle \int d\sigma_1 d\sigma_2  e^{i {\bf k}.
\left({\bf X}(\sigma_1,t_1) -{\bf X}(\sigma_2,t_2) \right)} \rangle \,.
\end{equation}
If one now assumes that the quantity $\left({\bf X}(\sigma_1,t_1)
 -{\bf X}(\sigma_2,t_2)\right)$ is Gaussianly distributed, with
mean zero and variance $\Gamma$, then
 it follows that
\begin{equation}
\label{gra}
\langle\Theta_{00}(k,t_1)\Theta^*_{00}(k,t_2)\rangle
=
\frac{1}{2} \int d\sigma_+ d\sigma_-  e^{-\frac {1}{6} k^2 \Gamma
(\sigma_-,t_1,t_2)}\,,
\end{equation}
where $\sigma_+=\sigma_1+\sigma_2$ and $\sigma_-=\sigma_1-\sigma_2$.
To estimate the variance $\Gamma$, 
one makes use of the idea that on
scales smaller than the correlation length $\xi t$ the string network
resembles a collection of straight line segments with velocity $v$,
which implies that
\begin{equation}
\label{eq-gamma1}
\Gamma(\sigma_-,t_1,t_2) = (1-v^2) \sigma_-^2 + v^2 (t_1^2-t_2^2)^2\,,
\end{equation}
for $|\sigma_-| < \xi t_1 / 2$ and on scales larger than $\xi t$
there are no correlations, so that
\begin{equation}
\label{eq-gamma2}
\Gamma(\sigma_-,t_1,t_2) = \infty\,,
\end{equation}
for $|\sigma_-| > \xi t_1 / 2$. 
In this picture, the length and number density of
string segments does not change, so that scaling behaviour
will have to be imposed on the correlators later by hand. 
Substituting for $\Gamma$ in (\ref{gra}), performing the integration over $\sigma_-$, and
using $\frac{1}{2}\int d\sigma_+ = V \xi^{-2}$ (assuming a length of
string $V \xi^{-2}$ per simulation volume $V$), one finds that 
\begin{equation}
\label{eq-equal-time}
\langle\Theta_{00}(k,t_1)\Theta^*_{00}(k,t_2)\rangle
 = \frac{V}{\sqrt{1-v^2}}\frac{1} 
{\xi t_1} 
\frac{2 \sqrt{6}}{k \xi t_1} {\rm{erf}} \left( \frac{k \xi t_1}
{2 \sqrt{6}}\right) e^{-\frac {1}{6} v^2 k^2 (t_1-t_2)^2}\,,
\end{equation}
where ${\rm{erf}} (x)$ is the error function,
\begin{equation}
{\rm{erf}} (x) = \int_{0}^{x} dx' e^{-x'^2}.
\end{equation}

The expression for the energy two-point correlator in 
(\ref{eq-equal-time}) has the wrong scaling behaviour, since changes in
string segment length and density have not been incorporated. 
It is possible introduce the right scaling behaviour into this
model by hand using 
\begin{equation}
\langle\Theta_{00}(k,t_1)\Theta^*_{00}(k,t_2)\rangle_{\rm sca}
= \frac {1}{\sqrt{ \xi t_1 \xi t_2}} \sqrt { P (k,t_1) P
(k,t_2) }   e^{-\frac {1}{6} v^2 k^2 (t_1-t_2)^2}\,,
\end{equation}
where $P(k,t)$ is $\xi t$ times 
the equal time energy correlator from 
(\ref{eq-equal-time}), that is,
\begin{equation}
P^\rho(k,t)=\xi t \langle\Theta_{00}(k,t)\Theta^*_{00}(k,t)\rangle\,.
\end{equation} 
Similar expressions can be derived for the all the other equal and
unequal time correlators.

This model has a number of shortcomings,
primarily because certain simplifying assumptions have been made in
order to make it possible to derive analytic expressions for the
two-point functions. By contrast, we need not work with simple analytic
forms, since we do not work with
the unequal time correlators directly. Instead, we
use numerical techniques to generate
histories for the source functions with the correct two-point
statistics.  
This has made it possible to improve and extend the 
model in a number of ways.  In particular, we include an improved treatment of
causality and scaling, and extract a different set of
components of the string
stress-energy. 
We briefly sketch these differences, before embarking on a
detailed mathematical description of our model.

\begin{itemize}
\item {\em Causality:}
One problem with the analytic model is that it does not
fully respect the constraints imposed by causality, which require that
there can be no correlations between source components at
space-time points whose past light cones do not intersect. In particular,
by assuming that the lag ${\bf X}(\sigma_1)-{\bf
X}(\sigma_2)$
between two-points on a
string segment is Gaussianly distributed, one assigns a non zero value to the 
probability of correlations existing on scales larger than the
causal horizon, making the model manifestly 
acausal. For this reason the oscillations, which should appear
generically in the two-point functions of causal theories
\cite{Ta}, are not present in the analytic expressions for the 
correlators.  
By contrast, our model is causal by design, because
we do not assume that this lag is Gaussianly distributed.

\item{\em Scaling behaviour:}
In ref. \cite{VHS}, the unequal time correlation functions were
simply multiplied by appropriate factors by hand in order to enforce the
correct scaling behaviour. 
We extend the model
by making the scaling form of the unequal time
correlators arise in a natural way,
as a consequence of the decay of the string
segments. It will be seen that a different super-horizon form for the
unequal time correlators arises as a result of this assumption.

\item{\em Choice of stress-energy components:}
Although we make use of the same basic picture of the string
network in order to calculate the two-point functions, we extract a
different set of components of the string stress-energy
tensor, calculating the others to maintain stress-energy conservation.
We extract the energy $\Theta_{00}$ and
the anisotropic stress $\Theta^S$, as opposed to $\Theta_D$, and since we also include tensor and
vector contributions in our calculation, we also compute vector and
tensor source components $\Theta^V$ and $\Theta^T$.
As discussed in ref. \cite{PSelTa}, this
particular choice of scalar components is very natural, as the remaining two
components of the
stress-energy tensor are found to be well behaved on integration of the
conservation equations, which is not necessarily true for the choice $\Theta_{00}$
and $\Theta_D$.
Also, taken in conjunction with $\Theta^V$
and $\Theta^T$, these components specify the super-horizon
perturbations in the most direct manner.

\end{itemize}

Having outlined the main differences between our model and that
described in ref. \cite{VHS}, we now proceed to set out the 
specific mathematical details.
For a general network of strings in an expanding universe, the stress
energy tensor has the form
\begin{equation}
\Theta_{\mu\nu}({\bf x},\tau)=\mu\int d \sigma
\left( \epsilon \dot{\bf X}_\mu\dot{\bf X}_\nu - \epsilon^{-1}  
{\bf X}'_\mu {\bf X}'_\nu \right)
\delta^{(3)}\left( {\bf x} - {\bf X}(\sigma,\tau)\right)\,,
\end{equation}
where $\mu$ is the string mass per
unit length, ${\bf X}$ are the coordinates of the string world
sheet, parameterized by conformal time $\tau$ and spatial variable
$\sigma$,
dot and prime represent differentiation with respect to $\tau$ and
$\sigma$ respectively, $\epsilon^2={\bf X}'^2/(1-\dot{\bf
X}^2)$, and $\delta^{(3)}({\bf x})$
denotes the Dirac delta function in three
dimensions.  
Note that now we are working in an expanding universe, each of the string segments will have size $\xi\tau$.
We are interested in histories for the Fourier transform of the string
stress-energy tensor, which is defined via 
\begin{equation}
\Theta_{\mu\nu}({\bf k},\tau)= \int d^3{\bf x}
e^{i{\bf k}.{\bf x}} \Theta_{\mu\nu}({\bf x},\tau)=
\mu\int d \sigma e^{i{\bf k}.{\bf X(\sigma,\tau)}} 
\left( \epsilon \dot{\bf X}^\mu\dot{\bf X}^\nu - \epsilon^{-1}  
{\bf X}'^\mu {\bf X}'^\nu \right)\,.
\end{equation}

Our conceptual `string network' consists
of a collection of straight line segments, each 
with an individual label $m$, which   
`decay' in a smooth
way, completely vanishing by some final time $\tau_f^m$. A history for the
evolution of the complete string stress-energy tensor is then
written as a sum of the histories for the stress-energy tensors of the
individual segments,
\begin{eqnarray} 
\Theta_{\mu\nu} ( {\bf k} ,\tau ) &=&
\sum_{m} \Theta^m_{\mu\nu}  ( {\bf k} ,\tau ) T^{\rm off}
\left(\tau,\tau_f,L_f
\right)  T^{\rm on} \left(\tau,\tau_f,L_i^1,L_i^2
\right)\,.
\end{eqnarray} 
The function $T^{\rm off}$ is a smooth segment decay function, chosen so that 
the segment starts to disappear at $L_f \tau^m_f$, and has
disappeared completely at $\tau^m_f$, with the additional features
that the stress-energy and its time derivative are continuous at
$L_f \tau^m_f$ and  $\tau^m_f$, which are necessary in order for the ODE solver to function properly.
With these properties in mind, we
chose the following form for $T^{\rm off}$:
\begin{equation}
\label{eqn_toff}
T^{\rm off}(\tau,\tau_f,L_f) = \left\{ \begin{array} {ll}
         1  & \ldots \tau < L_f\tau_f \\
         \frac{1}{2}+\frac{1}{4}(x^3-3x) & \ldots L_f \tau_f< \tau < \tau_f \\
         0  & \ldots \tau > \tau_f 
         \end{array} \right.
\end{equation}
where 
\begin{equation}
x=2 \frac{\ln (L_f \tau_f/\tau)} {\ln (L_f)} -1\,.
\end{equation}
Similarly,  $T^{\rm on}$ is a smooth segment appearance function, with
almost identical properties to  $T^{\rm off}$ except that it represents the 
smooth turning on of the segment at early times. By analogy to $T^{\rm off}$, we chose
\begin{equation}
T^{\rm on}(\tau,\tau_f,L_i^1,L_i^2) = \left\{ \begin{array} {ll}
         0  & \ldots \tau < L_i^1\tau_f \\
         \frac{1}{2}+\frac{1}{4}(3y-y^3) & \ldots L_i^1\tau_f < \tau <
L_i^2 \tau_f \\
         1  & \ldots \tau >L_i^2 \tau_f 
         \end{array} \right.
\end{equation}
where 
\begin{equation}
y=2 \frac{\ln (L_i^1 \tau_f/\tau)} {\ln (L_i^1/L_i^2)} -1\,.
\end{equation}
This function is only included for computational
efficiency, since it is possible to ignore any particular
string
segment at times earlier than $L_i^1 \tau_f$, provided $L_i^1$ and
$L_i^2$ are sufficiently small. We checked this by varying
the values of $L_i^1$ and $L_i^2$ and found that these variations
make very little
difference to the total stress-energy provided the values are small
enough.
This is 
because at any time the stress-energy tensor is dominated by strings
whose decay times lie in the near future. We choose values for
$L_i^1$ and $L_i^2$ which are small enough that
results are not changed by any further decrease.

During the generation of a particular string history, it is not
practical to keep track of every piece of string in our conceptual
simulation volume. This is because the number density of strings $n(\tau)$
scales like $\tau^{-3}$, so that to have of order one
string segment remaining by the final simulation time $\tau_0$, the number of
strings we would need to follow from the initial simulation time
$\tau_i$ would be of order $(\tau_0/\tau_i)^3$. In the case of a
mode tracked from before radiation-matter equality to the present day,
this would require us to follow of order $10^{12}$ strings.
Instead, since
the 
\begin{equation}
N_f=V \left[n (\tau_f) - n (\tau_f + d\tau_f) \right]
\end{equation} 
strings decaying between times $\tau_f$ and
$\tau_f+d\tau_f$ in our conceptual `simulation'
volume $V$ are randomly located, we can replace them by a
single string, whose amplitude is multiplied by $N_f^{1/2}$; the
power of $1/2$ coming from the fact that
random locations in real space correspond to random
phases in Fourier space, so that the amplitude of the Fourier
transform of a number of such
segments sums as a random walk for all $k\neq 0$.

The equation for a single source history then becomes
\begin{equation} 
\label{eqn-singsource}
\Theta_{\mu\nu} ( {\bf k} ,\tau )=
V^{1/2} \sum_{m} \left[n (\tau_f^{m-1}) -
n (\tau^m_f) \right]^{1/2} 
 \Theta^m_{\mu\nu}  ( {\bf k} ,\tau ) T^{\rm off}
\left(\tau,\tau_f,L_f
\right)  T^{\rm on} \left(\tau,\tau_f,L_i^1,L_i^2
\right)\,.
\end{equation} 
For each source history, we use $N_s$ individual string segments, with
values of $\tau_f$ equally spaced on a logarithmic scale 
between $\tau_i$ and
$F_{\rm max} \tau_0$, where $F_{\rm max}\tau_0$ must be later
than the 
final simulation time $\tau_0$ in order that all strings inducing
significant perturbations at time $\tau_0$ are included\footnote{The
effective total number of strings at any time is given by $ \sum
 \left[n (\tau_f^{m-1}) -
n (\tau^m_f) \right] T^{\rm off}
\left(\tau,\tau_f,L_f
\right)  T^{\rm on} \left(\tau,\tau_f,L_i^1,L_i^2
\right)$, and the normalization of $n(\tau)$ is chosen to ensure that
this quantity is equal to $(\xi\tau)^{-3}$.}.

The Fourier transform for each individual string segment is given by
\begin{eqnarray} 
\Theta_{\mu\nu} ^m ({\bf k},\tau)
 &=& \mu \int _{-\xi \tau/2}^{\xi \tau/2} d\sigma
e^{i {\bf k} . {\bf X_m}} \left( \epsilon \dot{X}_m^\mu 
 \dot{X}_m^\nu -  \epsilon ^{-1} {X'}_m^\mu 
 {X'}_m^\nu \right)\,,
\end{eqnarray}
where $\xi \tau$ is the length of the string segment at time $\tau$
and ${\bf X}_m(\sigma,\tau)$ are the coordinates of the string world sheet,
given by
\begin{eqnarray} 
 {\bf X_m}&=& 
 {\bf x_m} + \sigma{\bf \hat{X'}_m} + v_m \tau {\bf \hat{\dot{X}}_m}\,.
\end{eqnarray}
For each string segment, ${\bf x_m}$ is a random location (in
practice, we generate ${\bf k}.{\bf x_m}$ as a random number between
0 and $2\pi$), while ${\bf \hat{X'}_m}$ and ${\bf \hat{\dot{X}}_m}$ are
randomly oriented perpendicular unit vectors, such that, 
\begin{eqnarray} 
|{\bf \hat{X'}_m}| =| {\bf \hat{\dot{X}}_m}| &=& 1 \,,\\
{\bf \hat{X'}_m}. {\bf \hat{\dot{X}}_m} &=& 0\,.
\end{eqnarray}
The string velocity $v_m$ is a random number chosen from a Gaussian distribution with
mean zero and standard deviation $v$, truncated to prevent $|v_m|>1$.

Performing the integration over $\sigma$, and taking only the real
part,
we find that 
\begin{eqnarray} 
 \Theta^m_{00}&=& \frac{\mu}{\sqrt{1-v_m^2}}\xi \tau \rm{sinc} \left(
k \hat{X}'_3 \xi \tau /2 \right) \left( \cos ( {\bf k}.{\bf x_m})
\cos(k \hat{\dot{X}}_3 v_m \tau) + \sin ( {\bf k}.{\bf x_m})
\sin (k \hat{\dot{X}}_3 v_m \tau )
\right)\,,\\
\Theta^m_{ij}&=&\left[ v_m^2\hat{\dot{X}}_i\hat{\dot{X}}_j -
(1-v_m^2)\hat{{X'}}_i\hat{{X'}}_j \right] \Theta^m_{00}\,.
\end{eqnarray}
where ${\rm{sinc}}(x)=\sin(x)/x$ and the subscripts refer to the individual spatial components. 
For conciseness, we have now dropped the subscript $m$ on ${\bf X}$ here,
and in the following equations. 

As already noted there are two independent vector and tensor components of the stress-energy, which are sourcing the perturbations. However, each of these components will have the same two-point correlation statistics and hence we need only evolve one of 
each and multiply by the appropriate normalization once the power spectra are calculated. The anisotropic
stress, vector and tensor components are given in terms of the spatial
stresses $\Theta_{ij}$ by
\begin{eqnarray}
\Theta^S&=& (2\Theta_{33}-\Theta_{11}-\Theta_{22})/2\,,\\
\Theta^V&=&\Theta_1^V=\Theta_{13}\,,\\
\Theta^T&=&\Theta_{12}^T=\Theta_{12}\,.
\end{eqnarray} 
For each individual string segment, we find that 
\begin{eqnarray}
\Theta^S&=&\frac{1}{2}\left[ v_m^2\left( 3 \hat{\dot{X}}_3 \hat{\dot{X}}_3-1)\right) -
(1-v_m^2)\left( 3 {\hat{X}}'_3{\hat{X}}'_3-1)\right) \right] \Theta_{00} \,, \\
\Theta^V&=&\left[ v_m^2\hat{\dot{X}}_1 \hat{\dot{X}}_3 -
(1-v_m^2){\hat{X}}'_1{\hat{X}}'_3 \right] \Theta_{00} \,,\\
\Theta^T&=&\left[ v_m^2\hat{\dot{X}}_1 \hat{\dot{X}}_2 -
(1-v_m^2){\hat{X}}'_1{\hat{X}}'_2 \right] \Theta_{00}\,.
\end{eqnarray}
Integrating over the random orientation vectors, we find that for a
single string the super-horizon ratios are in agreement with
(\ref{superhor}) and 
since the total stress-energy tensor for the string network is just a
sum over the contributions from the individual segments, we
find that the super-horizon forms of the total stress-energy are also
in this ratio. However, we have already noted that the ratio of
$\Theta^S$ to $\Theta_{00}$ is not constrained in a similar way, and
is likely to be highly model dependent. For our model, we find that on
super-horizon scales,   
\begin{equation}
\langle |\Theta^S|^2\rangle : \langle |\Theta_{00}|^2 \rangle = 2-5v^2+15v^4 : 10\,,
\end{equation} 
if we assume that the velocities are Gaussianly distributed, rather than the truncated Gaussian which we use in practice. This limit, which has been used to make the problem analytically tractable, will be realized for small $v$.

Having worked out the energy and anisotropic stress, the remaining
scalar components follow by stress-energy conservation. By rearranging
these equations, we find the following differential equation for 
$\Theta_D$ in terms of $\Theta_{00}$ and $\Theta_{S}$
\begin{equation}
\label{thetaD-int}
\dot{\Theta}_{D}=- 2\frac{\dot{a}}{a} { \Theta_{D}}
-\frac{k^2}{3}\left(\frac{a}{\dot{a}}\left(\Theta_D-\dot{\Theta}_{00}\right)
-\Theta_{00}+2\Theta^S\right)= 0 \,,
\end{equation}
while $\Theta$ is just
\begin{equation}
\Theta=\frac{a}{\dot{a}} \left( \Theta_{D}-\dot{\Theta}_{00}
\right) -{\Theta}_{00}\,.
\end{equation}
In practice, we use the techniques described earlier to compute
histories  for
the components
$\Theta_{00}$, $\Theta^S$, $\Theta^V$ and $\Theta^T$. Values of each component
are stored for a set of times which are closely enough spaced
that a linear interpolation scheme can accurately reproduce the full
history for the function and its derivative.
These interpolated functions are then used as a set of driving terms
to the ODE solver in CMBFAST\cite{cmbfast}. In order to increase speed, the
evolution of $\Theta_D$ is only carried out 
for times satisfying $k \tau < x_{\rm max}$, where the parameter 
$x_{\rm max}$ is
chosen to be large enough that further increases do not affect the
results,  and for later times $\Theta_D$ is set to zero. This can be done
because in all 
the models we consider here perturbations in $\Theta_D$ are 
suppressed on scales much smaller than the horizon scale.
 
At this point we comment on the way in which stress-energy
conservation and
compensation are treated in our model. In constructing forms for
$\Theta_{00}$ and $\Theta^S$ we have only been thinking about the
behaviour of the long string, and not about the behaviour of the loops
and gravitational waves into which the long string decays. We have
ensured that stress-energy conservation is satisfied by only computing
two scalar components and using the conservation equations to work
out the other two. 

One way to treat the loops and gravity waves explicitly is to
consider the source $\Theta_{\mu\nu}$ to be the sum of two components,
a long string component $L_{\mu\nu}$ and a second fluid component
$S_{\mu\nu}$. We then model the rate at which energy and
momentum are being dumped 
from the long string into the second fluid, which in this case is loops and gravity waves, by introducing
two functions $g_0$ and $g_D$,
with $L_{\mu\nu}$ satisfying
\begin{eqnarray}
\dot{L}_{00} + \frac{\dot{a}}{a} \left( L_{00} + L \right)
-L_D  &=& -g_0 \,,\nonumber \\
\dot{L}_{D}+ 2\frac{\dot{a}}{a} { L_{D}}
+\frac{k^2}{3}\left(L+2L^S\right)&=& -g_D \,,
\end{eqnarray}
and $S_{\mu\nu}$ satisfying
\begin{eqnarray}
\dot{S}_{00} + \frac{\dot{a}}{a} \left( S_{00} +S \right)
-S_D  &=& g_0\nonumber \,,\\
\dot{S}_{D}+ 2\frac{\dot{a}}{a} { S_{D}}
+\frac{k^2}{3}\left(S+2S^S\right)&=& g_D \,.
\end{eqnarray}
Given a model for $L_{00}$, $L_D$, $L$ and $L^S$, such as the one described above, plus an equation of state for the second fluid, we can then compute the loop production functions and all the components of $S_{\mu\nu}$ and   
hence, we have the total, $\Theta_{\mu\nu}$. In the model described  here,
we have effectively done this by setting $S_{00}=S^S=0$. Although
this choice does not correspond to a particular, identifiable fluid, we
have found in studies of the CMB anisotropies created at the surface of last scattering,
that it gives results which are very similar to physical models for
$S_{\mu\nu}$, such as free-streaming massless particles. In
particular, the main conclusions of this paper and ref.\cite{ABRa}
will unchanged. However, more detailed Modeling will be required if
accurate predictions are required. The results of an in depth study of
this issue will be presented elsewhere \cite{ABRc}. 

\begin{figure}
\setlength{\unitlength}{1cm}
\begin{minipage}{8.0cm}
\leftline{\psfig{file=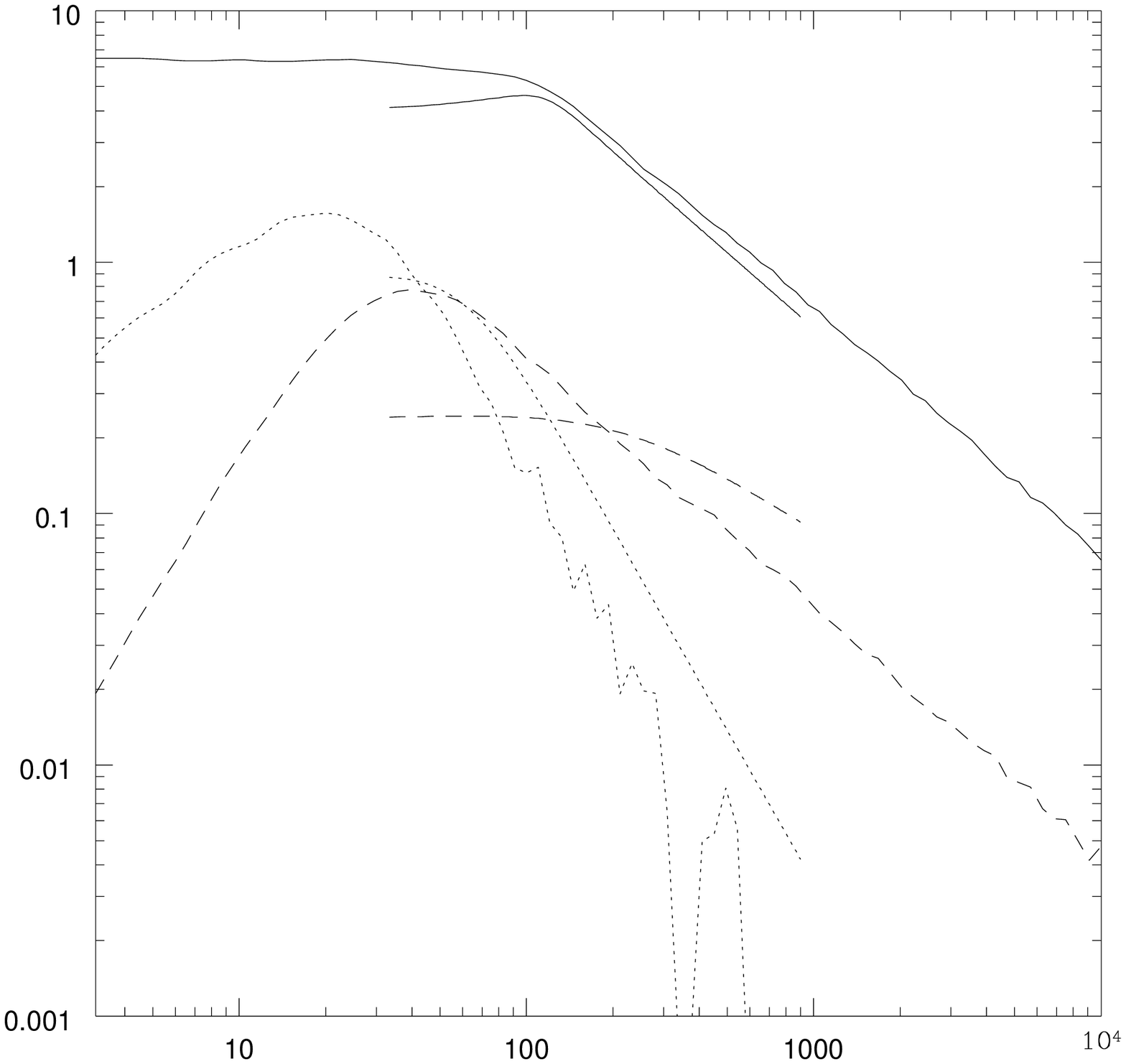,width=3.4in}}
\end{minipage}\hfill
\begin{minipage}{8.0cm}
\rightline{\psfig{file=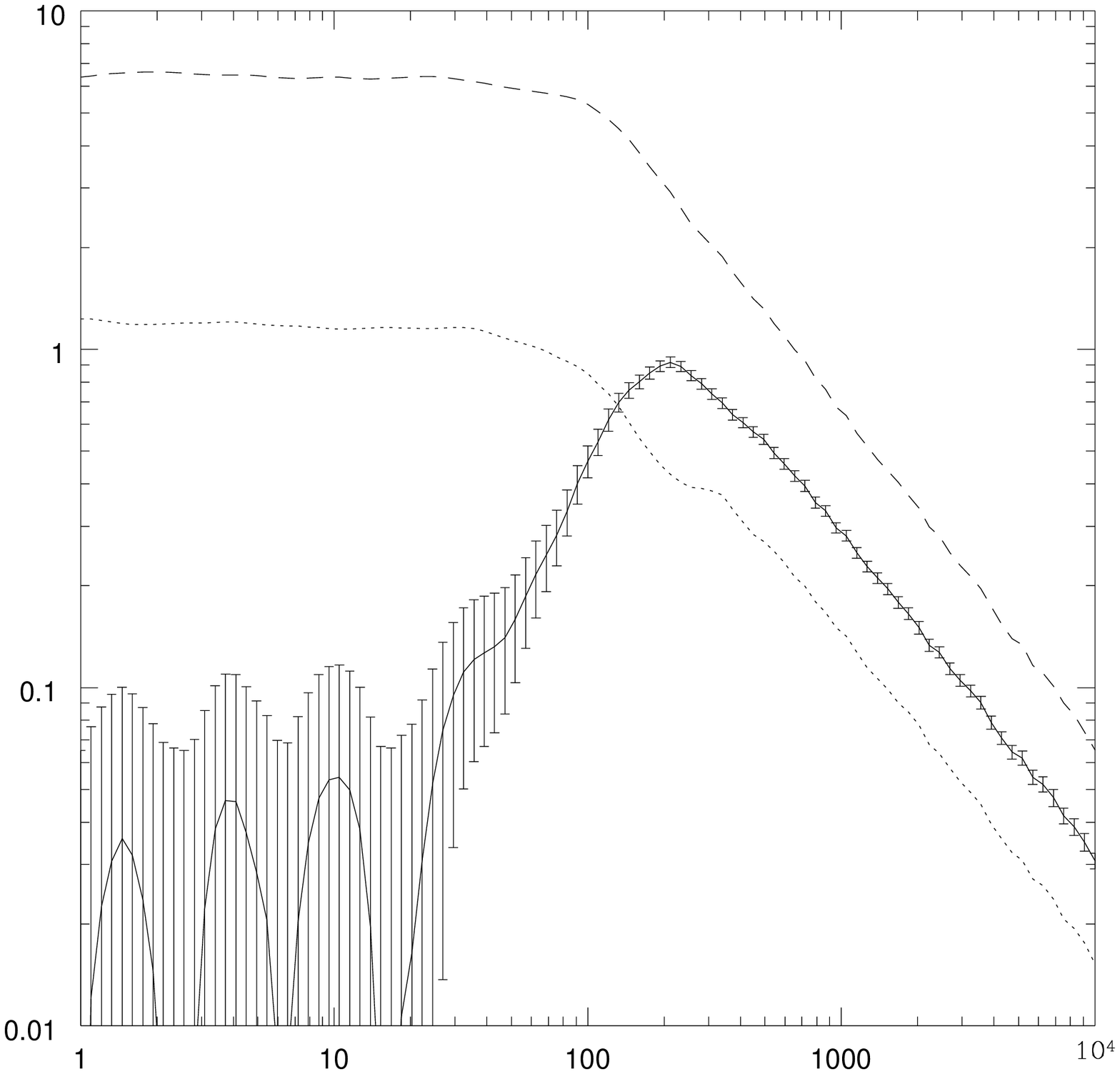,width=3.4in}}
\end{minipage}
\caption{The left hand graph shows 
equal time two-point correlation functions from our standard
scaling model: $\tau\langle\Theta_{00}(k,\tau)\Theta^*_{00}(k,\tau)\rangle$
(solid line), $\tau^3\langle\Theta_{D}(k,\tau)\Theta^*_{D}(k,\tau)\rangle/k^2$ (dashed line),
$\tau^2\langle\Theta_{00}(k,\tau)\Theta^*_{D}(k,\tau)\rangle/k$ (dotted line).
Two-point functions
measured in Minkowski space 
string simulations are shown for
comparison, using the same line types, but truncated to
illustrate the approximate range probed by the simulations. 
The right hand graph also shows
equal time two-point correlation functions from our standard
scaling model:
$\tau\langle\Theta_{00}(k,\tau)\Theta^*_{00}(k,\tau)\rangle$ (dash-line),
$\tau\langle\Theta^{S}(k,\tau)\Theta^{S*}(k,\tau)\rangle$ (dotted-line),  
$\tau\langle\Theta_{00}(k,\tau)\Theta^{S*}(k,\tau)\rangle$
(solid-line).  
On each graph, the
$x$-axis is
$\tau$ and $k$ is 0.1 ${\rm Mpc}^{-1}$.} 
\label{twotime}
\end{figure}

We now present a sample of two-point functions calculated using these
techniques. In the left hand graph of Fig. \ref{twotime} 
we show equal time two-point functions for $\Theta_{00}$ and $\Theta_{D}$, together with fitting functions 
for the same two-point functions as measured in the simulations of
ref. \cite{VHS}. The noisy curves are those computed in our model,
using 8000 realizations, while the smooth truncated curves are those
of ref.\cite{VHS}.
The two-point functions for $\Theta_D$ in this graph are obtained by 
integrating (\ref{thetaD-int}) for each history. In order to
make sensible comparisons between our expanding universe calculations and the Minkowski space simulations, we compare our conformal lengths
and times with their physical lengths and times.
 Firstly, it should be noted that the dynamic range
probed by the simulations is small, whereas within the framework of
the model the dynamic range can be extended arbitrarily. 
Within the range probed by the simulations, the 
model appears to give two-point functions in good agreement; the one exception
being the limiting behaviour of
the $\Theta_D$ self correlator. However, it should be noted that the
simulations only probe $\Theta_D$ for the long string, not the loops
and gravitational radiation which the long string spits off.
In fact, the fitting function for
the  $\Theta_D$ self correlator has a super-horizon form which is
inconsistent with causality and stress-energy conservation, since $\langle\Theta_D\Theta^*_D\rangle\sim k^2$ rather than $k^4$. Our model
on the other hand fully respects stress-energy conservation with $\langle\Theta_D\Theta^*_D\rangle\sim k^4$, so it is
not surprising that there is some level of discrepancy between the limiting
forms of the functions for this particular component. 

The exact forms of the two-point correlators within our
model depends on
the choice of string parameters $v$ and $\xi$. We find that optimal
agreement between our two-point functions and those of ref.\cite{VHS} is obtained when we input
values of $v$ and $\xi$ which are slightly different to those which are
actually measured in the simulations. For Fig. 
\ref{twotime} we use $v=0.35$, $\xi=0.15$. In this
respect, our model does slightly worse than that reported in ref.\cite{VHS}, which manages to achieve a miraculously good fit to the amplitude
and form of the energy equal time
cross correlator using exactly the values of the
parameters $v=0.6$ and $\xi=0.15$ which were measured in
simulations. In spite of this, the limiting behaviour of the two-point
functions has the correct form in our model and for some choice of
the parameters $v$ and $\xi$ we are able to obtain a good fit to the
correlators measured in the simulations.

\begin{figure}
\setlength{\unitlength}{1cm}
\begin{minipage}{8.0cm}
\leftline{\psfig{file=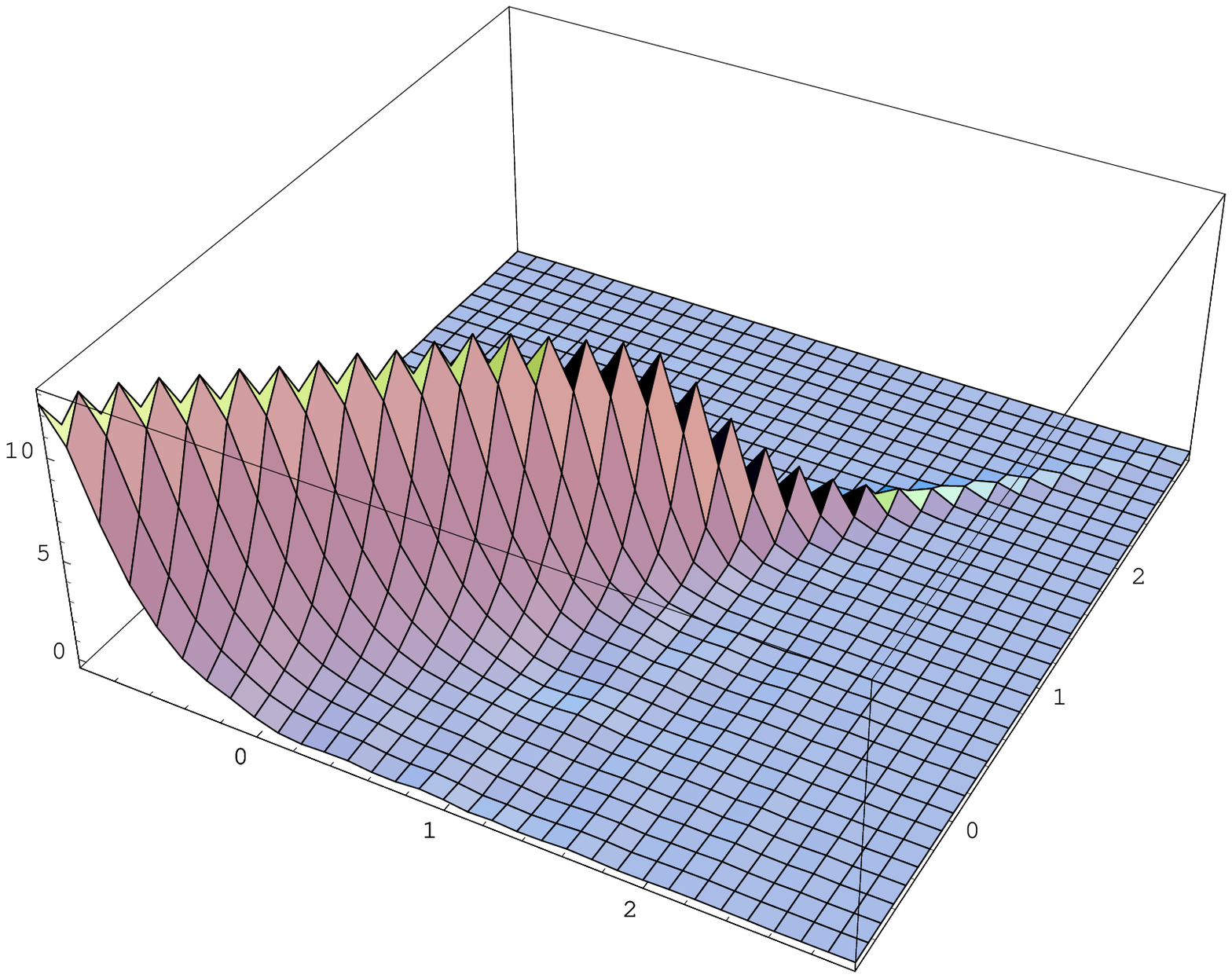,width=3.4in}}
\end{minipage}\hfill
\begin{minipage}{8.0cm}
\rightline{\psfig{file=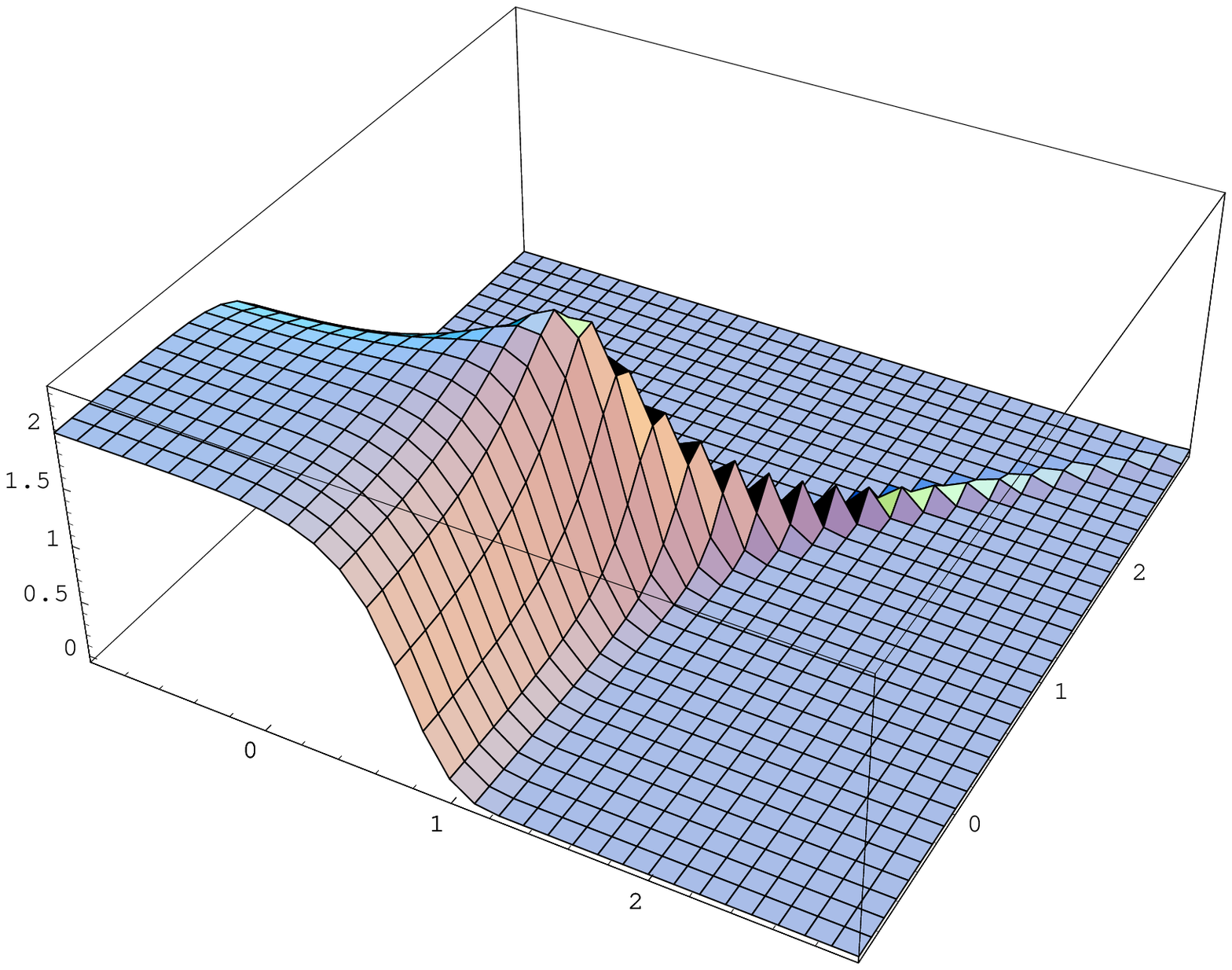,width=3.4in}}
\end{minipage}
\caption{The left hand graph shows the unequal time correlation function
$\langle\Theta_{00}(k,\tau_1)\Theta^*_{00}(k,\tau_2)\rangle$ from our standard
scaling model.
The z-axis is $
\left(\tau_1\tau_2\right)^{1/2}\langle\Theta_{00}\Theta^*_{00} 
\rangle$, the 
x-axis is  $\rm{Log}_{10} (k\tau_1)$ and the  y-axis is  
$\rm{Log}_{10} (k\tau_2)$. The right hand graph shows the same 
plot created using the model from ref.[30].}
\label{twotime-all}
\end{figure}

In the left hand graph of Fig. \ref{twotime-all} we show equal time two-point functions for
$\Theta_{00}$, $\Theta^S$ and their cross correlator, along with
$2\sigma$ error-bars, computed using
8000 realizations. We see that far outside the horizon 
the cross-correlator is
relatively noisy, but its behaviour is consistent with a power law of
$k^2$ everywhere except
inside the horizon, where it is of order the two self
correlators. In fact, it is easy to show analytically within the 
framework of the model that the cross
correlator must go like $k^2$ outside the horizon in the limit of a 
large number of realizations and this behaviour clearly manifests
itself in the range $\tau=20$ to $\tau=100$. 
We should note that the noisy behaviour of the cross correlator far
outside the horizon does
not appear to have too large an effect on the matter and CMB power spectra, for which the ensemble average has a relatively small variance even for only 40
realizations. 

In Fig. \ref{twotime-all} shows the unequal time correlation function
for the energy and the corresponding
plot from ref.\cite{VHS}. It can be seen that the
sub-horizon form of the unequal time correlators is similar in both
models. However,
we see that our unequal time
correlators have a 
distinctly different form on super-horizon scales. We quantify this
difference by using the function $U(k,\tau_1,\tau_2)$ defined in terms of the violation of the factorization relation (\ref{incohfactorize}) as
\begin{equation}
U(k,\tau_1,\tau_2) 
=\frac {\langle P(k,\tau_1) P^*(k,\tau_2) \rangle} 
{\langle P(k,\tau_1)P^*(k,\tau_1) \rangle^{1/2} \langle P(k,\tau_2)P^*(k,\tau_2)
\rangle^{1/2} }\,,
\end{equation}
for some arbitrary function $P(k,\tau)$, where $\tau_1$ and $\tau_2$ are the two times in question, with
$\tau_2>\tau_1$.
In our model, only those strings which are present both at $\tau_1$
and $\tau_2$ can contribute to the cross correlator and hence only those strings present at the later time $\tau_2$ can
contribute, implying that $\langle
P(k,\tau_1) P^*(k,\tau_2) \rangle \propto \langle P(k,\tau_2) P^*(k,\tau_2)
\rangle$.  Hence, we find 
\begin{equation}
U(k,\tau_1,\tau_2) 
\propto\sqrt{\frac {\langle P(k,\tau_2)P^*(k,\tau_2)\rangle}
      {\langle P(k,\tau_1)P^*(k,\tau_1) \rangle  }}\,,
\end{equation}
which outside the horizon gives
\begin{equation}
U(k,\tau_1,\tau_2)\propto \left(\frac{\tau_1}{\tau_2}\right) ^{1/2}\,.
\end{equation}
On the other hand in ref.\cite{VHS}, the super-horizon  fall-off of
the unequal time correlators is modelled as an
exponential decay, with
\begin{equation}
U(k,\tau_1,\tau_2) = e^{-(\tau_1-\tau_2)^2/\tau_c^2}
\end{equation}
where the coherence time $\tau_c$ grows like $k^{-1}$ outside the horizon.
This behaviour gives a good fit on the sub-horizon scales which their
simulations primarily probe. However, on super-horizon scales the
power-law fall-off evident in our model must eventually dominate. 

In summary, therefore, we have outlined methods for creating source histories 
based on a model with two parameters, the rms \footnote{As mentioned
earlier the distribution of strings has been truncated to prevent
strings moving faster than the speed of light. This prevents $v$ from
being exactly the rms value, the difference from the rms value being
minimized for small $v$.} speed of the strings $v$ and the persistence
length $\xi$, which are measured in simulations. 
In doing this we have been forced to introduce various `system'
parameters, to allow the problem to be solved in a finite time on a
discrete system, such as a computer. The value of each of these
parameters was chosen, so that further increases or decreases toward
the continuum value resulted in no change in the two-point
functions. In particular, for results presented in this paper, we used
 $x_{\rm max}=1000$, $L_i^1=0.001$, $L_i^2=0.01$, $N_S=200$
and $F_{\rm max}=10.0$. We have also introduced the parameter $L_f$,
quantifying the rate at which string segments are turned off. 
Unlike the systems parameters, $L_f$ clearly has some degree of
physical significance. However, in section \ref{mod-model}
we demonstrate that the
dependence of the results on the value of $L_f$ is relatively weak, and we choose
to use the value $L_f=0.5$ for the rest of our computations.

\subsection{Power spectra for the standard scaling model}

\begin{figure}
\setlength{\unitlength}{1cm}
\begin{minipage}{8.0cm}
\leftline{\psfig{file=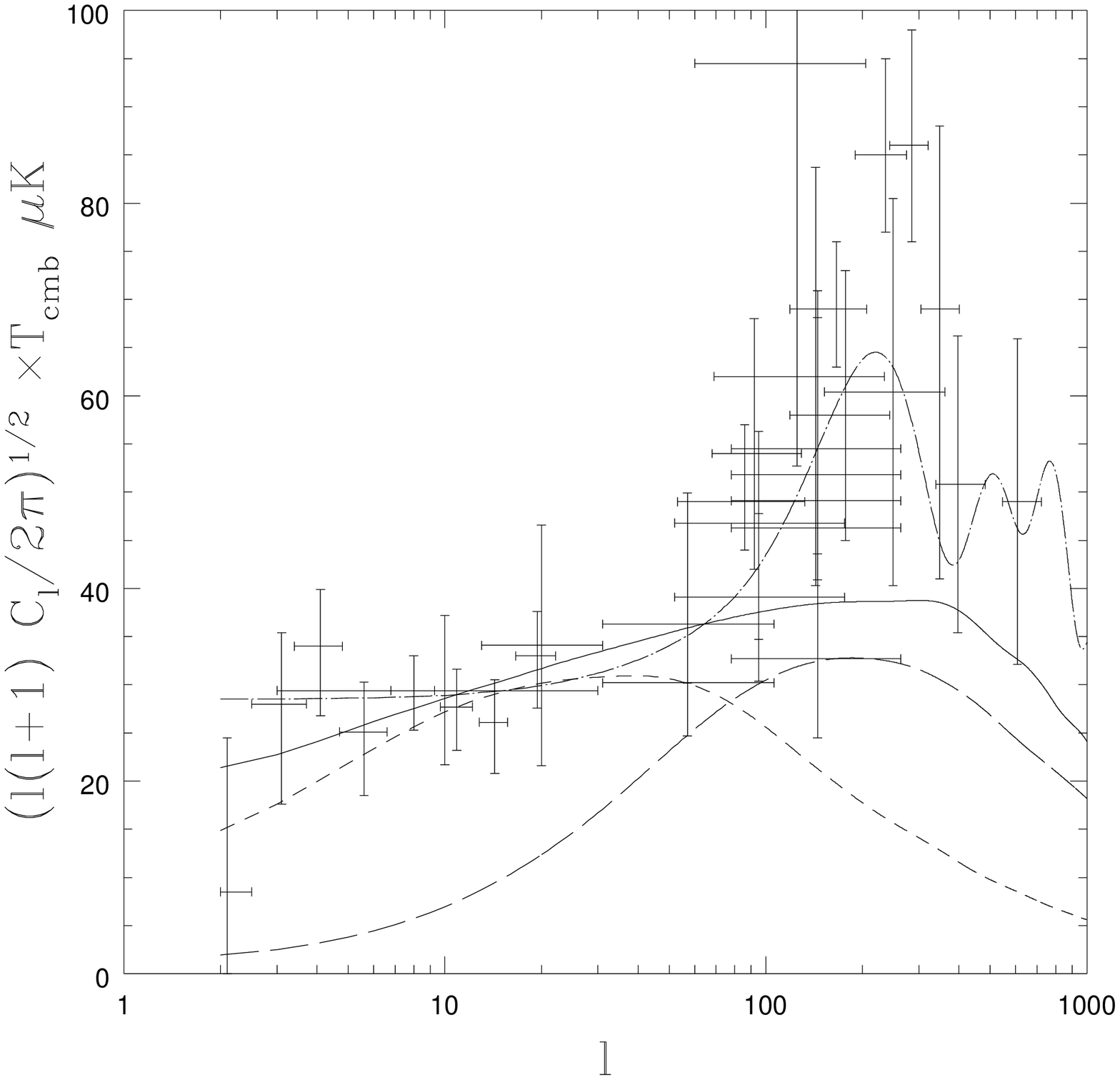,width=3.4in}}
\end{minipage}\hfill
\begin{minipage}{8.0cm}
\rightline{\psfig{file=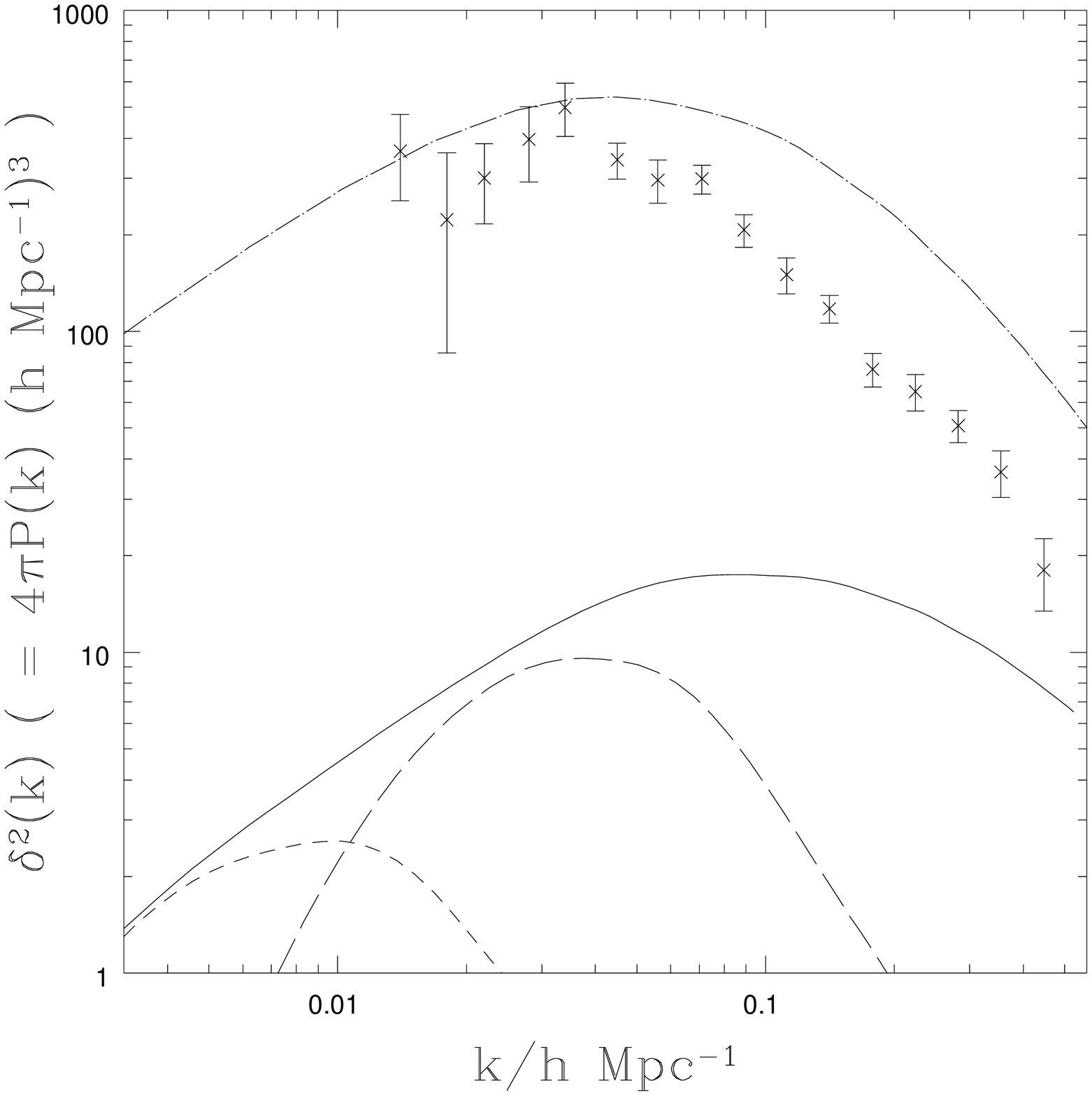,width=3.4in}}
\end{minipage}
\caption{The COBE normalized angular power spectrum of CMB
anisotropies, (left-hand graph) and the matter power spectrum
(right-hand graph) for the standard string model (solid curve).  The
contributions from the defects during two time windows ($1300 < z <100$
-- Long dashed, $100 < z < 1.6$ -- short dashed) are also
included for comparison.
Standard CDM (dot-dash curve) and observational data (data points) are
included for comparison.} 
\label{standard}
\end{figure}

We define the standard scaling model to be one which uses the above
two-point functions with the model parameters $v=0.65$ and $\xi=0.3$,
measured in expanding universe simulations \footnote{Although, note
the earlier comment, that we find better agreement with the two-point
functions measured 
in flat space simulations for slightly different
values of $v$ an $\xi$ when we use our causal, stress-energy
conserving model. We have decided to use the calculated values from
expanding universe simulations as our standard since they are likely
to be more relevant for our model.} and an assumption of perfect
scaling 
from defect formation to the present day. Also, we must specify a
particular cosmogony and we do this by analogy to what has become
called Standard Cold Dark Matter, that is, a flat background
$(\Omega_{\rm tot}=1)$ spacetime comprising 95\% collisionless
cold dark matter and 5\% baryons $(\Omega_c=0.95, \Omega_b=0.05)$,
with a Hubble constant at the present day of $H_0=50{\rm km}\,{\rm
sec}^{-1}{\rm Mpc}^{-1}$. 
Fig.~\ref{standard} shows the resulting
power spectra, normalized to COBE, for the CMB and CDM (solid
lines) compared with the standard adiabatic scenario based on
inflation (dot-dashed line) and the published data points with
error-bars based on the assumption of Gaussianity \cite{tegmark,PD}.  

\begin{figure}
\setlength{\unitlength}{1cm}
\begin{minipage}{8.0cm}
\leftline{\psfig{file=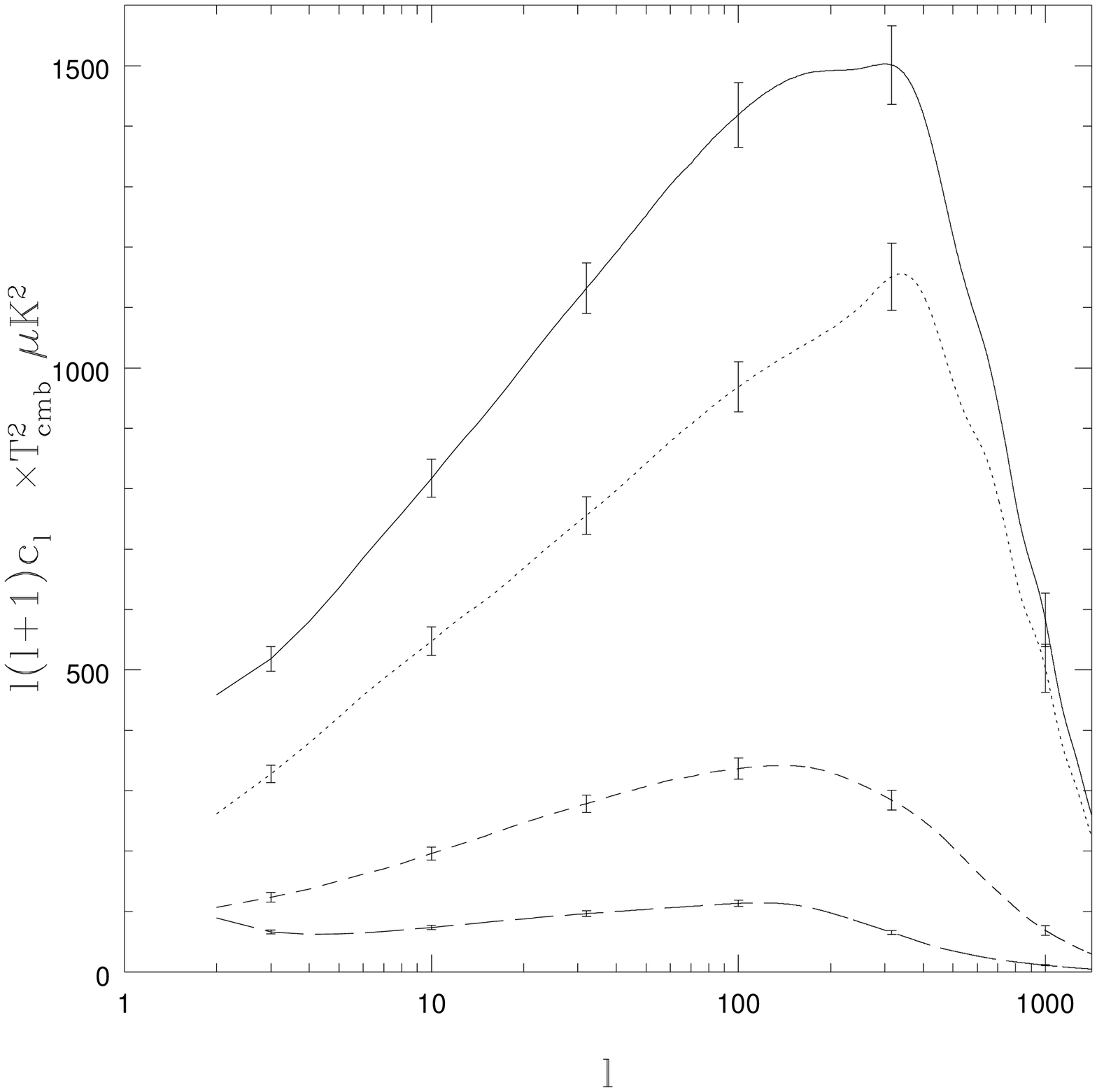,width=3.4in}}
\end{minipage}\hfill
\begin{minipage}{8.0cm}
\rightline{\psfig{file=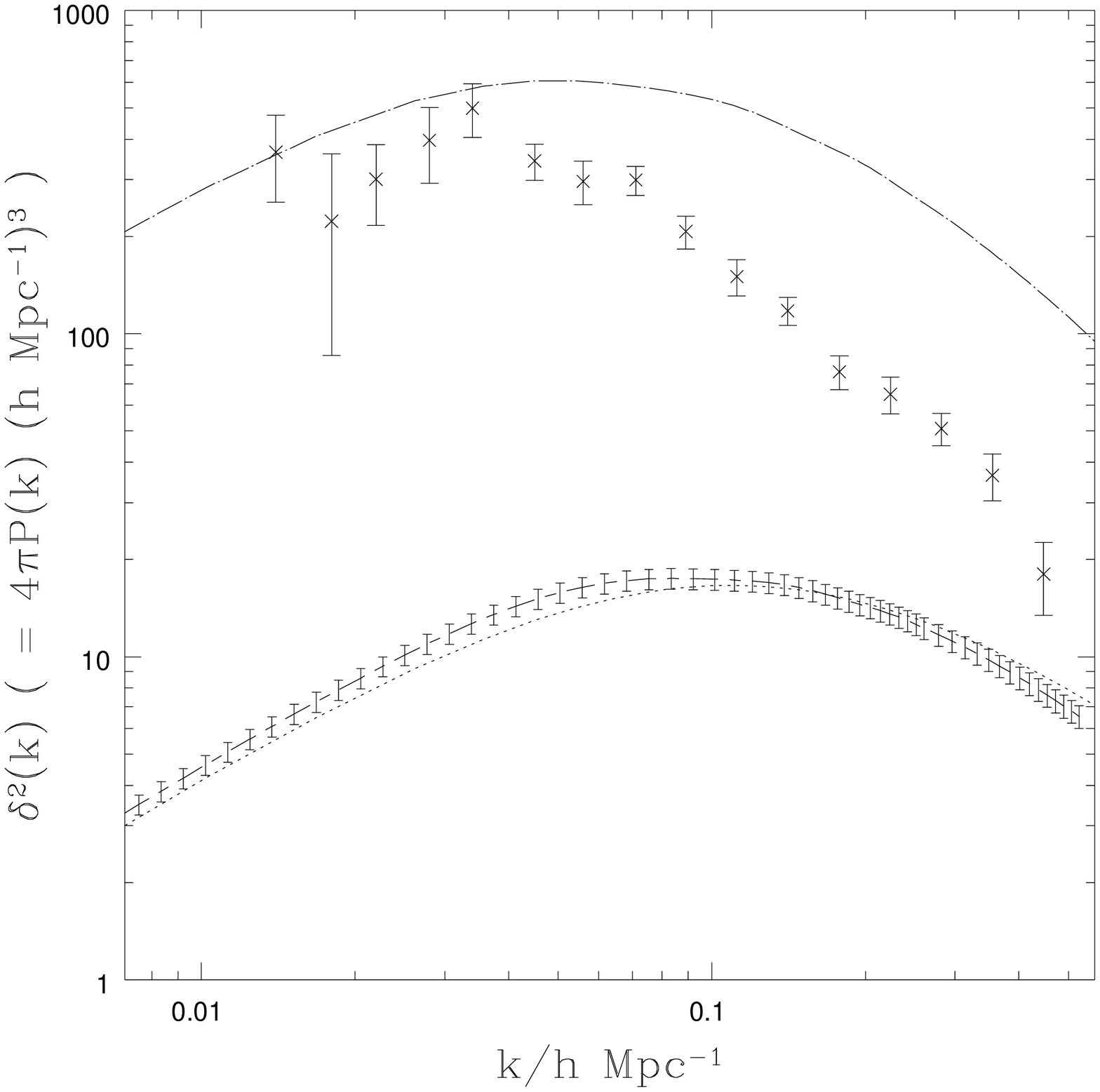,width=3.4in}}
\end{minipage}
\caption{The COBE normalized angular power spectrum of CMB
anisotropies, (left-hand graph) and the matter power spectrum
(right-hand graph) for the standard string model. The total anisotropy
(solid line) is illustrated, along with the partial contributions from
scalar (dotted), vector (short-dash) and tensor(long-dash) 
components in the case of the CMB
anisotropies. The errorbars show $1\sigma$ statistical uncertainties 
derived from the
finite number of realizations in our calculations. Observational data
and the prediction for standard CDM (dot-dash curve) are also
illustrated in the case of the matter power spectrum. The dotted line
on the matter spectrum shows the prediction of Ref. \protect\cite{AS},
as discussed in the text.}
\label{svt_split}
\end{figure}

The CMB angular power spectrum appears to have no discernible Doppler
peak for two reasons: firstly, there is a substantial ISW component to the scalar, vector and tensor anisotropies. The split into
the different components is illustrated for this model in
Fig.~\ref{svt_split} and we see that  the scalars are larger than the
vectors, with the tensors further suppressed relative to the
other two. More precisely, we find that the contributions to the
angular power spectrum are in the ratio, 
\begin{equation}
C_l^S:C_l^V:C_l^T = 3 : 1 : 0.4 \,,
\end{equation}
at $l=10$. Although the difference between our models and those
presented refs.\cite{PSelTa,PSelTc} are only at the level of a factor of
two or so, it is still  worth noting the discrepancy as a direction
for future work. We suggest that this is due to a difference in super-horizon ratio of $\langle |\Theta^S|^2\rangle$ and $\langle |\Theta_{00}|^2\rangle$, already discussed in an earlier section.

And secondly, the component of the angular power spectrum created at
the surface of last scattering is incoherent, with the secondary
Doppler peaks being cancelled out by decoherence as suggested in
refs.\cite{ACFM,MACFa,MACFb}. This leads to a further  suppression of
the amplitude in the ensemble average, relative to the large angular
scales, since we are averaging high peaks and low troughs. We should note that although the comparison with the published
CMB data does not appear to be good, the plotted error-bars are only at
the level of one sigma and deviations from non-Gaussianity may require
even larger errorbars, particularly for experiments with small sky
coverage. We expect the situation to be much clearer when the new CMB
data begins to arrive in the very near future. 

However, the situation seems to be much more clear-cut in the case of
the CDM power spectrum. Once normalized to COBE, the linear power
spectrum of the CDM 
appears to fit the data extremely badly, with the predicted curve
lying much further outside the observational errorbars than in the
case of the 
CMB angular power spectrum. Again the errorbars are based on
an assumption of Gaussianity and consideration of a non-Gaussian
theory will no doubt require us to increase their size, but the level
of disagreement is much larger than seems likely in any of the
realistic scenarios, which are thought to be only mildly non-Gaussian
on these scales. If we assume for the moment that we can compare the
theoretical curves with the data in this very naive way, resolution of
the absence of power on any particular scale requires us to postulate
a bias between the CDM and the data. While the idea of a  bias between the CDM and baryonic matter, probed by the catalogues of galaxies and
clusters of galaxies which make up the dataset, is not uncommon, large
values $(>2)$ are thought to be unrealistic. 

The bias required on a scale of $Rh^{-1}${\rm Mpc}, where now
$H_0=100h\,{\rm km}\,{\rm sec}^{-1}{\rm Mpc}^{-1}$, can be estimated  by
calculating the fractional matter over-density in a ball of radius
$Rh^{-1}${\rm Mpc}, 
\begin{equation}
\sigma^{DM}_R=4\pi\int {dk\over k} k^3P(k) |W(kR)|^2 \,,
\end{equation}
where the window function $W(x)$ is given by
\begin{equation}
W(x)={3\over x^3}\left(\sin x-x\cos x\right)\,,
\end{equation}
and comparing it to $\sigma_R$ calculated from a hypothetical curve which
fits the data, that is, $b_{R}=\sigma_R/\sigma_R^{DM}$. The scale
$R=8$ corresponds to $\sigma_R \approx 1$, that is, scales turning non-linear at the present day, and is the most
common scale on which comparisons are made.
For standard CDM $\sigma_8=1.2$ for $h=0.5$, while
the value favoured by observations is $\sigma_8=0.5$, illustrating the
celebrated excess of power on small scales for this model. 
When we perform the same calculation for the string model,  we get
$\sigma_8=0.31$ and hence the bias on these scales is $b_8\approx
1.5$. Such a value is around the limit of what is thought to be
possible, but given the uncertainties involved is not totally
unreasonable. 

This comparison does, however, ignore the obvious woeful absence of
power on much larger scales. In order to quantify this we choose
$R=100$ since (1) it is unlikely that scales of $100h^{-1}{\rm Mpc}$ are
affected by non-linear gravitational evolution, (2) the
distribution of galaxies is likely to be more Gaussian than on smaller
scales and (3) such scales are above the neutrino free-streaming scale
and so the introduction of a hot dark matter  (HDM) component cannot
be used to modify the shape of the spectrum. We estimate that the
standard string model requires a bias of $b_{100}=5.4$ to reconcile it
with the data. Since the chances of either the actual Universe\cite{GF,WSDK}
or the physical model\cite{abel,wandelt} having such a
bias seem remote, we conclude that the standard
string model is in serious conflict with the current observations
on scales around $100h^{-1}$Mpc. 

The COBE normalization of our standard scaling model also allows us to
calculate a value for the dimensionless quantity
$G\mu$, where $\mu$ is the string mass per unit
length, and $G$ is Newton's gravitational constant.
For our standard scaling model, we find that 
$G\mu=2.0\times 10^{-6}$, which is very close
the value $G\mu=1.7\times 10^{-6}$
obtained from
calculations of large angle CMB
anisotropies using high resolution local string
simulations\cite{ACDKSS}.
Although these values are in good agreement,
we stress 
that the purpose of our work has not been to compute $G\mu$ to high
precision and we expect that variations in details of the string
evolution which 
we have not attempted account for in our model, such as the
amount of small scale structure, could give 
rise to variations in $G\mu$. Instead,
we are primarily interested in the relative normalization of
anisotropies on different scales, in particular between COBE and
$\sigma_{100}$, which can be obtained without knowledge of the
absolute value of $G\mu$. In
future sections we present a number of variations on our standard
scaling model, each giving rise to a different value for the string
mass per unit 
length. To emphasize the fact that we do not intend to use our model to obtain
precise predictions for the absolute value of $G\mu$ in realistic
cosmic string scenarios, we quote all subsequent values in terms of
the ratio $\mu/\mu_{S}$, where $\mu_{S}$ is the value obtained for
our standard scaling model. We should note that large values of
$\mu/\mu_S$ are likely to be excluded by the absence of residuals in
timing measurements for milli-second pulsars\cite{CBS}. 

As one final point, in Fig.\ref{svt_split} we include the prediction for the
matter power spectrum for strings in a background of CDM computed 
in ref.\cite{AS}. The dotted line on the matter graph shows the
prediction from the I model described in this reference, normalized to
give the same value for $\sigma_8$ as our standard scaling model. We
note that the predictions are in extremely close agreement, although
in the current work the amplitude is no longer arbitrary,  but fixed
by normalization of the associated CMB fluctuations to COBE.

\subsection{Modifications to model parameters}
\label{mod-model}

\begin{figure}
\setlength{\unitlength}{1cm}
\begin{minipage}{8.0cm}
\leftline{\psfig{file=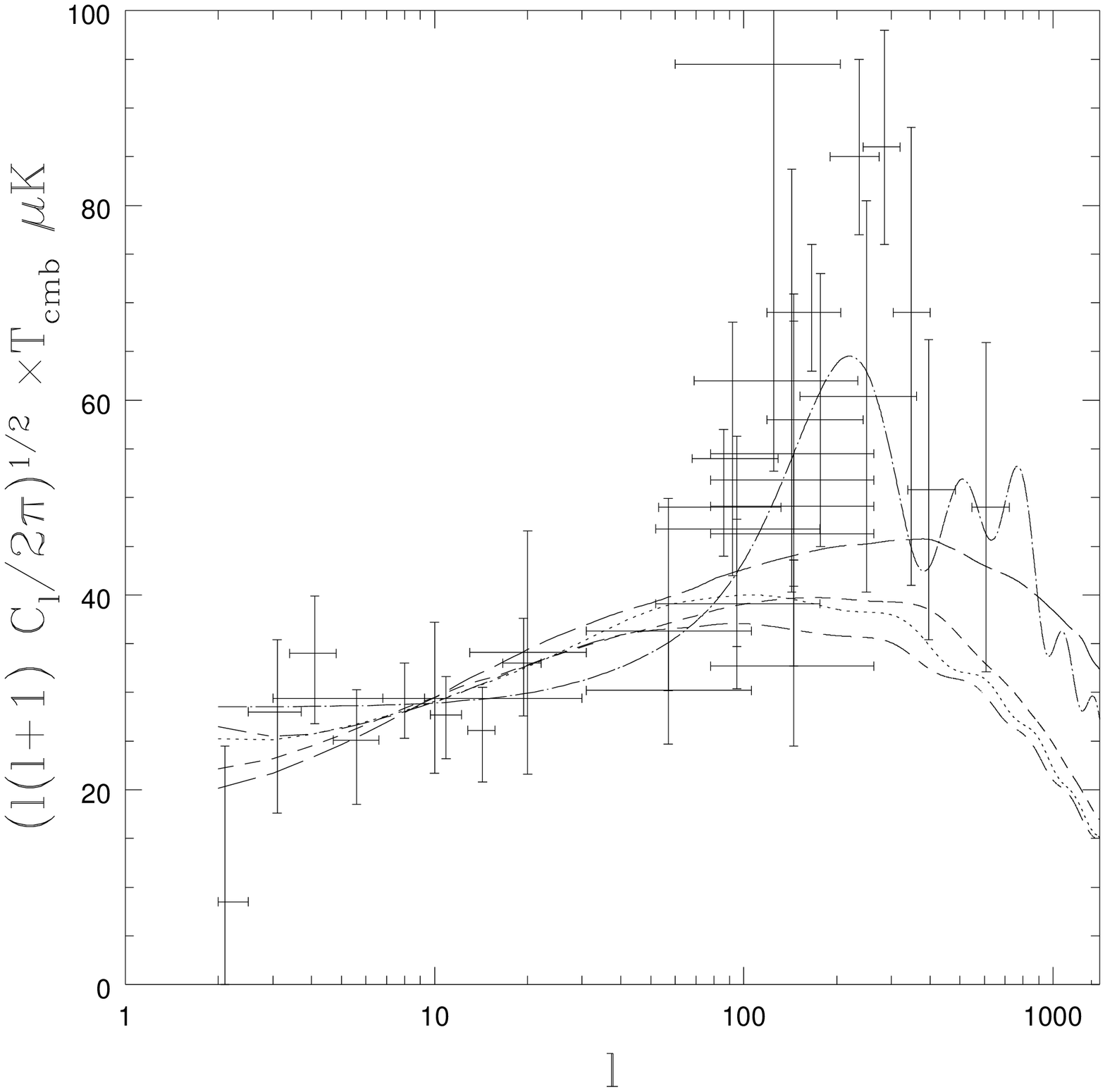,width=3.4in}}
\end{minipage}\hfill
\begin{minipage}{8.0cm}
\rightline{\psfig{file=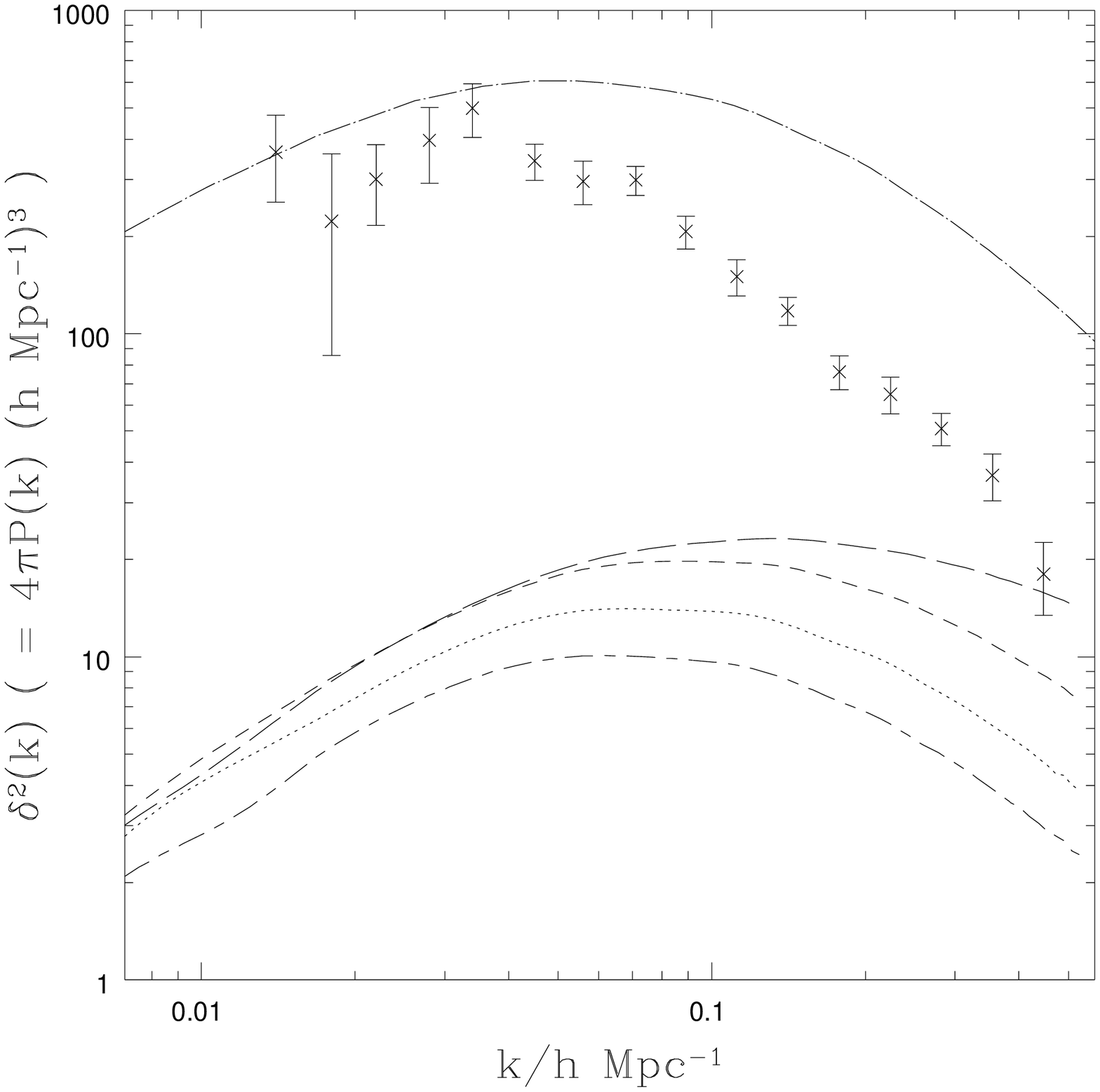,width=3.4in}}
\end{minipage}
\caption{Varying $\xi$, the string coherence length:  We plot the COBE
normalized angular power spectrum of CMB 
anisotropies, (left-hand graph) and the matter power spectrum
(right-hand graph) for various values of the parameter $\xi$, with
$v=0.65$.($\xi=2.0$ --
long-short-dash line, $\xi=1.0$ --
dotted line, $\xi=0.3$ --
short-dash line, $\xi=0.03$ --
long-dash line).
Observational data
and the prediction for standard CDM (dot-dash curve) are included for
comparison.}
\label{xi}
\end{figure}

\begin{figure}
\setlength{\unitlength}{1cm}
\begin{minipage}{8.0cm}
\leftline{\psfig{file=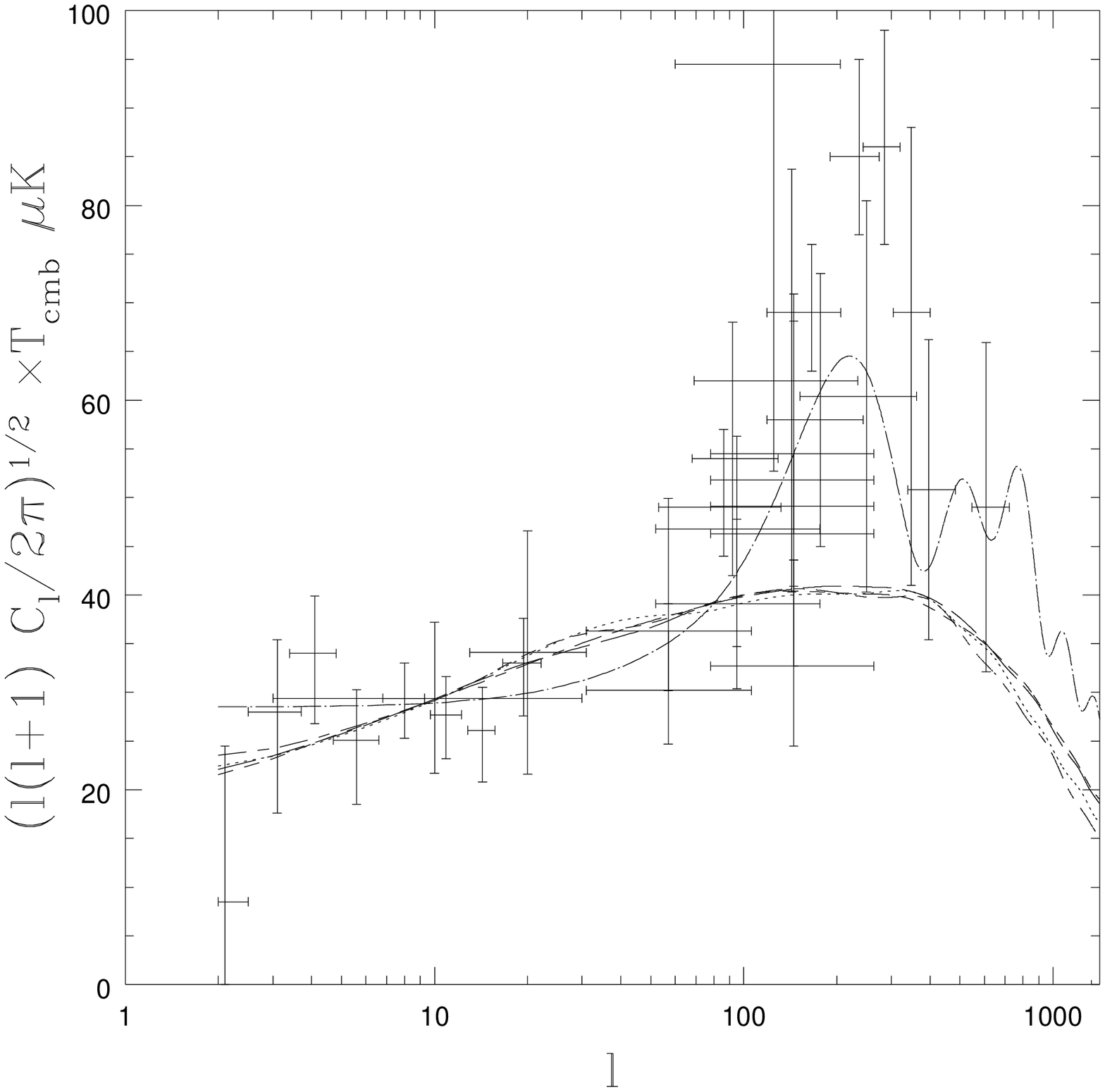,width=3.4in}}
\end{minipage}\hfill
\begin{minipage}{8.0cm}
\rightline{\psfig{file=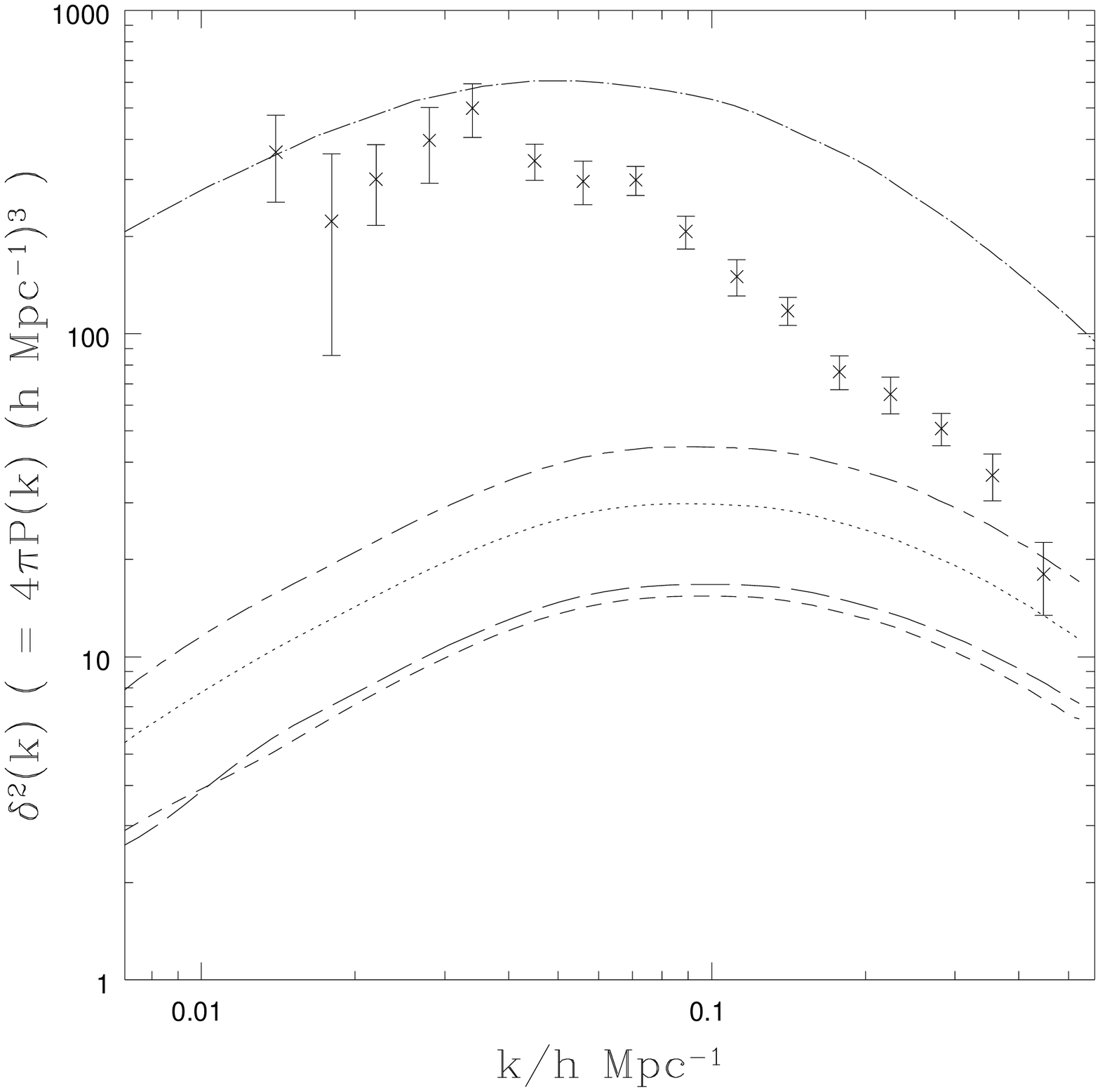,width=3.4in}}
\end{minipage}
\caption{Varying $v$, the string velocity: We plot the COBE normalized
angular power spectrum of CMB anisotropies, (left-hand graph) and the
matter power spectrum 
(right-hand graph) for various values of the parameter $v$, with
$\xi=0.3$.($v=0.0$ --
long-short-dash line, $v=0.3$ --
dotted line, $v=0.65$ --
short-dash line, $v=0.99$ --
long-dash line).
Observational data
and the prediction for standard CDM (dot-dash curve) are included for
comparison.}
\label{v}
\end{figure}

In order to test the robustness of the conclusions about the standard
model, we have repeated the calculations for different values
of $\xi$ and $v$.  These variations could either represent
fundamentally different types of cosmic strings to those modelled in
simulations, for example, those with superconducting currents, or any
possible systematic uncertainties in the measurement of these
quantities. Fig.~\ref{xi} shows the
power spectra for different values of $\xi$ with $v=0.65$, while
Fig.~\ref{v} shows the power spectra for different values of 
$v$ with $\xi=0.3$, with the values of $b_{100}$,
$b_8$, and $\mu/\mu_{S}$ summarized in table \ref{tab-one}. All other
parameters are as in the standard scaling model. 

The first thing to note is that, although our choice of values spans
the possible parameter space, none of the models does significantly
better than the standard scaling case as far as the value of $b_{100}$
is concerned and there is also no discernible Doppler peak in the CMB
spectrum. However, at a more microscopic level there are differences
which follow the trends of previous calculations, giving us confidence
that the model is reproducing intuitive results.  

As $\xi$ decreases one is reducing the scale at which the
two-point functions turnover from white-noise, with $\xi=2.0$ being
close the causal limit \cite{RW}. It was predicted in ref.\cite{MACFa}
that this would lead to the contribution to the CMB anisotropy from
the surface of last scattering being peaked at smaller scales, which
is indeed what we see in Fig.~\ref{xi}, although this feature is
slightly masked out by the large ISW component. This can be
seen more clearly in the CDM power spectrum which turns over at a
scale $k\sim\xi k_{\rm eq}$. One would of course expect a reduction in
$\xi$ to result in a substantially higher peak, but this is partially
counter balanced by an almost equivalent increase in the ISW
component, the only direct evidence for such an effect being that smaller
values of $\xi$ require a smaller value of $G\mu$. This change in
$G\mu$ can be understood from the formula $\rho_s\sim\mu/ \xi^2$; if
the value of $\xi$ decreases the value of $\mu$ must also decrease to
keep the string density and hence the amplitude of the CMB anisotropy the same. 

The lack of dependence on $v$ is less intuitive. One might think that
in the limit of $v\rightarrow 0$, the network would become more
coherent, since the strings are not moving, creating perturbations in
the same place. However, such a limit does not correspond in any way
to the standard picture of string evolution. If the strings are moving
slowly or are stationary, then reconnection will take place only very
infrequently and the scaling regime will be difficult to attain. But
scaling is implicit in our model, being put in by hand, so even if
they are not moving the strings will have to decay in some way, which is
essentially random, introducing decoherence. The only discernible effect of changing $v$ is a weak dependence of
the amplitude of the matter power spectrum, believed to be due to the
dependence of the relative amplitudes of $\Theta_{00}$ and $\Theta^S$
on $v$ already discussed. Also interesting, however, is the apparent
independence of the turnover of the matter power spectrum, suggesting
that the corresponding turnover in the two-point functions is also
independent of $v$. Once again the values of $b_8$, $b_{100}$ and
$\mu/\mu_s$ are summarized in table \ref{tab-one}. 

\begin{table}
\centering
\begin{tabular}{||ccc|ccc||}
\hline
Description&Figure&Line type&$b_8$&$b_{100}$&$\mu/\mu_S$ \\
\hline \hline
Standard Scaling\footnote{ We should note that the standard scaling
model is referred to a number of times in the following tables and an
observant reader will notice that the values quoted are slightly
different. The values quoted here are for 400 realizations and all the
others are for 100 realizations.}   &\ref{standard}&Solid  &    1.61 &
5.36 & 1.0 \\ \hline  
\hline\hline
$\xi=2.0$&\ref{xi}&Long-short dash      &    2.46 &    7.31 &    5.56 \\ \hline
$\xi=1.0$&&Dotted               &    2.16 &    6.57 &    2.88 \\ \hline
$\xi=0.3$&&Short-dash           &    1.74 &    5.81 &    1.0 \\ \hline
$\xi=0.03$&&Long-dash            &    1.45 &    5.67 &    0.25 \\ \hline
\hline
$v=0.0$&\ref{v}&Long-short dash      &    1.04 &    3.39 &    1.93 \\ \hline
$v=0.3$&&Dotted               &    1.27 &    4.13 &    1.64 \\ \hline
$v=0.65$&&Short-dash           &    1.74 &    5.81 &    1.00 \\ \hline
$v=0.9$&&Long-dash            &    1.66 &    5.58 &    0.92 \\ \hline
\hline
$L_f=0.01$&\ref{l}&Long-short dash      &    1.45 &    4.79 &    0.40 \\ \hline
$L_f=0.1$&&Dotted               &    1.52 &    5.04 &    0.45 \\ \hline
$L_f=0.3$&&Short-dash           &    1.64 &    5.45 &    0.56 \\ \hline
$L_f=0.5$&&Solid-dash           &    1.74 &    5.81 &    1.0 \\ \hline
$L_f=0.7$&&Long-dash            &    1.83 &    6.13 &    1.48 \\ \hline
$L_f=0.9$&&Dot-short dash       &    1.91 &    6.40 &    2.12 \\ \hline
\hline
$h=0.3$&\ref{h}&Long-short dash      & 2.68  &  6.78  &    0.46 \\ \hline
$h=0.4$&&Dotted               &  2.08  & 6.18  &    0.71 \\ \hline
$h=0.6$&&Short-dash           & 1.52  & 5.57  &    1.32 \\ \hline
$h=0.7$&&Long-dash            & 1.36  & 5.39  &    1.67 \\ \hline
\hline
$\Omega_b=0.01$&\ref{b}&Long-short dash   &  1.64 & 5.76  &   1.00 \\ \hline
$\Omega_b=0.03$&&Dotted               & 1.64  & 5.78   &    1.00 \\ \hline
$\Omega_b=0.1$&&Short-dash           & 1.87  & 5.86   &    1.00 \\ \hline
$\Omega_b=0.2$&&Long-dash            & 2.17  & 5.94  &    1.00 \\ \hline
\hline
``Best of all worlds'' &\ref{best}&Long-short dash&0.3 & 1.56 & 0.03 \\ \hline
\hline
\end{tabular}
\vskip 10pt
\caption{Table of biases and values of $\mu/\mu_S$ for the standard
scaling model, and the simplest variations described  
in this paper. Each model is labelled by the figure and
line type where it appears.}
\label{tab-one}
\end{table}
%\multicolumn{4}{c||}{Physical parameters}&
%\multicolumn{6}{c||}{Systems parameters}\\

\begin{table}
\centering
\begin{tabular}{||ccc|ccc||}
\hline
Description&Figure&Line type&$b_8$&$b_{100}$&$\mu/\mu_S$ \\
\hline \hline
$\tau_T=100$&\ref{mu1}&Long-short dash      &    1.19 &    4.84 &    1.00 \\ \hline
$\tau_T=400$&&Dotted               &    0.86 &    3.96 &    1.00 \\ \hline
$\tau_T=1000$&&Short-dash           &    0.81 &    3.29 &    0.98 \\ \hline
$\tau_T=5000$&&Long-dash            &    0.93 &    3.09 &    0.84 \\ \hline
\hline
$\chi=2$&\ref{mu2}&Long-short dash      &    1.19 &    4.84 &    1.00 \\ \hline
$\chi=5$&&Dotted               &    0.86 &    3.96 &    1.00 \\ \hline
$\chi=10$&&Short-dash           &    0.81 &    3.29 &    0.98 \\ \hline
$\chi=20$&&Long-dash            &    0.81 &    3.29 &    0.98 \\ \hline
\hline
$\tau_T=100$,$L_T=0.1$&\ref{mu3}&Long-short dash      &    0.64 &    4.01 &    1.00 \\ \hline
$\tau_T=100$,$L_T=0.8$&&Dotted               &    0.37 &    2.59 &    1.00 \\ \hline
$\tau_T=1000$,$L_T=0.1$&&Short-dash           &    0.35 &    1.63 &    0.91 \\ \hline
$\tau_T=1000$,$L_T=0.8$&&Long-dash            &    0.46 &    1.60 &    0.68 \\ \hline
\hline
$\chi=100$,varying $n$&\ref{n}&Long-short dash      &    0.72 &    3.95 &    0.98 \\ \hline
$\chi=400$,varying $n$&&Dotted               &    0.38 &    2.37 &    0.97 \\ \hline
$\chi=100$,varying $\mu$&&Short-dash           &    0.34 &    1.67 &    0.91 \\ \hline
$\chi=400$,varying $\mu$&&Long-dash            &    0.54 &    1.81 &    0.57 \\ \hline
\hline
$\alpha=0.25$&\ref{pow}&Long-short dash      &    0.56 &    2.66 &    8.16 \\ \hline
$\alpha=0.5$&&Dotted               &    0.20 &    1.33 &   59.96 \\ \hline
$\alpha=0.75$&&Short-dash           &    0.08 &    0.66 &  375.14 \\ \hline
$\alpha=1.0$&&Long-dash            &    0.03 &    0.28 & 1887.57 \\ \hline
\hline
``Best of all worlds'' &\ref{best}&Solid dash&0.12 & 0.82 & 0.02 \\ \hline
\hline
\end{tabular}
\vskip 10pt
\caption{Table of biases and values of $\mu/\mu_S$ for each of the models
with deviations from scaling.
Each model is labelled by the figure and
line type where it appears.}
\label{tab-dev}
\end{table}
%\multicolumn{4}{c||}{Physical parameters}&
%\multicolumn{6}{c||}{Systems parameters}\\

\begin{table}
\centering
\begin{tabular}{||ccc|ccc||}
\hline
Description&Figure&Line type&$b_8$&$b_{100}$&$\mu/\mu_S$ \\
\hline \hline
C1&\ref{co}&Long-short dash      &    0.38 &    1.84 &    -- \\ \hline
C2&&Dotted               &    1.29 &    3.20 &    -- \\ \hline
C3&&Short-dash           &    6.18 &   19.19 &    -- \\ \hline
C4&&Long-dash            &    1.71 &    3.62 &    -- \\ \hline
\hline
$\nu=0$&\ref{w}&Long-short dash      &    2.10 &    6.14 &    1.23 \\ \hline
$\nu=-1$&&Dotted               &    2.73 &    9.36 &    0.70 \\ \hline
$\nu=0.5$&&Short-dash           &    1.64 &    4.99 &    1.27 \\ \hline
$\nu=\infty$&&Long-dash            &    2.11 &    7.91 &    -- \\ \hline
\hline
$\epsilon=2$&\ref{e}&Long-short dash      &    2.06 &    4.43 &    7.28 \\ \hline
$\epsilon=5$&&Dotted               &    1.85 &    4.78 &    2.86 \\ \hline
$\epsilon=10$&&Short-dash           &    1.70 &    4.86 &    1.65 \\ \hline
$\epsilon=20$&&Long-dash            &    1.63 &    5.01 &    1.23 \\ \hline
\hline
$q=0$&\ref{sg}&Long-short dash      &  1.74  & 5.81  &    1.00 \\ \hline
$q=-1$&&Dotted               &  1.42  & 4.68  &    0.72 \\ \hline
$q=-1.25$&&Short-dash           & 1.52  & 5.04  &    0.77 \\ \hline
$q=-1.5$&&Long-dash            & 1.35  & 4.16  &   -- \\ \hline
\hline
$q=0$&\ref{sg2}&Long-short dash      &    0.54 &    2.55 &    0.07 \\ \hline
$q=-1.0$&&Dotted               &    0.35 &    1.64 &    0.06 \\ \hline
$q=-1.25$&&Short-dash           &    0.40 &    1.89 &    0.06 \\ \hline
$q=-1.5$&&Long-dash            &    0.19 &    1.00 &    -- \\ \hline
\hline
\end{tabular}
\vskip 10pt
\caption{Table of biases and values of $\mu/\mu_S$ for each of the
models where our standard source has been further modified.
Each model is labelled by the figure and
line type where it appears. Note that we have not calculated $\mu/\mu_S$ for coherent models, which are not based on string-like two-point functions.}
\label{tab-all}
\end{table}

\begin{figure}
\setlength{\unitlength}{1cm}
\begin{minipage}{8.0cm}
\leftline{\psfig{file=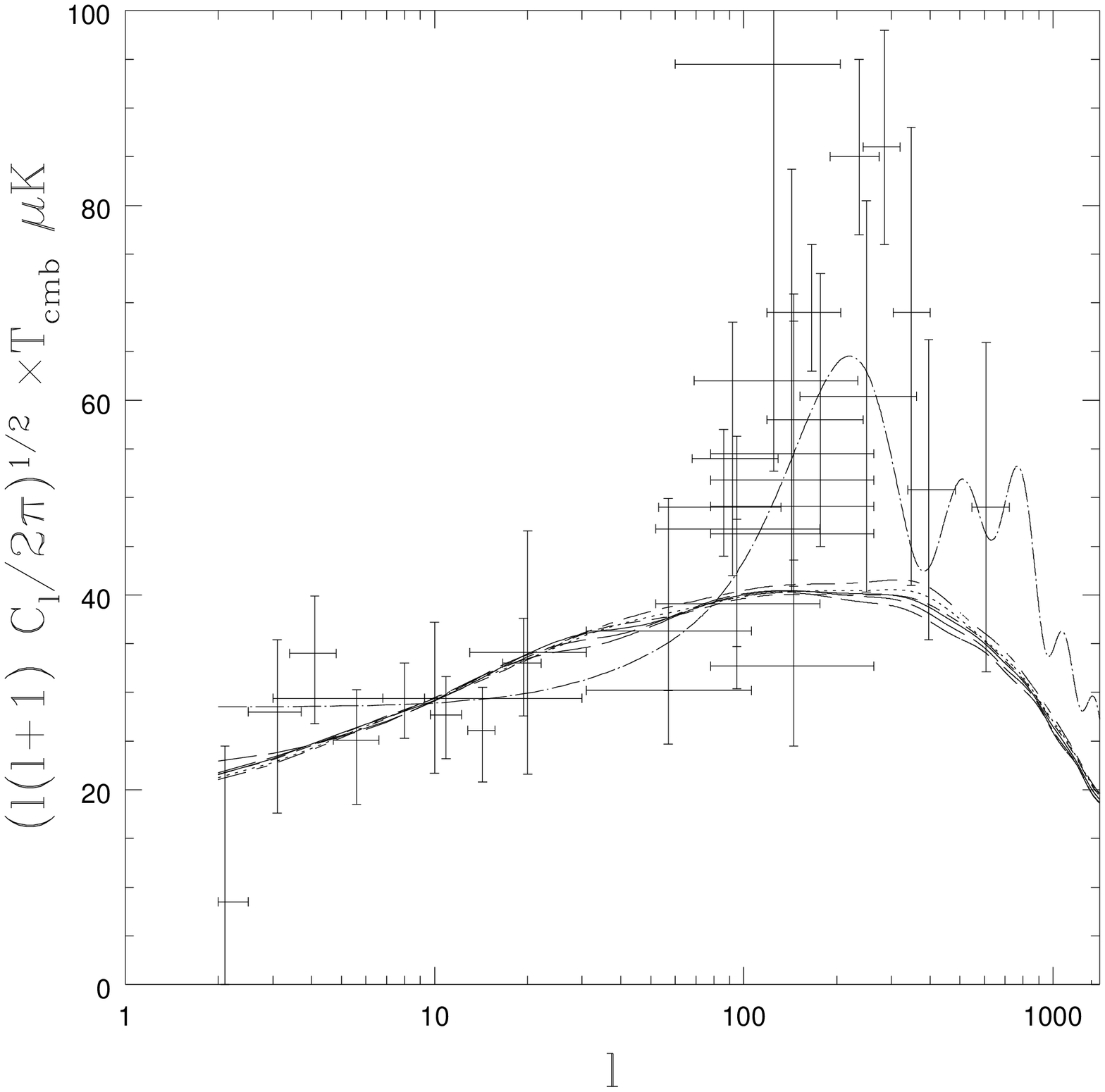,width=3.4in}}
\end{minipage}\hfill
\begin{minipage}{8.0cm}
\rightline{\psfig{file=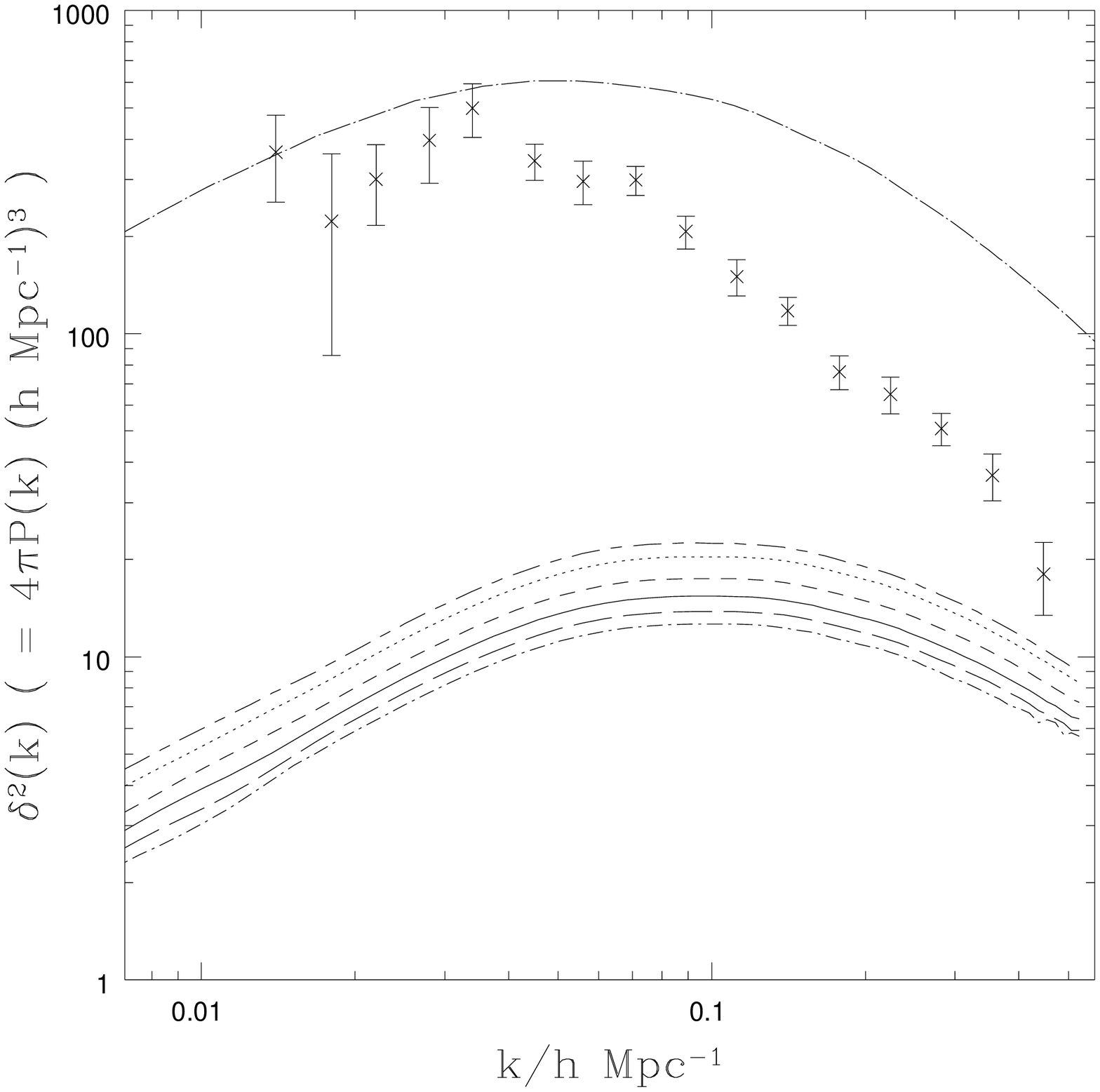,width=3.4in}}
\end{minipage}
\caption{Varying $L_f$, the string `turn off' parameter: We plot the
COBE normalized angular power spectrum of CMB 
anisotropies, (left-hand graph) and the matter power spectrum
(right-hand graph) for various values of the parameter $L_f$, with
$\xi=0.3$, $v=0.65$.($L_f=0.01$ --
long-short-dash line, $L_f=0.1$ --
dotted line, $L_f=0.3$ --
solid line, $L_f=0.5$ --
short-dash line, $L_f=0.7$ --
long-dash line, $L_f=0.9$ --
dot-short-dash line).
Observational data
and the prediction for standard CDM (dot-long-dash curve) are included for
comparison.}
\label{l}
\end{figure}

We have also tested the dependence of our results 
on the parameter $L_f$, representing
the rate at which the strings are turned off. CMB and matter power
spectra for various values of $L_f$ are illustrated in
Fig.~\ref{l}. The results have a weak dependence on $L_f$,
with smaller values of $L_f$ giving slightly better values for
$b_{100}$ than larger values. We note that further decreases in $L_f$
below $0.01$ do not change the results further. Since the dependence on
$L_f$ is only weak, we choose an intermediate value of $L_f=0.5$ for
the remaining calculations in this paper (with the exception of fig
\ref{best} in which we illustrate the possible improvement to $b_{100}$
which could result from exploiting all conceivable uncertainties in
our model). The fact that the results
depend minimally on $L_f$ also represents evidence to suggest that the
results will not depend strongly on the exact way in which the decay of long
string is treated. 

\subsection{Modifications to cosmological parameters}

\begin{figure}
\setlength{\unitlength}{1cm}
\begin{minipage}{8.0cm}
\leftline{\psfig{file=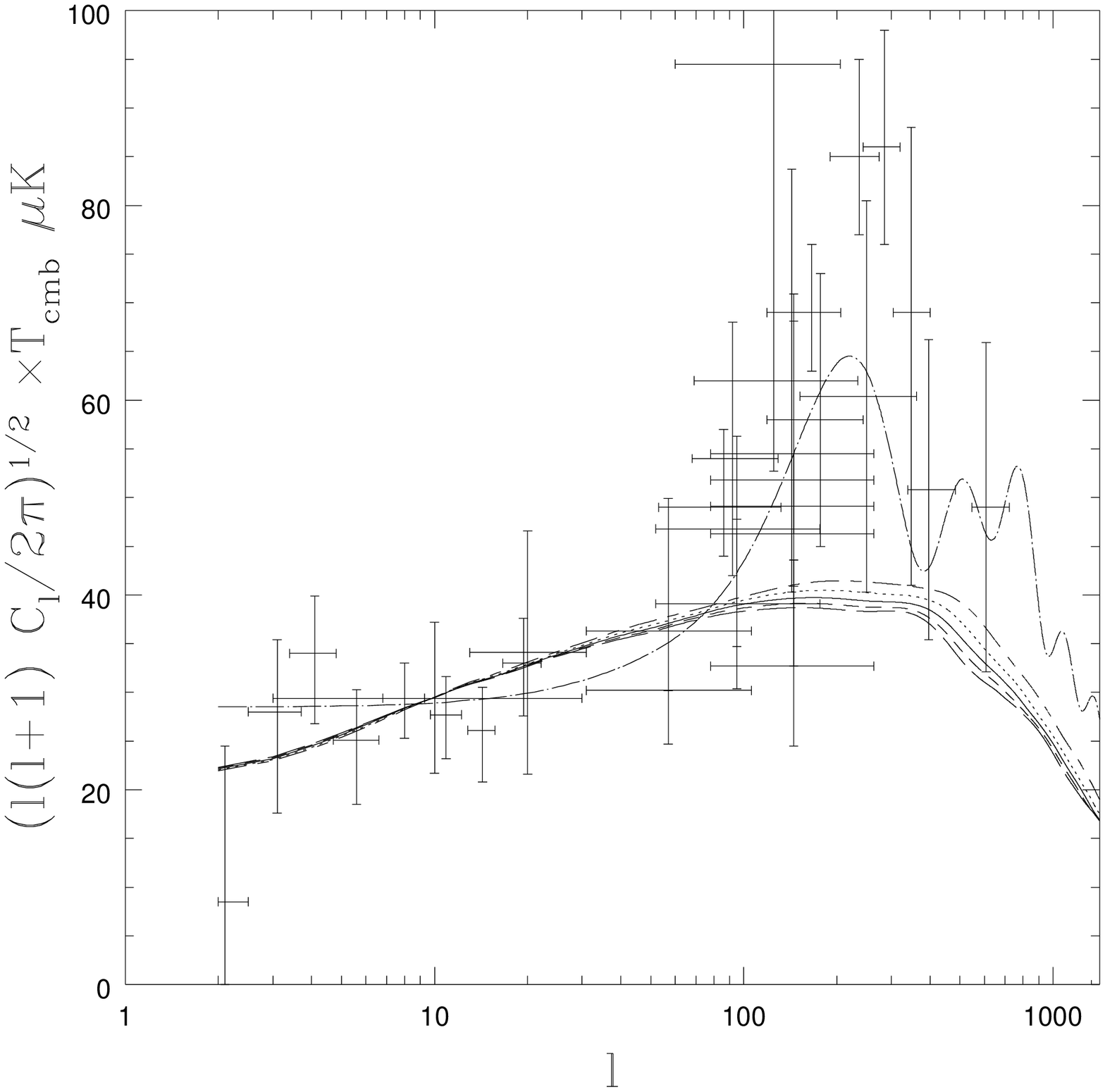,width=3.4in}}
\end{minipage}\hfill
\begin{minipage}{8.0cm}
\rightline{\psfig{file=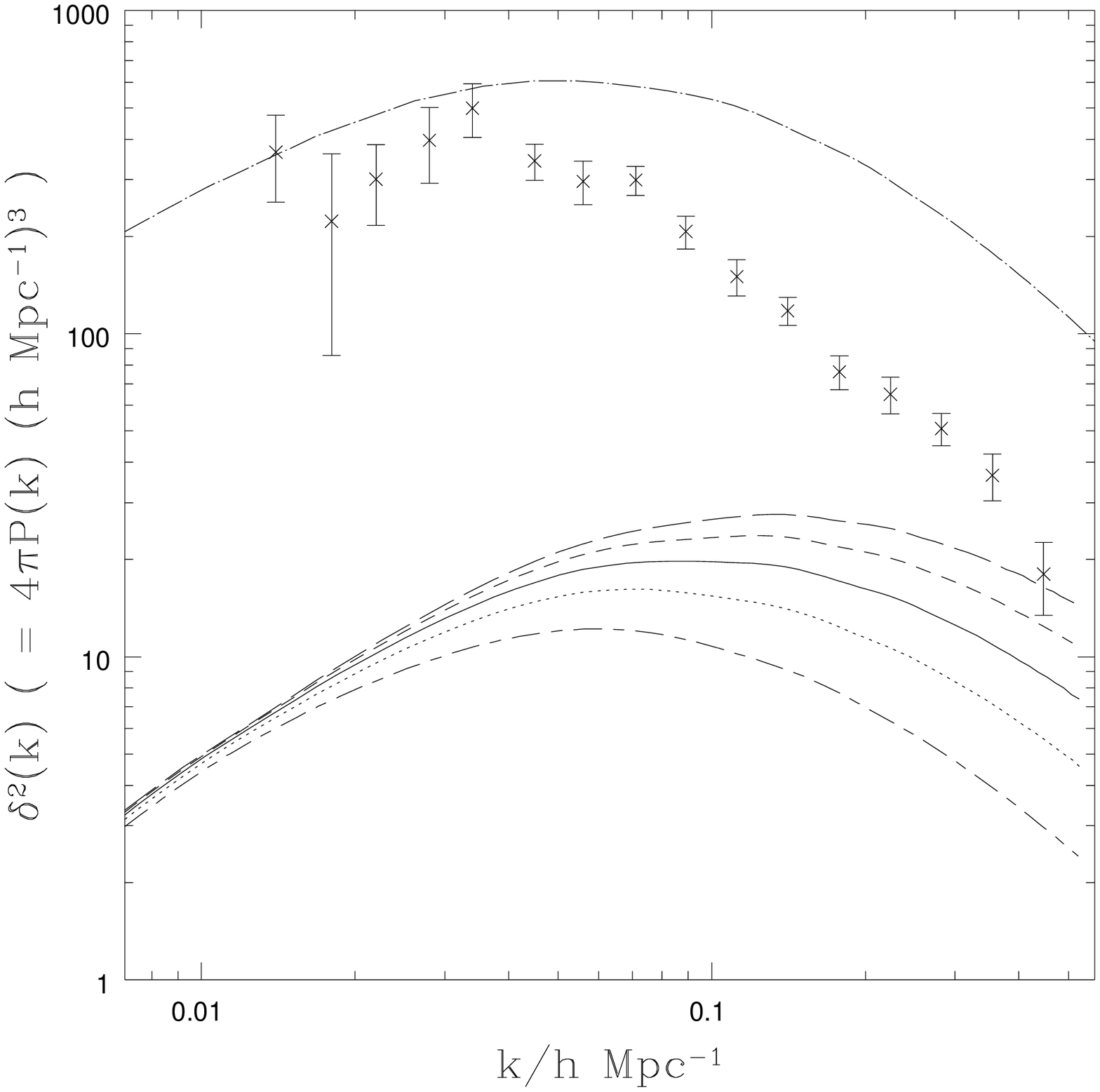,width=3.4in}}
\end{minipage}
\caption{Varying $h$: We plot the COBE normalized angular power
spectrum of CMB 
anisotropies, (left-hand graph) and the matter power spectrum
(right-hand graph) for various values of the cosmological expansion
rate $h$, with $\Omega_b=0.05$. ($h=0.3$ --
long-short-dash line, $h=0.4$ --
dotted line, $h=0.5$ --
solid line, $h=0.6$ --
short-dash line, $h=0.7$ --
long-dash line).
Observational data
and the prediction for standard CDM (dot-dash curve) are included for
comparison.}
\label{h}
\end{figure}

\begin{figure}
\setlength{\unitlength}{1cm}
\begin{minipage}{8.0cm}
\leftline{\psfig{file=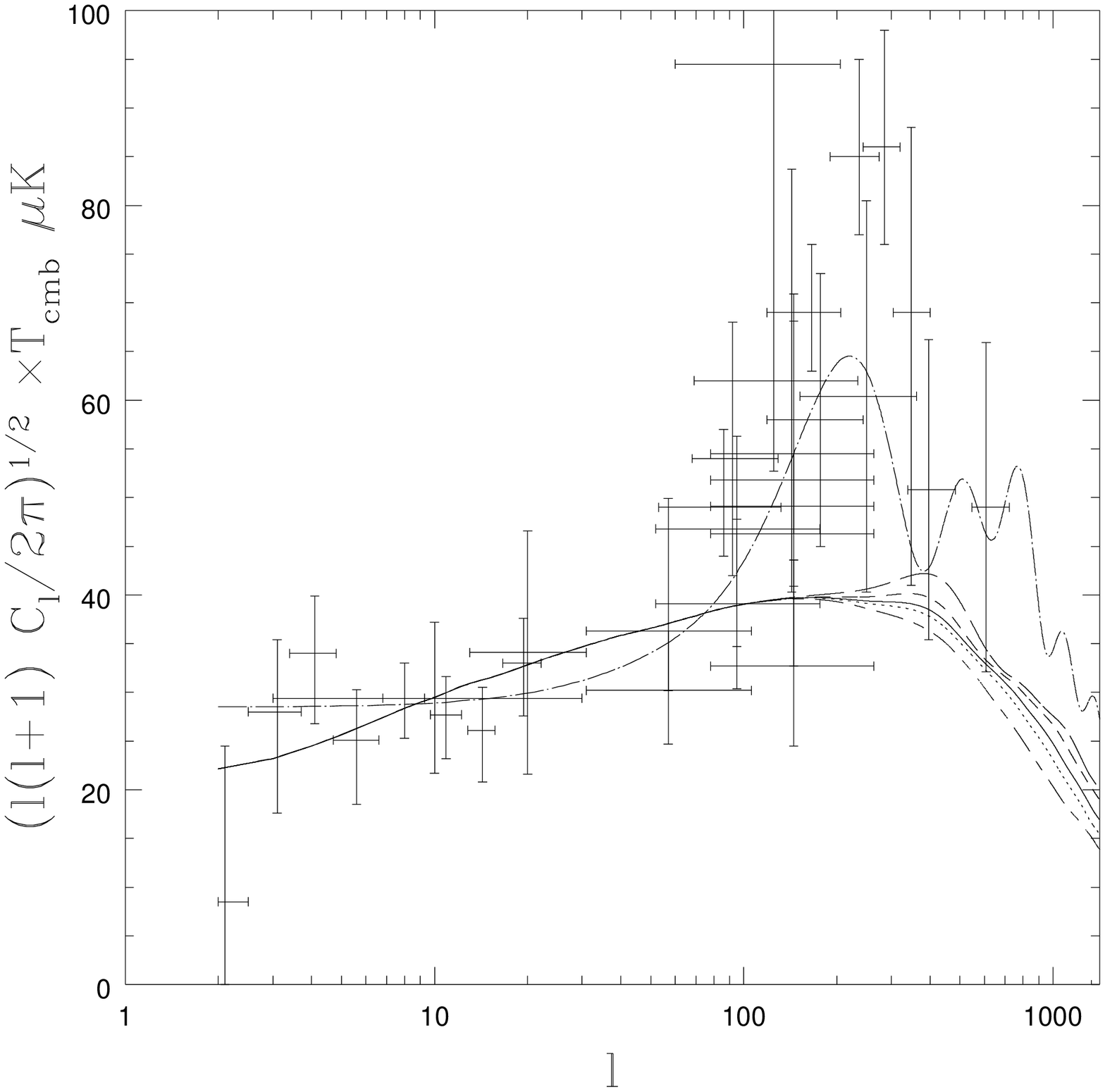,width=3.4in}}
\end{minipage}\hfill
\begin{minipage}{8.0cm}
\rightline{\psfig{file=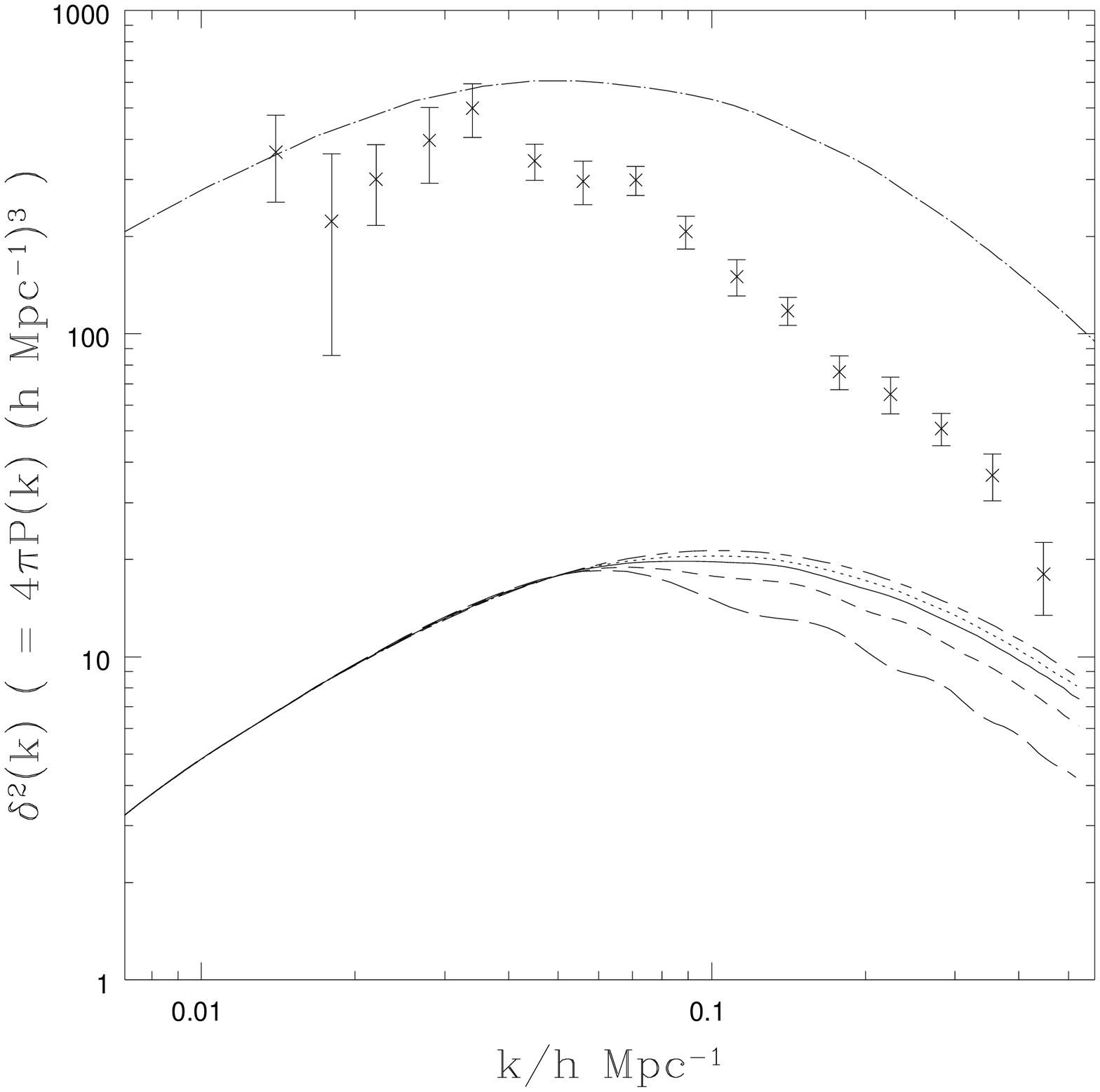,width=3.4in}}
\end{minipage}
\caption{Varying $\Omega_b$:  We plot the COBE normalized angular
power spectrum of CMB 
anisotropies, (left-hand graph) and the matter power spectrum
(right-hand graph) for various values of the cosmological expansion
rate $\Omega_b$, with $h=0.5$. ($\Omega_b=0.01$ --
long-short-dash line, $\Omega_b=0.03$ --
dotted line, $\Omega_b=0.05$ --
solid line, $\Omega_b=0.1$ --
short-dash line, $\Omega_b=0.2$ --
long-dash line).
Observational data
and the prediction for standard CDM (dot-dash curve) are included for
comparison.}
\label{b}
\end{figure}

We must also consider the possibility of different cosmogonies, since
most cosmological parameters are not constrained to better than a
factor two. It is simple to change the Hubble constant and also the
relative content of baryons and CDM. The resulting spectra are
presented in Fig.~\ref{h} for $h=0.3$ to $h=0.7$ and in
Fig.~\ref{b} for $\Omega_b=0.01$ to $\Omega_b=0.2$, keeping
$\Omega_{\rm tot}=1$ and using the standard scaling model for the two-point functions. Once again, no model significantly improves the value of
$b_{100}$ (see table \ref{tab-one}). 

The CMB angular power spectrum is largely unaffected by the changes in
cosmological parameters that we have tried. However, we do see that
the shape of the CDM power spectrum is modified by changes in $h$, via
the time of equal-matter radiation and the well-known shape parameter $\Gamma\approx\Omega h$. This fixes the
position of the turnover in the power spectrum, with larger values of
$h$ leading to a turnover at smaller scales. We also notice slight
oscillations in the power spectrum for larger values of $\Omega_b$. In
these models the oscillations that are present in the photon-baryon
fluid are transferred to the CDM, but in contrast to an inflationary
model, they are damped out by the effects of decoherence. 

We should comment on the apparent absence of any marked dependence on
these cosmological parameters, since the CMB spectrum for $l>100$ is very
strongly dependent on them in the case of the
standard adiabatic scenario, and  this dependence has been suggested
as a way of making extremely accurate estimations of many cosmological
parameters using satellite experiments. To understand this difference,
we should remember that the anisotropies created at the surface of
last scattering by acoustic waves are incoherent, leading an absence
of secondary peaks, effectively washing out the strong dependence on $\Omega_b$ and $H_0$. And more importantly, any contribution from the
last scattering surface seems to be swamped by the ISW effect, which
is less sensitive to changes in the cosmogony. 

One modification to the standard scenario which is often used to allow
the standard cold dark matter model to obtain a better fit to the galaxy data
is to introduce a small amount of hot dark matter in the form of
neutrinos. A similar, procedure would also allow the standard string
model to fit the shape of the observed power spectrum on scales below
the neutrino free streaming scale $\lambda\approx 20h^{-1}$Mpc, but we
anticipate that this will not be as efficient on larger scales, and in
particular we expect the introduction of HDM to have little bearing on
the $b_{100}$ problem. Nonetheless, we plan to investigate the
implications of making such a changes in future work. 

\subsection{Summary}

\begin{figure}
\setlength{\unitlength}{1cm}
\begin{minipage}{8.0cm}
\leftline{\psfig{file=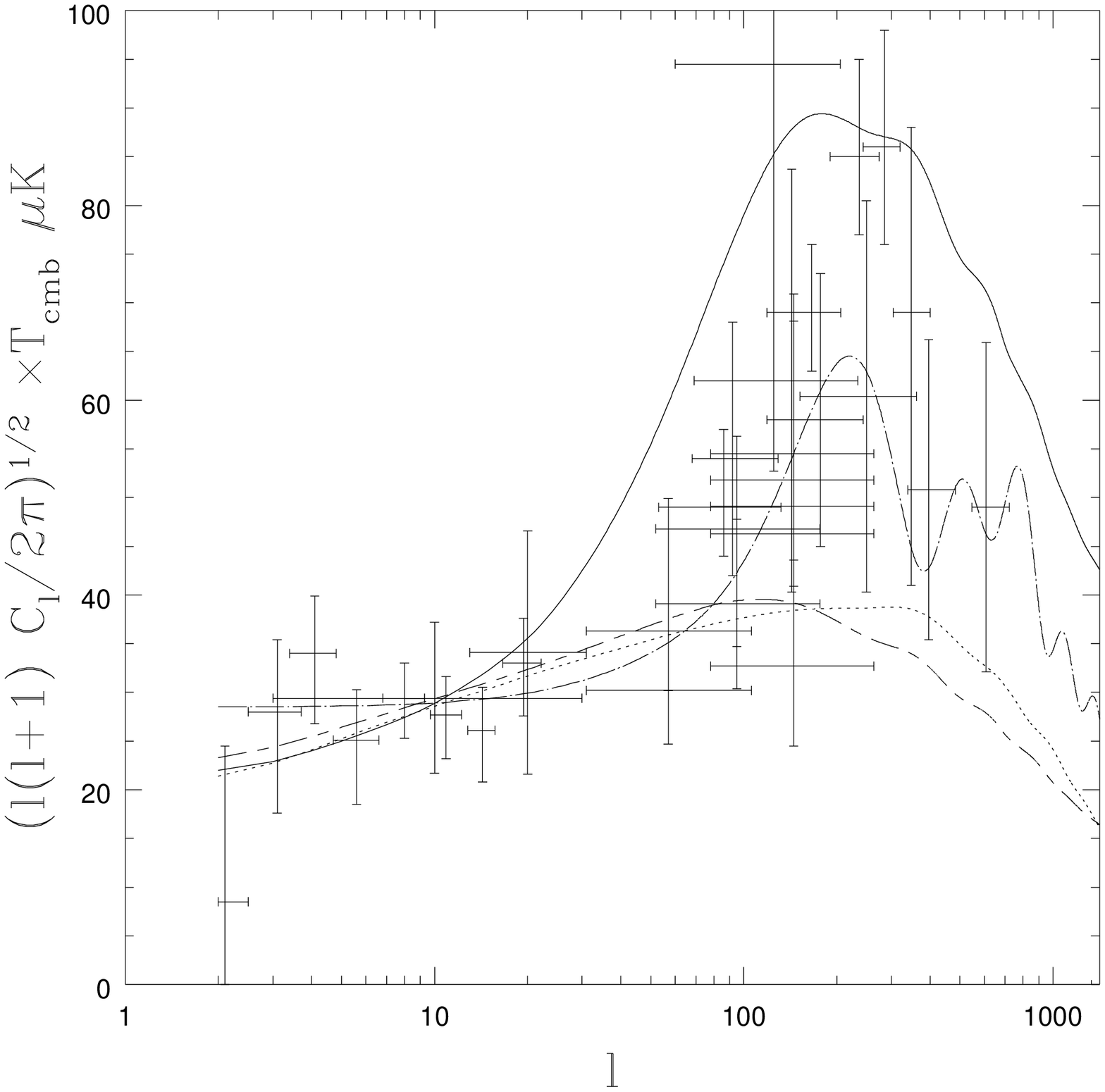,width=3.4in}}
\end{minipage}\hfill
\begin{minipage}{8.0cm}
\rightline{\psfig{file=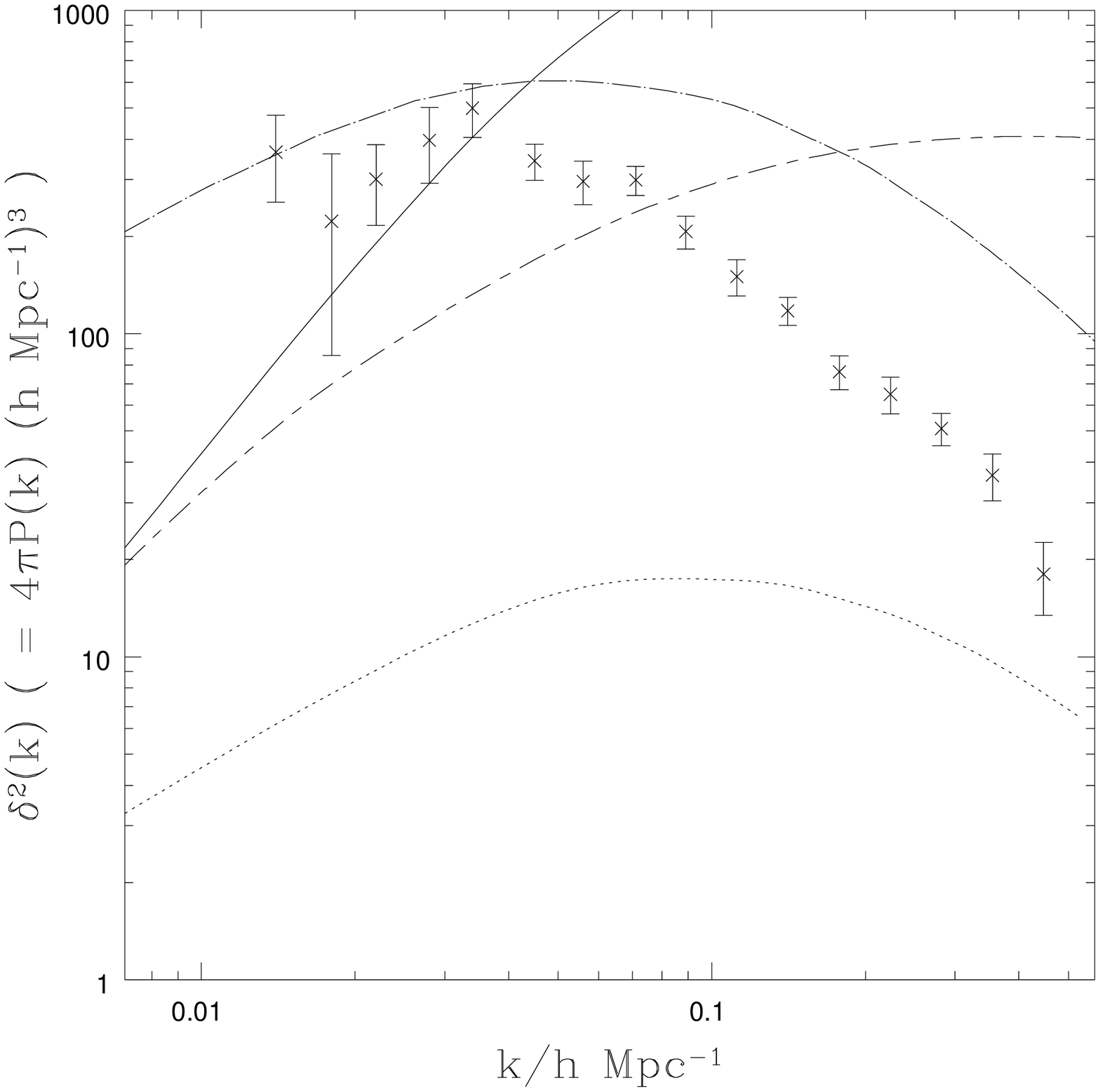,width=3.4in}}
\end{minipage}
\caption{The ``best of all worlds'' models: The COBE normalized angular
power spectrum of CMB anisotropies, (left-hand graph) and the matter
power spectrum (right-hand graph) for the standard scaling model
(dotted line) and two
``best of all worlds'' models. In the first of these models,
(long-short dash line), the values
of each of the parameters $\xi$, $v$, $h$, $\Omega_b$ and $L_f$ have been
pushed as far as possible in the direction which favours high
$\sigma_{100}$, while a standard scaling law is used. The second model,
(solid line) is identical, except that it also incorporates a fairly 
plausible
deviation from scaling over the matter-radiation transition.
Observational data
and the prediction for standard CDM (dot-dash curve) are included for
comparison.}
\label{best}
\end{figure}

In summary, we have scanned the range of each of the parameters
in our model, while maintaining perfect scaling, and we have found
that none of these simple variations are capable of significantly
reducing the deficit of power on scales around
$100h^{-1}$Mpc. However, some of the parameters do marginally improve
the situation, in some cases reducing the required $b_{100}$ from
$\approx 5$ to $\approx 3$. One might postulate, therefore, that
modifications to all these parameters simultaneously might lead to a
more substantive amount of power on 100$h^{-1}$Mpc scales, and indeed
this is the case. To illustrate this, we performed a run with
$\xi=0.0001$, $v=0$, $h=0.7$, 
$\Omega_b=0.01$, and $L_f=0.01$, which we describe as the `best of all
worlds' model, and the result is presented in  
Fig.~\ref{best}. We see that the situation is improved, but still the
bias required, $b_{100}=1.6$, is not unity and we now find a large
excess of power on scales around $8h^{-1}$Mpc. Also, the CMB angular
power spectrum appears to be worse fit to the data.  
While this model serves as a useful caveat to our
arguments, we believe that pushing the model this far is not realistic
within the current understanding of defect models. Nonetheless, it may
serve as impetus for future model building.

Except for the caveat described above, the minimal dependence of
$b_{100}$ on the wide variations in these parameters is already strong
evidence to suggest that the $b_{100}$ 
problem will be a feature of most scaling defect models. In the next
two sections we further test this idea by examining the results of
further modifications and generalizations of the standard model, with
all deviations being described as perturbations from the standard
scaling model.

We should note that there are two simple variations of the cosmogony which we have ignored in the section on cosmological parameters, namely an open universe
($\Omega <1$) or the introduction of cosmological constant
$(\Omega_m+\Omega_{\Lambda}=1)$. We anticipate that these variations
will lead to modifications to scaling similar to those described in
the next section \cite{M,VB,Fer,ACM}, and might lead to more acceptable
values of $b_{100}$. An in depth investigation of this problem is the
subject of ongoing research \cite{us} (see also ref.\cite{paul}). 

\section{Variations from standard scaling}
\label{dev}
\subsection{Motivation and implementation}

In the previous section, we introduced the $b_{100}$ problem for
scaling defect models and showed that it is apparently robust for a range
of different parameters. However, we have also noted that this
scaling
assumption has only been tested using defect
simulations with a very small
dynamic range. Hence, in the spirit of testing of the standard model,
we should allow for the
 possibility of modifications to scaling. In fact, this
is the  most obvious resolution to the $b_{100}$ problem since the
scaling assumption is what
relates the contributions from defects on different scales.  Deviating
from scaling would  allow us to effectively tilt the power
spectrum. A similar approach has become a popular solution to the excess power on small scales in standard CDM models based inflation, but in our case we want to create more power on large scales. 

This point can be illustrated most effectively by reconsidering
Fig. \ref{standard} which shows the results for the standard model and
also the partial contributions from the strings present in during
$z=1300$ to $z=100$ (short dashed curve) and between $z=100$ and
$z=1.6$ (long dashed curve). These two curves give us intuition about
when the perturbations relevant for COBE normalization and $b_{100}$
are laid down. For example, if one could create an imbalance between
the strings present during these two time windows, one could hope to
change their relative amplitude and reconcile the current data points
with COBE normalization. This is possible if one modifies the scaling
picture, with most graphic illustration being the total removal of
the string network around $z=100$.  

The first type of deviation we consider is motivated by
the mild shift in the behaviour of a string network which is observed in
simulations under going a radiation-matter transition.
Typically, quantities such as the string velocity, persistence length,
string density, and level of small scale structure
are seen to undergo a small change at around
radiation-matter equality. In general, this shift tends to make
strings move slower and be less dense. 

One simple way to implement this step-type transition is to allow the
effective energy per unit length to change, with the mass per unit length
being a factor $\chi$   
larger before the transition  than after it.  We generate 
histories for a source with normal scaling behaviour and then for each
history, we multiply the value of each source component at each time $\tau$
by some factor $f_R(\tau)$.
Since we require our source histories to behave
smoothly, we implement a smooth shift in the value of $f_R$ using
\begin{equation}
\label{eqn-fr}
f_R(\tau)=1+(\chi-1)T^{\rm off} (\tau,\tau_T,L_T)\,,
\end{equation}
where $T^{\rm off}$ is the same smoothly varying function which was
used for turning the string segments off, defined in 
(\ref{eqn_toff}), but now $L_T\tau_T$ is
the starting time of the transition, and $\tau_T$ is the end time.

We have also tried other ways of implementing a 
transition, such as introducing a shift in the time dependence of the
number of strings per unit volume. 
In the standard case, we have $n(\tau)=(\xi\tau)^{-3}$, which we
modify  by setting $n(\tau)=f_R(\tau)(\xi\tau)^{-3}$,
where $f_R$ is the smoothly varying transition function given in 
(\ref{eqn-fr}). 
We find that the net results from these two ways of implementing the
transition are   
very similar and so in the results section, we concentrate on the former, 
simpler case. 

The second type of deviation we consider is a deviation
in the scaling exponent. We implement such a transition by altering the dependence of 
the number of strings per unit volume on time.
In the standard scaling picture, there is roughly one piece of string
per correlation length cubed. Since the correlation length is
proportional to the horizon size $\tau$, we find that the number
density of strings as a function of time is, $n(\tau)\propto \tau^{-3}$, which we modify by setting $n(\tau)\propto \tau^{-(3+2\alpha)}$,
with
$\alpha=0$ being the standard value.
Using (\ref{eqn-singsource}) we see that the  power spectra of the
$\Theta_{\mu \nu}$'s depend on 
$ n(\tau) $
and hence the time dependence of
$\Theta_{00}$ (which behaves like the square root of the power
spectrum) outside the horizon is now
$\Theta_{00}=\tau^{-(\frac{1}{2}+\alpha)}$

\begin{figure}
\setlength{\unitlength}{1cm}
\begin{minipage}{8.0cm}
\leftline{\psfig{file=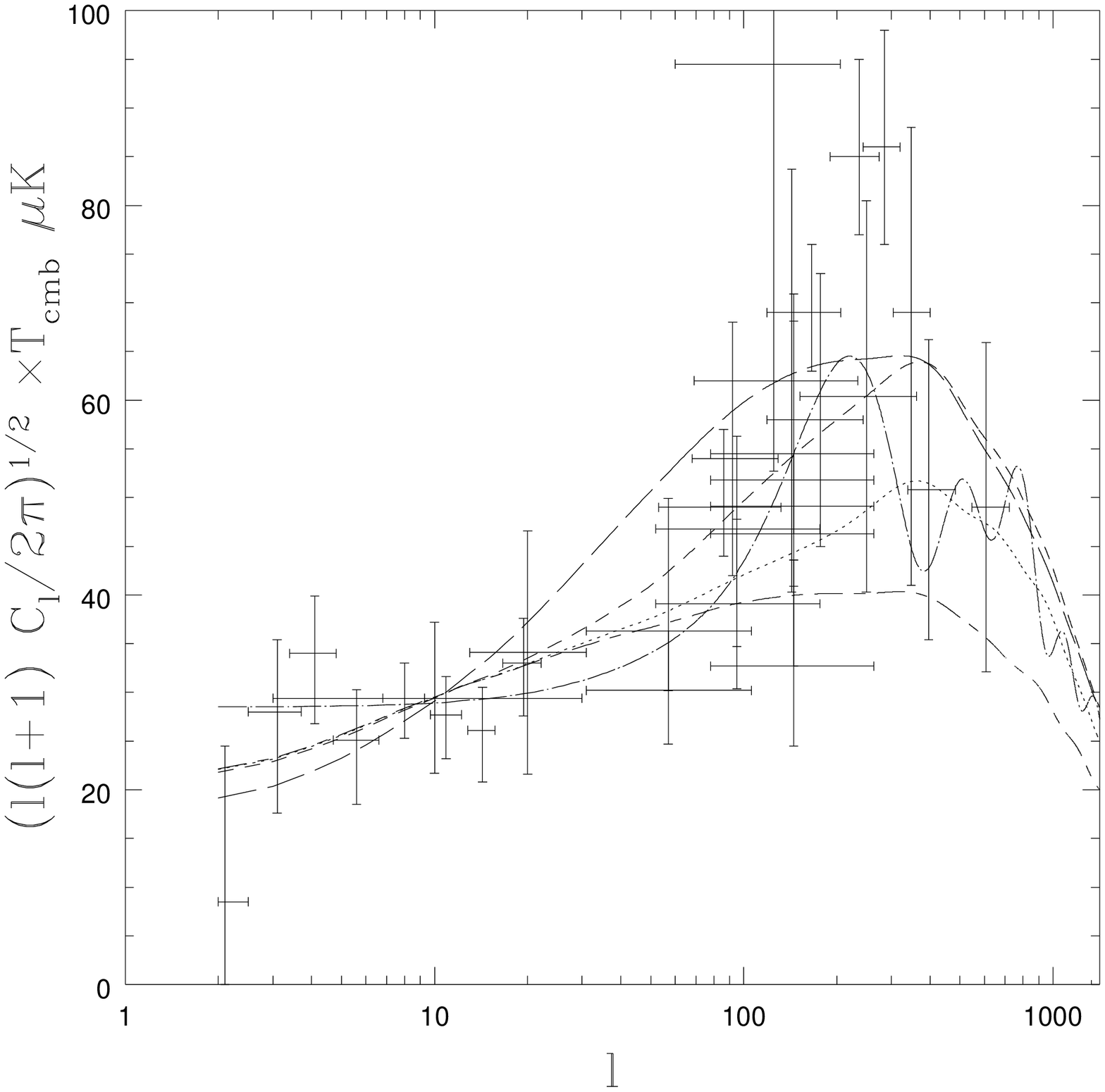,width=3.4in}}
\end{minipage}\hfill
\begin{minipage}{8.0cm}
\rightline{\psfig{file=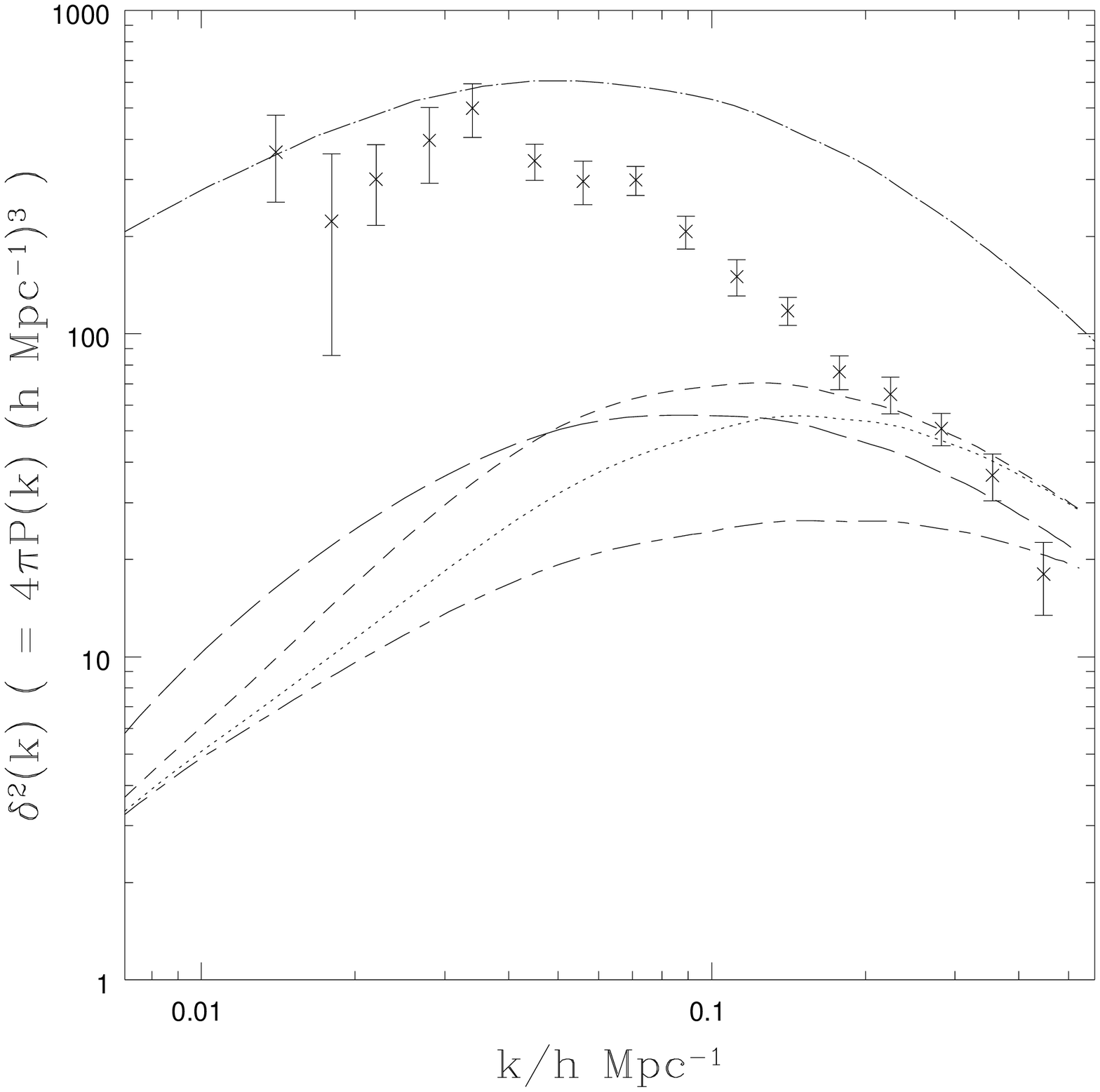,width=3.4in}}
\end{minipage}
\caption{Varying $\tau_T$, the end time for the modelled non-scaling
radiation-matter transition:  We plot the
COBE normalized 
angular power spectrum of CMB 
anisotropies,(left-hand graph) and the matter power spectrum
(right-hand graph) for a matter radiation transition with 
various values of the final time $\tau_T$, with
$L_T=0.1$ and $\chi=2$.($\tau_T=100$ --
long-short-dash line, $\tau_T=400$ --
dotted line, $\tau_T=1000$ --
short-dash line, $\tau_T=5000$ --
long-dash line).
Observational data
and the prediction for standard CDM (dot-dash curve) are included for
comparison.}
\label{mu1}
\end{figure}

\begin{figure}
\setlength{\unitlength}{1cm}
\begin{minipage}{8.0cm}
\leftline{\psfig{file=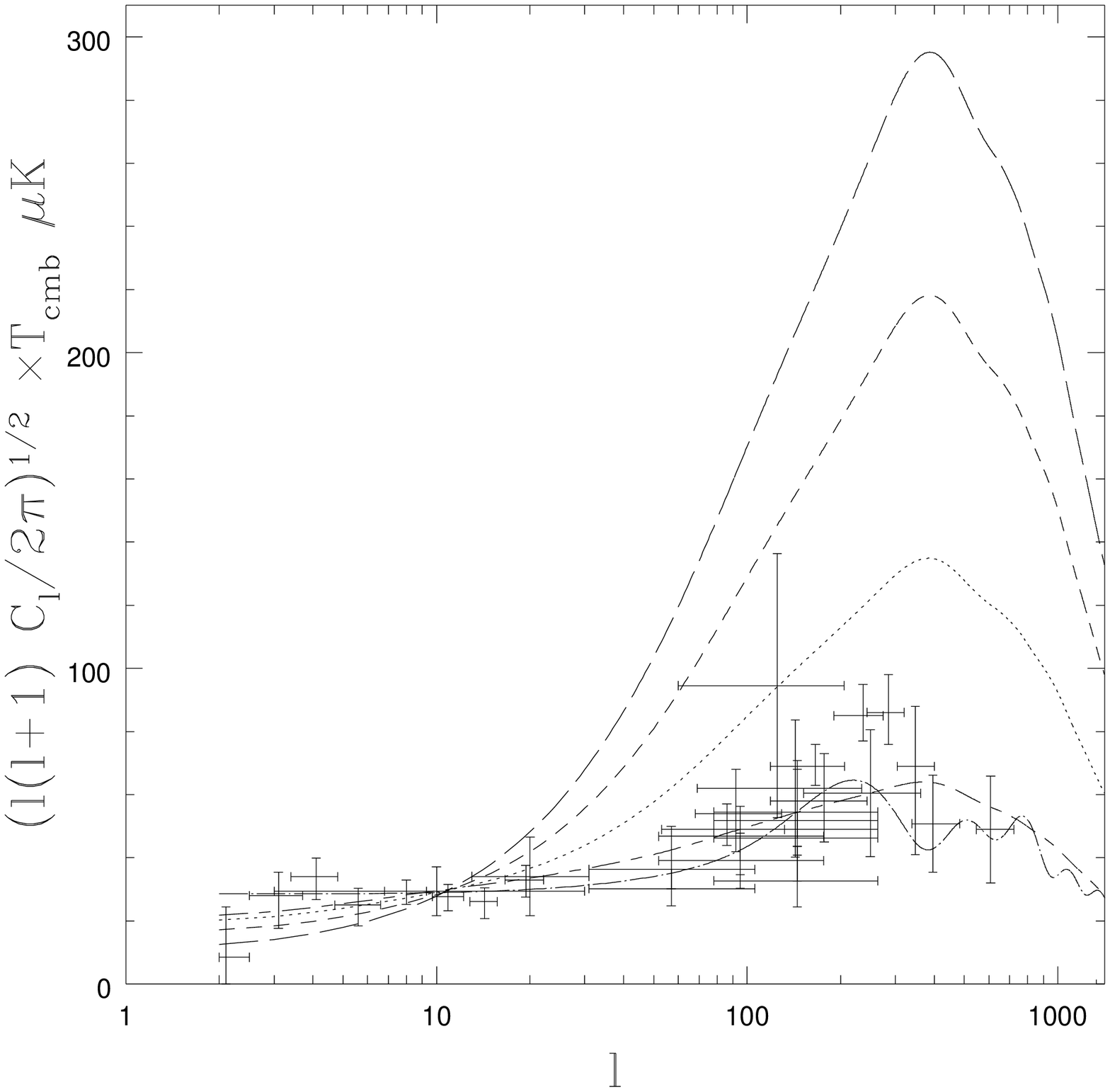,width=3.4in}}
\end{minipage}\hfill
\begin{minipage}{8.0cm}
\rightline{\psfig{file=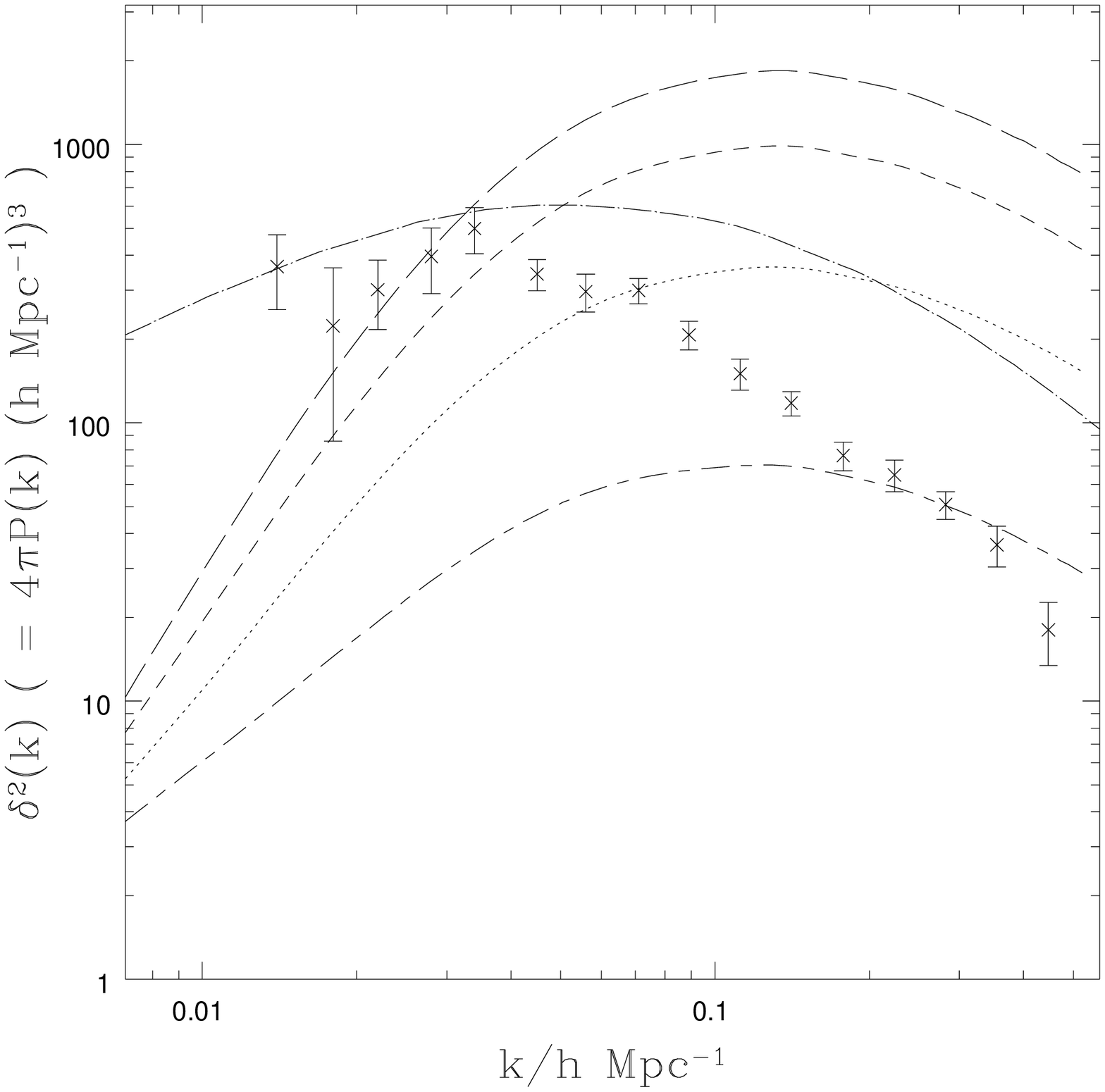,width=3.4in}}
\end{minipage}
\caption{Varying $\chi$, the factor by which $\mu_{\rm eff}$ is assumed to
be larger before the radiation-matter transition:  We plot the COBE normalized
angular power spectrum of CMB 
anisotropies, (left-hand graph) and the matter power spectrum
(right-hand graph) for a matter radiation transition with 
various values of the factor $\chi$, with
$\tau_T=1000$ and $L_T=0.1$.($\chi=2$ --
long-short-dash line, $\chi=5$ --
dotted line, $\chi=10$ --
short-dash line, $\chi=20$ --
long-dash line).
Observational data
and the prediction for standard CDM (dot-dash curve) are included for
comparison.}
\label{mu2}
\end{figure}

\begin{figure}
\setlength{\unitlength}{1cm}
\begin{minipage}{8.0cm}
\leftline{\psfig{file=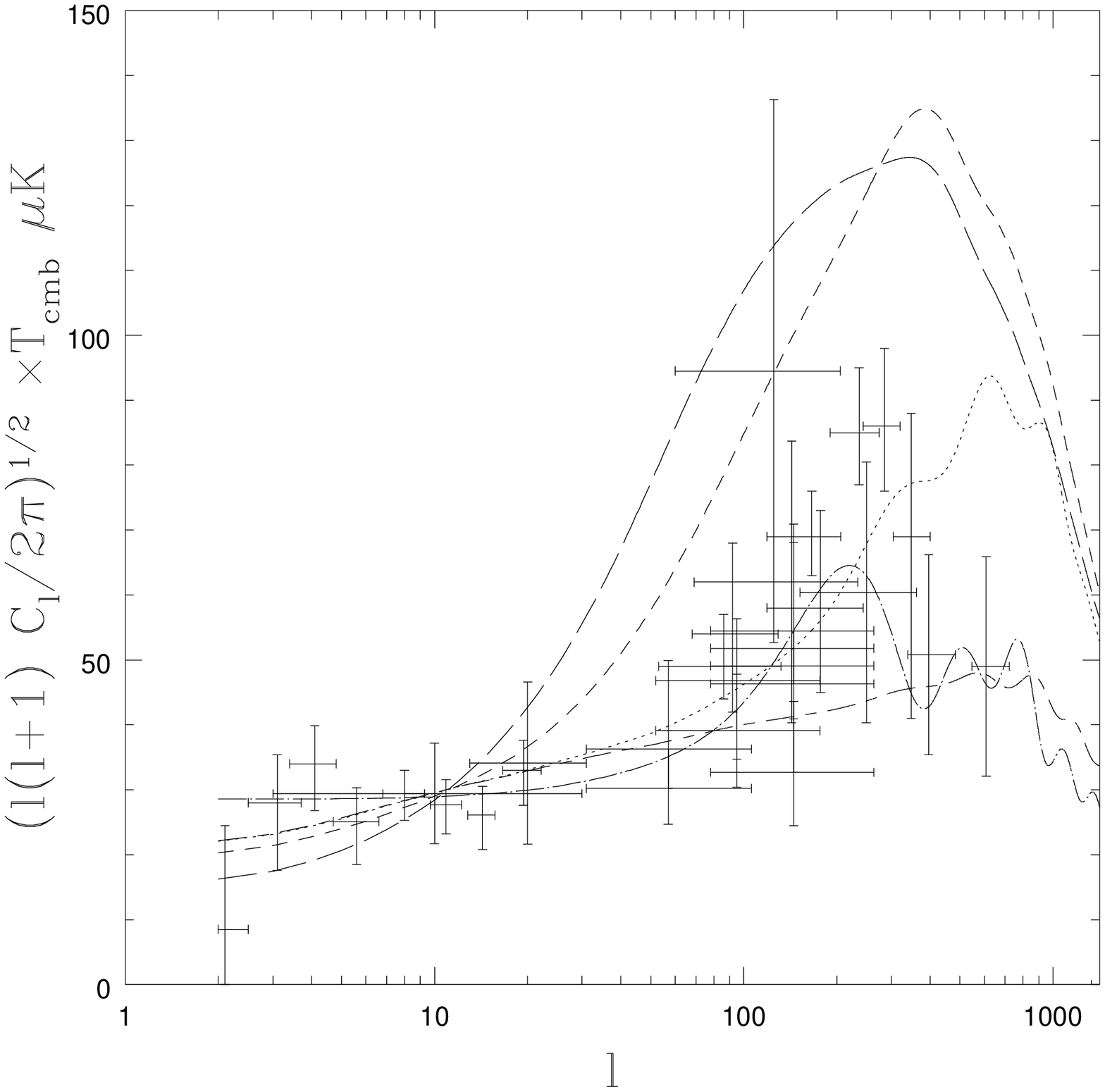,width=3.4in}}
\end{minipage}\hfill
\begin{minipage}{8.0cm}
\rightline{\psfig{file=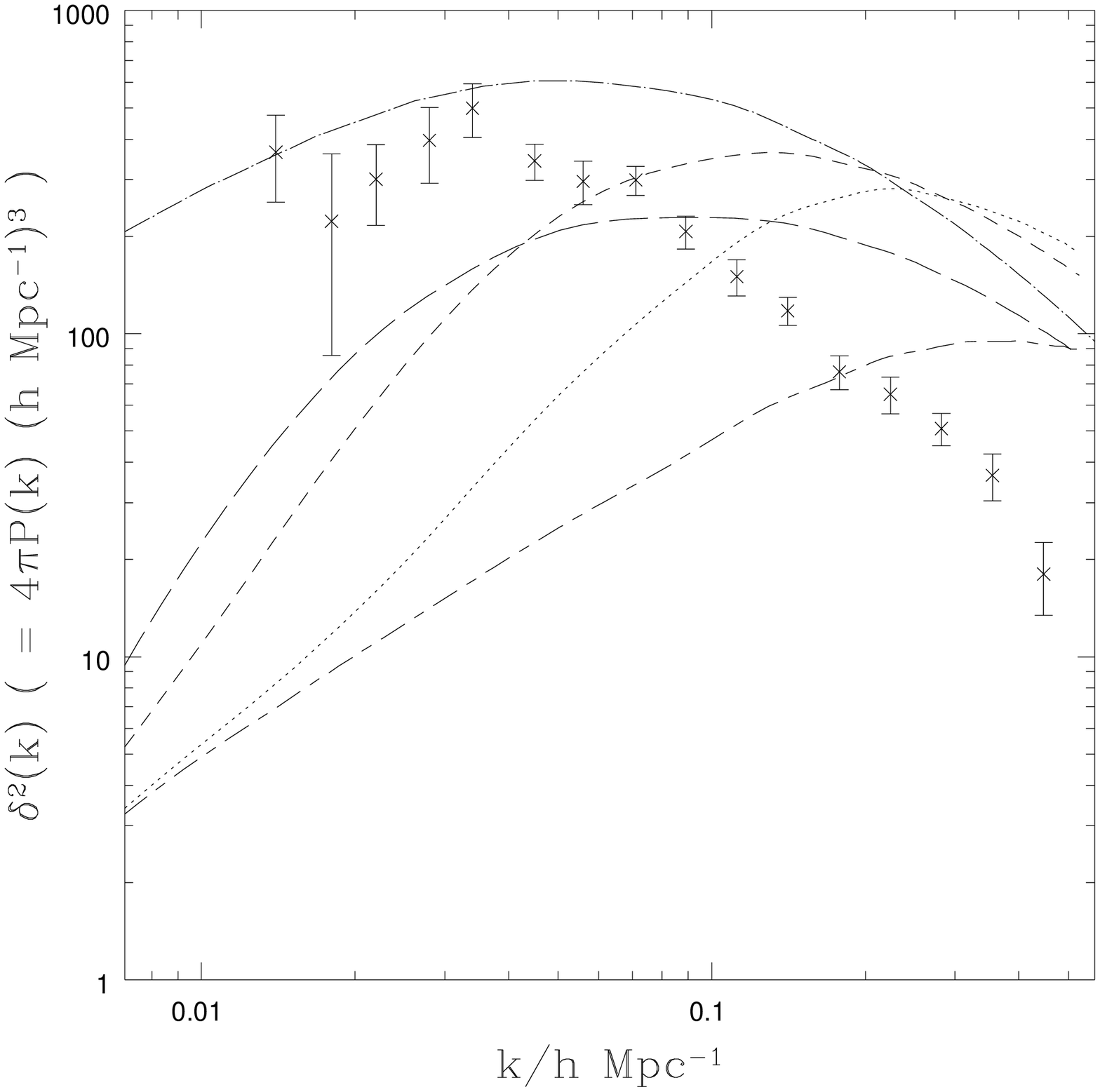,width=3.4in}}
\end{minipage}
\caption{Varying $L_T$, the start time (in units of the end time)
of the non-scaling radiation-matter transition modelled in the
defect sources:  We plot the COBE normalized angular power spectrum of CMB
anisotropies, (left-hand graph) and the matter power spectrum
(right-hand graph) for matter radiation transition models with 
different values of the length $L_T$, with amplitude $\chi=5$ 
($\tau_T=100$, $L_T=0.1$ --
long-short-dash line, $\tau_T=100$, $L_T=0.8$ --
dotted line, $\tau_T=1000$, $L_T=0.1$ --
short-dash line, $\tau_T=1000$,  $L_T=0.8$ --
long-dash line).
Observational data
and the prediction for standard CDM (dot-dash curve) are included for
comparison.}
\label{mu3}
\end{figure}

\begin{figure}
\setlength{\unitlength}{1cm}
\begin{minipage}{8.0cm}
\leftline{\psfig{file=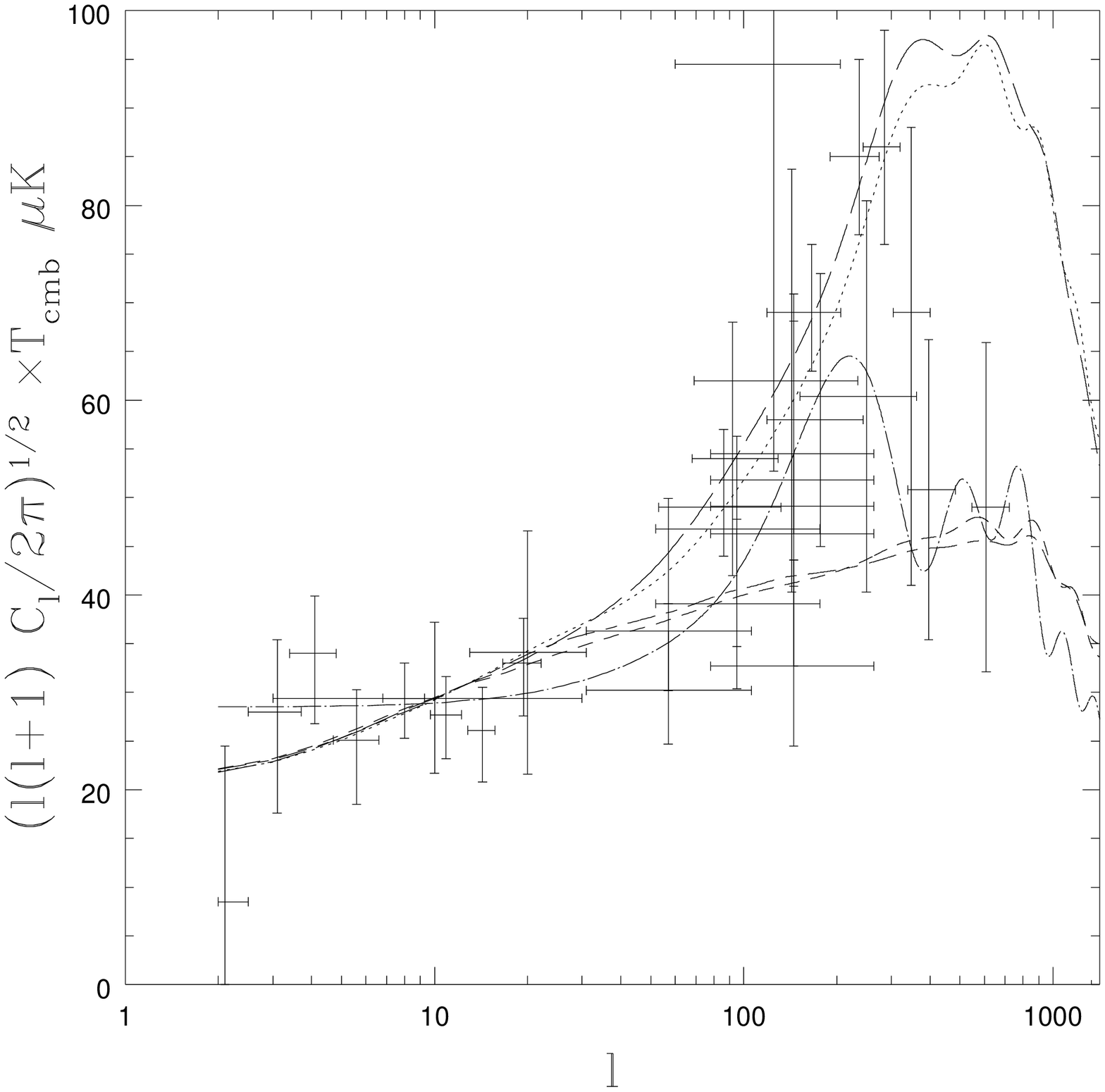,width=3.4in}}
\end{minipage}\hfill
\begin{minipage}{8.0cm}
\rightline{\psfig{file=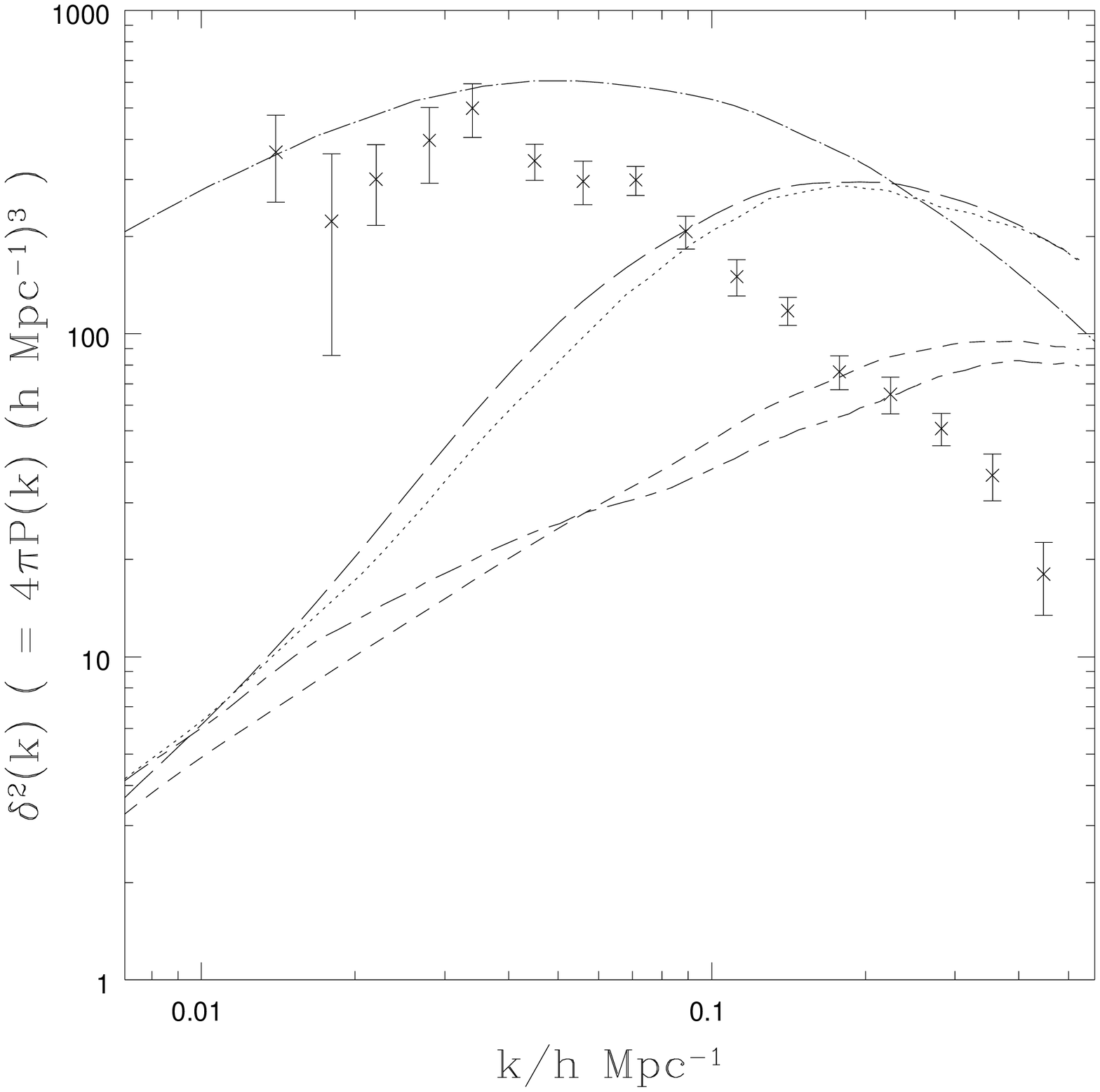,width=3.4in}}
\end{minipage}
\caption{Different types of non-scaling radiation-matter transitions:
We plot the COBE normalized angular power spectrum of CMB
anisotropies, (left-hand graph) and the matter power spectrum
(right-hand graph) for two different implementations of the matter
matter radiation transition, each with $\chi=5$ and $L=0.1$.
($\tau_T=100$, varying $n$ --
long-short-dash line, $\tau_T=400$, varying $n$ --
dotted line, $\tau_T=100$, varying $\mu$ --
short-dash line, $\tau_T=400$, varying $\mu$ --
long-dash line).
Observational data
and the prediction for standard CDM (dot-dash curve) are included for
comparison.}
\label{n}
\end{figure}

\begin{figure}
\setlength{\unitlength}{1cm}
\begin{minipage}{8.0cm}
\leftline{\psfig{file=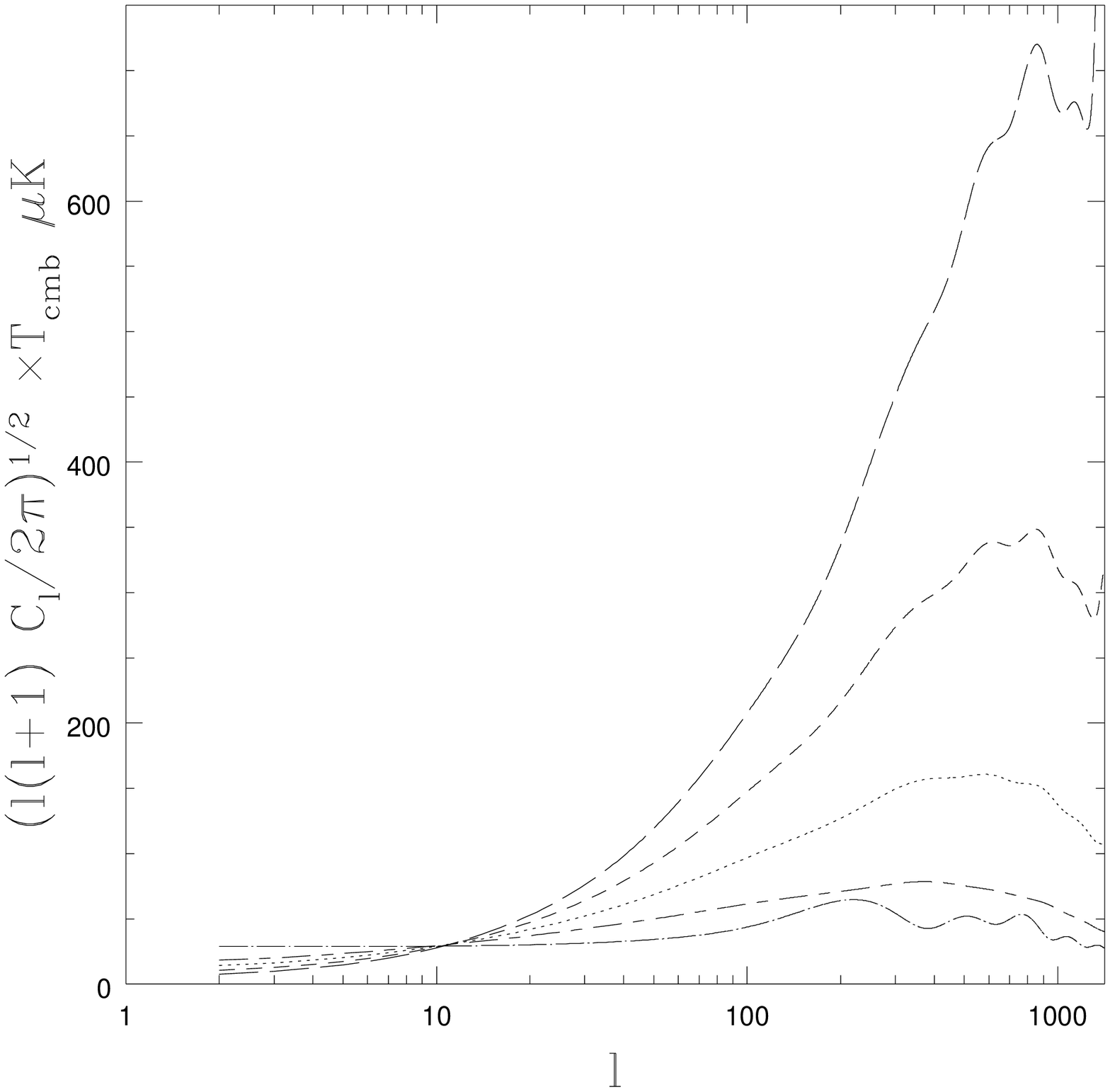,width=3.4in}}
\end{minipage}\hfill
\begin{minipage}{8.0cm}
\rightline{\psfig{file=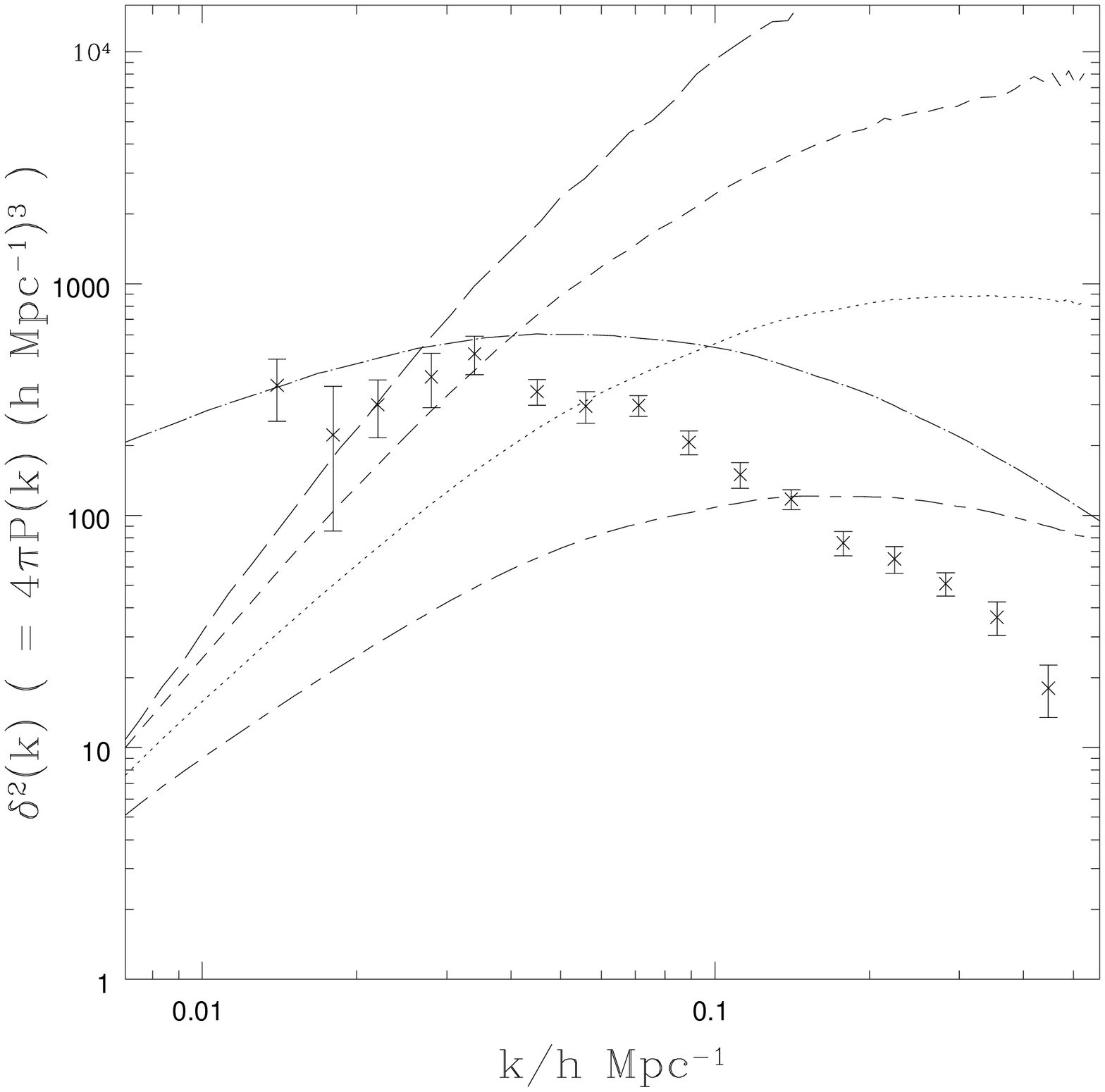,width=3.4in}}
\end{minipage}
\caption{Power law deviations from scaling:  We plot the COBE
normalized angular power spectrum of CMB 
anisotropies, (left-hand graph) and the matter power spectrum
(right-hand graph) for a power law deviation from scaling with 
various values of the parameter $\alpha$. ($\alpha=0.25$ --
long-short-dash line, $\alpha=0.5$ --
dotted line, $\alpha=0.75$ --
short-dash line, $\alpha=1.0$ --
long-dash line).
Observational data
and the prediction for standard CDM (dot-dash curve) are included for
comparison.}
\label{pow}
\end{figure}

\subsection{Illustrative examples}

Figs. \ref{mu1}, \ref{mu2} and \ref{mu3} illustrate the results for radiation-matter transition runs (implemented by varying $\mu$) with various 
choices of the parameters $\chi$, $\tau_T$ and $L_T$.
 The first (Fig. \ref{mu1}) shows mild transitions, with an
amplitude $\chi=2$, each lasting for a factor of 10 in conformal time $\tau$, that is, $L_T=0.1$.
Each curve shows results for a different choice of final time
$\tau_T$. Initially, we see that as the time of the transition is moved later,
the peak in the matter power spectrum gets higher and is shifted to
larger scales. However, as discussed in the standard scaling section
illustrating the two time windows, we find that as the time of the
transition is increased beyond $\tau_T=1000$, the height of the peak in
the matter power spectrum actually falls again. This is because a very
late transition tends to boost the perturbations on COBE scales as well
as those on scales relevant to the large scale matter power
spectrum. It is clear that a transition occurring as late
as today is equivalent to having no transition at all. None of these
reasonable, mild transitions ($\chi=2$) significantly improves the
$b_{100}$ problem. 

In the second figure (Fig. \ref{mu2}), we stick to a transition time
of $\tau_T=1000$, which as we have discussed gives us the best chance
of introducing a shift between the COBE and large scale matter
normalization. Using a transition length of $L_T=0.1$, we vary the
amplitude $\chi$ of the transition. A value of $\chi=10$ improves things significantly on scales of $100h^{-1}$Mpc and
$\chi=20$ does better still, but we find
that further increases only affect the features of the matter spectrum
on smaller scales. It is interesting that there is a limiting value
for the relative COBE/$\sigma_{100}$ normalization,
and that this limiting value
happens to fit the observations very well. We note however that the
cases which fit the large scale matter data require implausibly large
values of the transition amplitude $\chi$, for which no precedent has
been seen in studies of defect evolution. Furthermore, extremely
drastic additional alterations would have to be made to the model in
order to fit the small scale matter and CMB data. 

In the third figure (Fig.~\ref{mu3}), we illustrate the dependence on the
length $L_T$ of the transition, for two different choices of the final
time $\tau_T$, with $\chi$ fixed to be 5. We see that the length of the
transition does not strongly influence the $\sigma_{100}$
normalization.  
In fact, we have been unable to find any region of our transition
parameter space capable of fitting $\sigma_{100}$, where the choices of
parameter values is more plausible than those illustrated in
Fig. \ref{mu2}  

Fig.~\ref{n} illustrates the comparison between a radiation-matter
transition in which $\mu$ is varied, with one in which $n$ is
varied. The first pair of curves shows a transition with $L_T=0.1$ and
$\tau_T=100$. In the case where $\mu$ is varied, we choose $\chi=5$
(short-dash line) whereas in the case that $n$ is varied, we
choose $\chi=25$ (long-short dash line). The reason for this
difference in amplitudes is that increasing $\mu$ by a factor of $\chi$
increases the power spectrum of the perturbations by a
factor of $\chi^2$, while the same increase in $n$ affects the power
spectrum of perturbations by a factor of $\chi$. We see that the
resulting spectra for this pair of models are very similar.
The second pair of curves illustrate a transition with $L_T=0.1$ and
$\tau_T=400$, for the same choices of $\chi$ as above. Again,
the curves are very similar for each of the transition models. Hence,
the resulting spectra of perturbations does not seem to be strongly
dependent on the way in which the transition is implemented.

Fig.~\ref{pow} shows deviations in the scaling exponent of
various degrees. We see that models where the deviation from scaling
is significant enough to bring about a substantial increase in the
amount of power in the matter spectrum on scales around $100h^{-1}$Mpc
are so extreme that they completely miss the small scale matter and
CMB data. We find that further increases in the scaling exponent
beyond $\alpha=1.0$ do not affect the $\sigma_{100}$ normalization,
only giving rise to significant differences in the resulting power
spectra on smaller scales. As in the case of the radiation-matter
transition, it is interesting to note that there is a limiting value
of $\sigma_{100}$ normalization with increasing alpha, and that this
limiting value happens to pass through the large scale matter data.

As one final point, Fig.~\ref{best} shows the results of a relatively
mild deviation from scaling (varying $\mu$, $\chi=2$,
$\tau_T=10\tau_{eq}$,
$L_T=0.8$) where all of the other standard parameters have been pushed
as far as possible in a direction which favours a large value for
$b_{100}$ as in the `best of all worlds model' ($\xi=0.0001$, $v=0$, $h=0.7$,
$\Omega_b=0.01$, and $L_f=0.01$) discussed earlier. In this case, $b_{100}$ is of order one. However, in obtaining a
reasonable value for $b_{100}$, the resulting model totally fails to fit
the shape of the matter power spectrum, with an extreme excess of
power on smaller scales. Although we do not believe this
limit will allow the resurrection of the standard defect scaling
defect scenario, it does present a possible road of attack for the 
construction of more exotic models which could fit all of the data.

To summarize, we have presented results for a number of models showing
deviations from scaling. None of these models is able to fit the large
scale matter data
without extreme corrections to the standard scaling picture, or
forcing all other model parameters in the direction which favours
large $b_{100}$. Even
in the cases where the value of $b_{100}$ is reasonable, a very
considerable amount of further work would need to be carried out in
order to make the model fit small scale  matter and CMB data. 

\section{Further modifications to the model}

In the previous sections we have discussed possible variations from 
our standard scaling
string model, and we showed that only extreme
deviations from the standard scaling model can significantly rectify
the problem.  In section III we showed how
the $b_{100}$ problem is relatively robust to changes in the model parameters
$v$, $\xi$ and $L_f$ (as well as the cosmological parameters $\Omega_b$ and
$h$). Although all of the variations we have considered take  
place within the context of our string model, the robustness to these 
changes already provides evidence to suggest that the results will be
similar for other types of defect. For instance, the independence of
the results on the parameter $\xi$ suggests that results will be
similar for other types of defect for which the two-point functions cut
off at different sub-horizon values. In this section we further test
this idea by introducing further modifications to the forms of our two-point functions by hand, in order to see how extreme these
modifications must become before the $b_{100}$ situation is
significantly improved.

One change which might alleviate the problem is to alter the relative
strength of 
various components of the stress-energy.
Recent work \cite{DS}, which makes use of a
coherent approximation to model source behaviour, has suggested that
defect models with a highly suppressed anisotropic stress might give
rise to more acceptable values of the matter bias factor
$b$ and indeed we have already mentioned that the ratio of $\Theta^S$
and $\Theta_{00}$ is model dependent. In order to investigate this
possibility in the context of our work, we make a  
simple modification to our standard scaling model. 
We  multiply
the energy $\Theta_{00}$ by $\nu$, where in models with values of $|\nu|<1$
the significance of the anisotropic stress is boosted relative
to that of the energy, while in models with $|\nu|>1$, the energy 
is boosted. We should note that the ratio of the anisotropic stress to
vector and tensor components is unaltered, as required by isotropy and causality.

\begin{figure}
\setlength{\unitlength}{1cm}
\begin{minipage}{8.0cm}
\leftline{\psfig{file=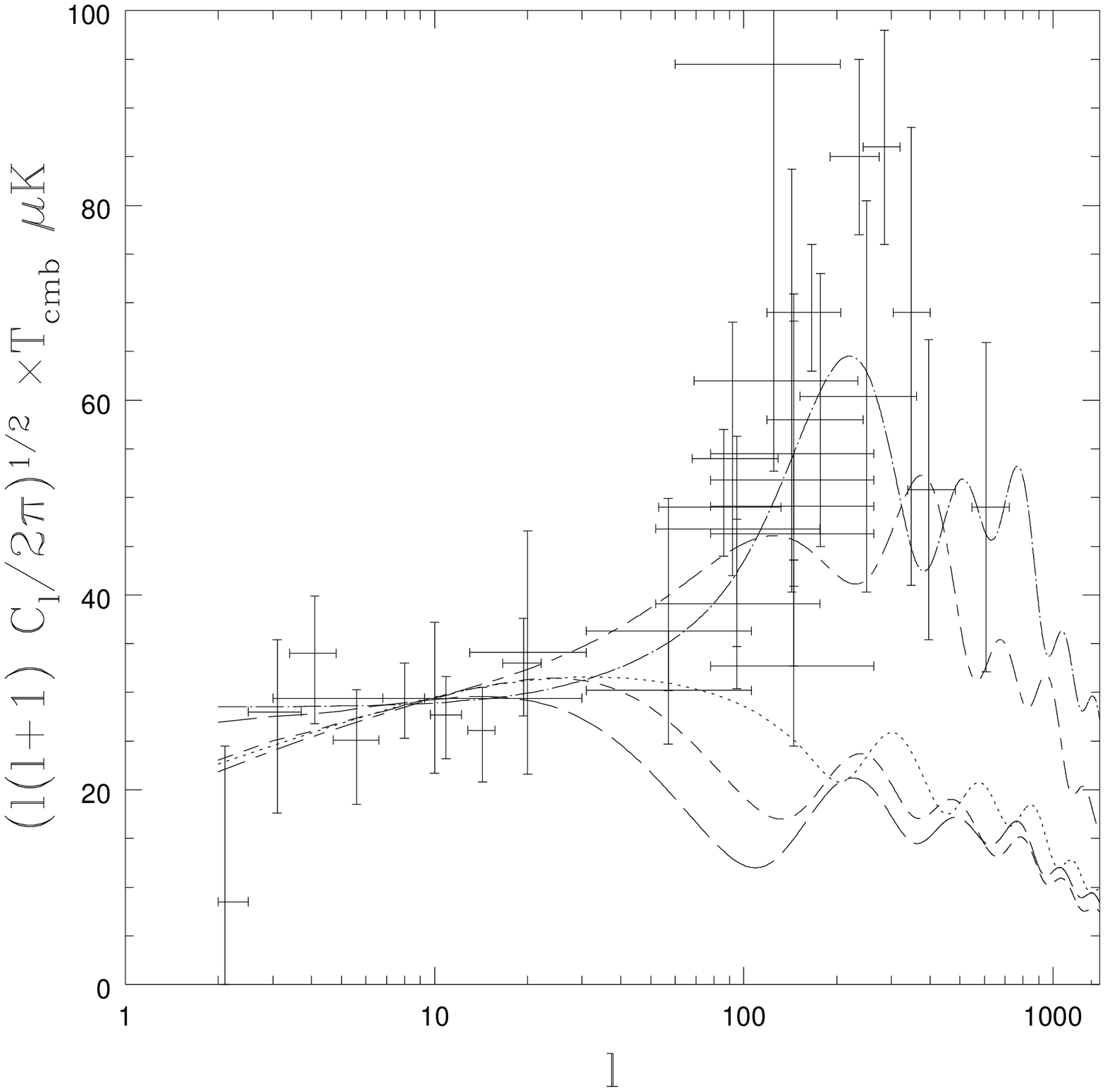,width=3.4in}}
\end{minipage}\hfill
\begin{minipage}{8.0cm}
\rightline{\psfig{file=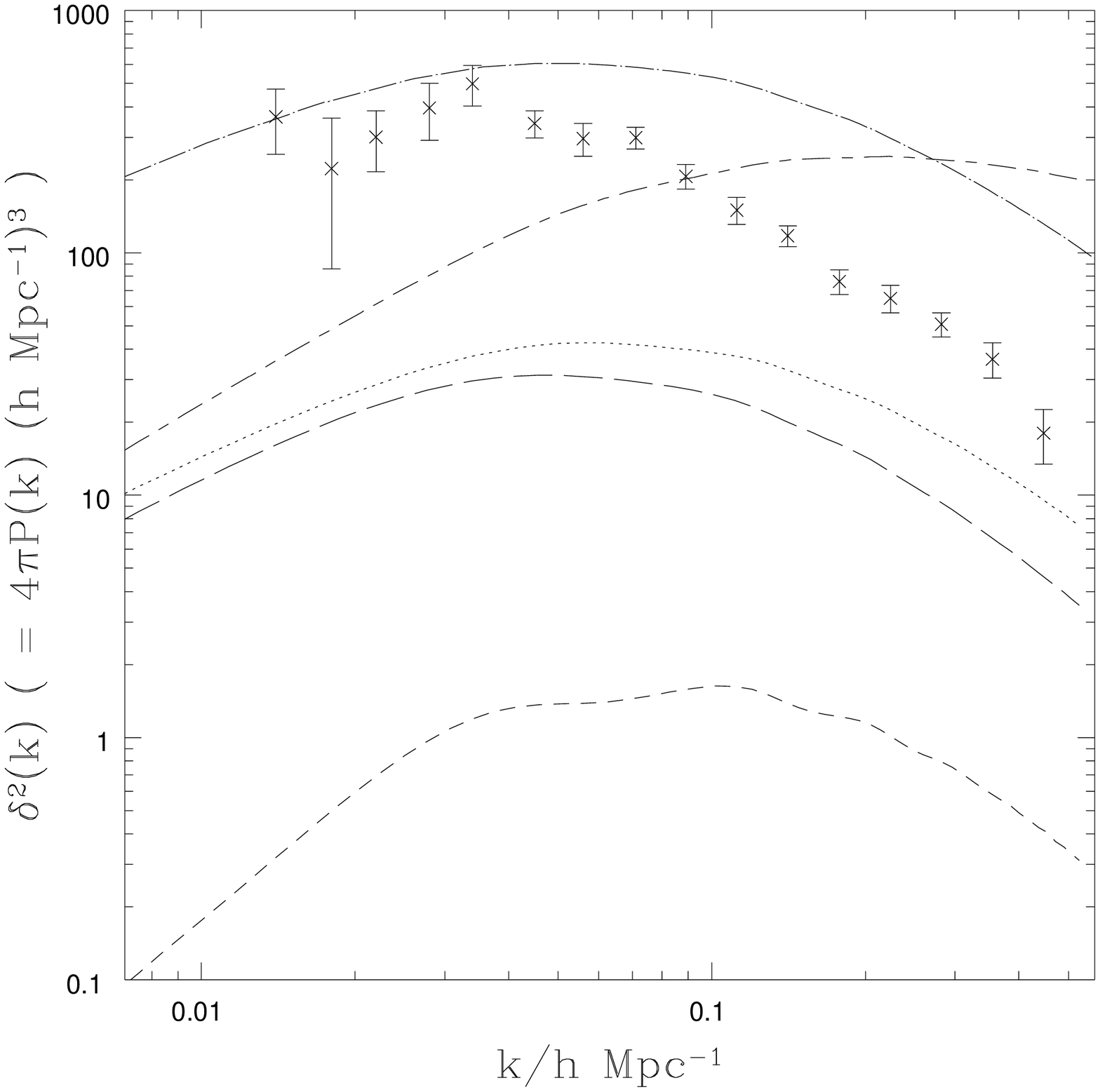,width=3.4in}}
\end{minipage}
\caption{Various coherent limits:  We plot the COBE normalized angular
power spectrum of CMB 
anisotropies, (left-hand graph) and the matter power spectrum
(right-hand graph) for a range of coherent models discussed in the
text.
(Non-causal models: C1 --
long-short-dash line, C2 --
dotted line.  Causal Models:  C3 --
short-dash line, C4 --
long-dash line).
Observational data
and the prediction for standard CDM (dot-dash curve) are included for
comparison.}
\label{co}
\end{figure}

Before presenting results for our full string model, we discuss the
effect of these changes in the context of some simple coherent
models\footnote{Since in each of these models, either
$\Theta_{00}$ or $\Theta^{S}$ is zero, super horizon constraints on
the cross-correlator $\langle\Theta_{00}\Theta^{S}\rangle$ place no
constraints on the super horizon behaviour of $\Theta^S$ (see section 
\ref{cohere}).}. 
In Fig.~\ref{co} we present the CMB and matter power spectra for four such models.
The first model (which we call C1) has
\begin{eqnarray}
\Theta_{00}&=& 
 \left\{ \begin{array} {ll}
         \tau^{-1/2}  & \quad k\tau \le 5 \,,\\
         0 & \quad k\tau > 5 \,,
         \end{array} \right.\nonumber \\
\Theta_{S}&=&0 \,,
\end{eqnarray}
while for the second model (C2)
\begin{eqnarray}
\Theta_{00}&=&0\,, \nonumber \\
\Theta^S &=& \left\{ \begin{array} {ll}
         \tau^{-1/2}  & \quad k\tau \le 5 \,,\\
         0 & \quad k\tau > 5\,. 
         \end{array} \right.
\end{eqnarray}
For this pair of models, the version with suppressed anisotropic
stress has a more acceptable value of $b_{100}$ than the version with
suppressed energy. However, this pair of models is not exactly causal,
because the real space two-point correlation functions do not exactly
vanish outside the horizon.
We also present results for the following pair of models, which do
exactly satisfy this condition. They are (C3), with
\begin{eqnarray}
\Theta_{00}&=&\tau^{-1/2} \frac{\sin (k\tau)}{k\tau}\,,  
 \nonumber \\
\Theta_{S}&=&0 \,,
\end{eqnarray} 
and (C4) with
\begin{eqnarray}
\Theta_{00}&=&0\,,
 \nonumber \\
\Theta_{S}&=&\tau^{-1/2} \frac{\sin (k\tau)}{k\tau}\,.  
\end{eqnarray} 
The sources in the
second pair of models are designed to exhibit sub-horizon 
decay at a similar value of
$k\tau$ as the first pair of models we have discussed.
However, we see that in this explicitly causal case, the model with
suppressed energy actually gives a more acceptable $b_{100}$ than the
model with suppressed anisotropic stress. These results represent a
surprising contrast to those of the case which is not exactly causal,
implying that exact imposition of causal constraints is necessary for
physically meaningful results.  Activity outside the causal
horizon presents opportunities to seed standard `adiabatic' perturbations
(which will naturally make the $b_{100}$ look better).  We suspect our
slightly acausal model has done this to a remarkable degree.

\begin{figure}
\setlength{\unitlength}{1cm}
\begin{minipage}{8.0cm}
\leftline{\psfig{file=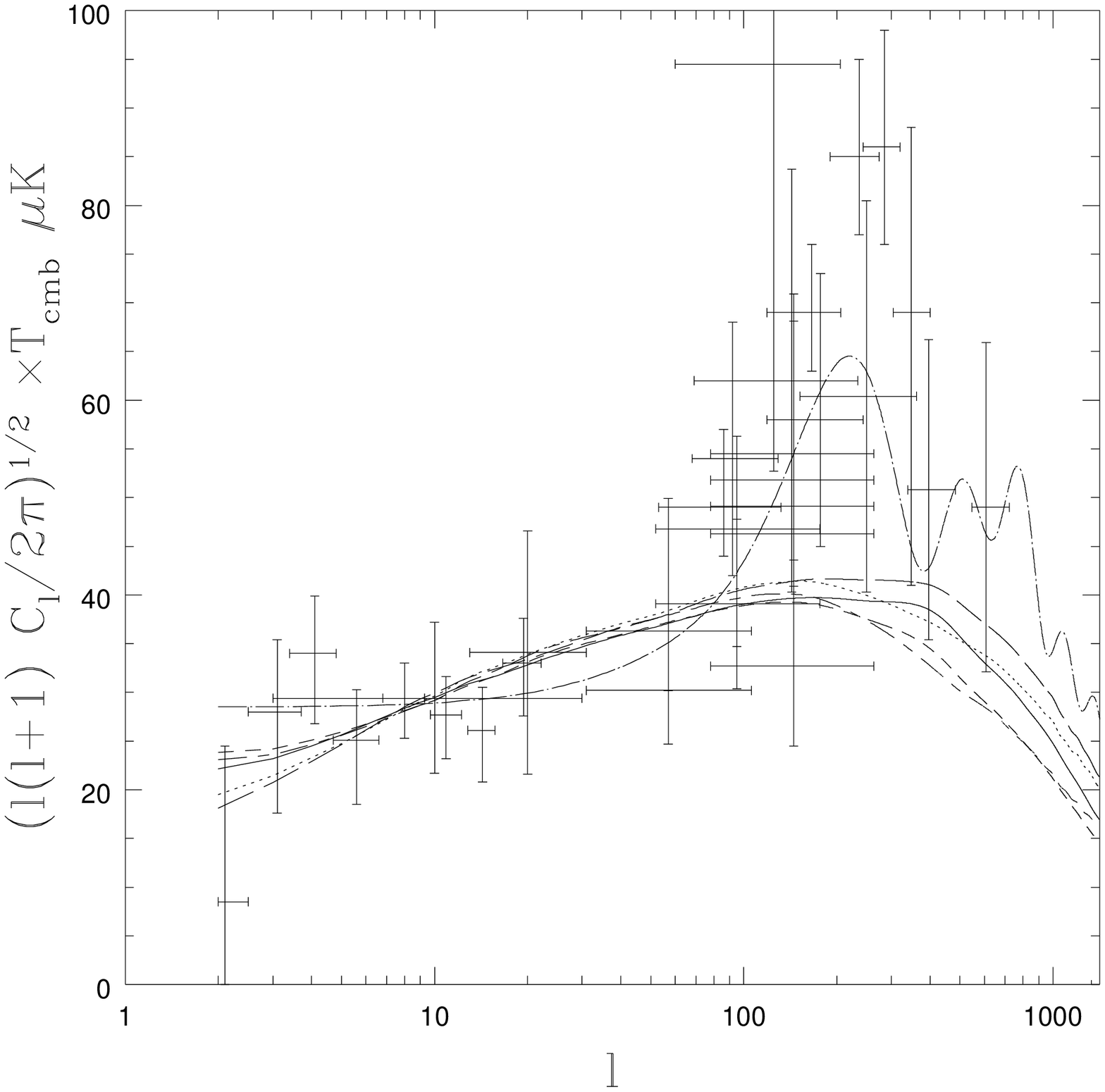,width=3.4in}}
\end{minipage}\hfill
\begin{minipage}{8.0cm}
\rightline{\psfig{file=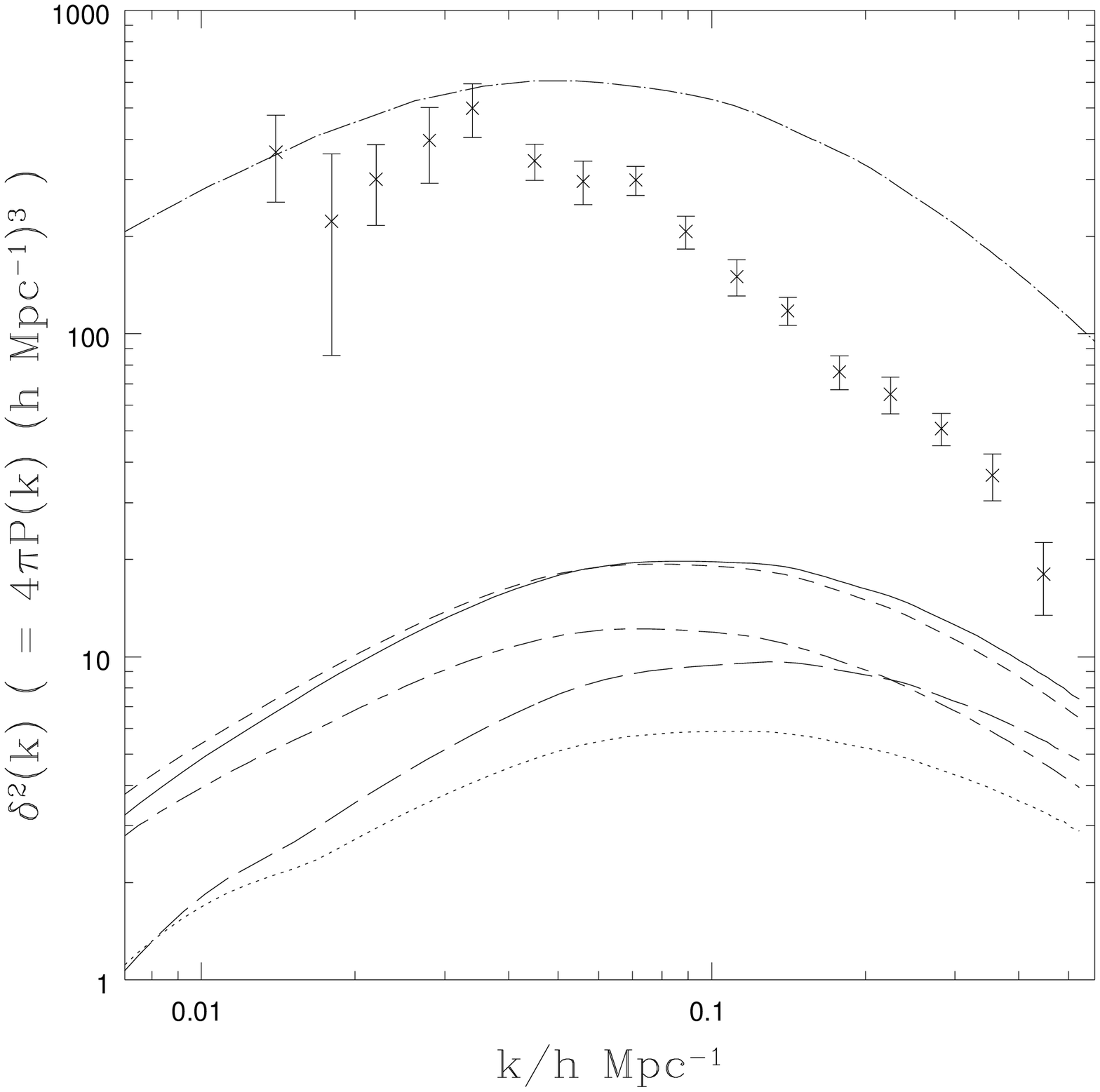,width=3.4in}}
\end{minipage}
\caption{Varying $\nu$, the enhancement factor for $\Theta_{00}$ which
artificially modifies the ratio between 
energy and anisotropic stress:  The 
COBE normalized angular power spectrum of CMB anisotropies, (left-hand
graph) and the matter power spectrum (right-hand graph) for standard
scaling subject to various values of 
the parameter $\nu$.
($\nu=1$ -- solid line -- $\nu=0$ --
long-short-dash line, $\nu=-1$ --
dotted line, $\nu=0.5$ --
short-dash line, $\nu=\infty$ --
long-dash line).
Observational data
and the prediction for standard CDM (dot-dash curve) are included for
comparison.}
\label{w}
\end{figure}

In Fig.~\ref{w} we present the results for different values of the
energy factor $\nu$, modifying our standard string two-point
functions. As in the causal coherent case above, we see that
$\sigma_{100}$ is actually better for models with suppressed energy
rather than suppressed anisotropic stress and in fact no value of $\nu$ would
give rise to a significant improvement in the $b_{100}$ problem (see table \ref{tab-all}).

\begin{figure}
\setlength{\unitlength}{1cm}
\begin{minipage}{8.0cm}
\leftline{\psfig{file=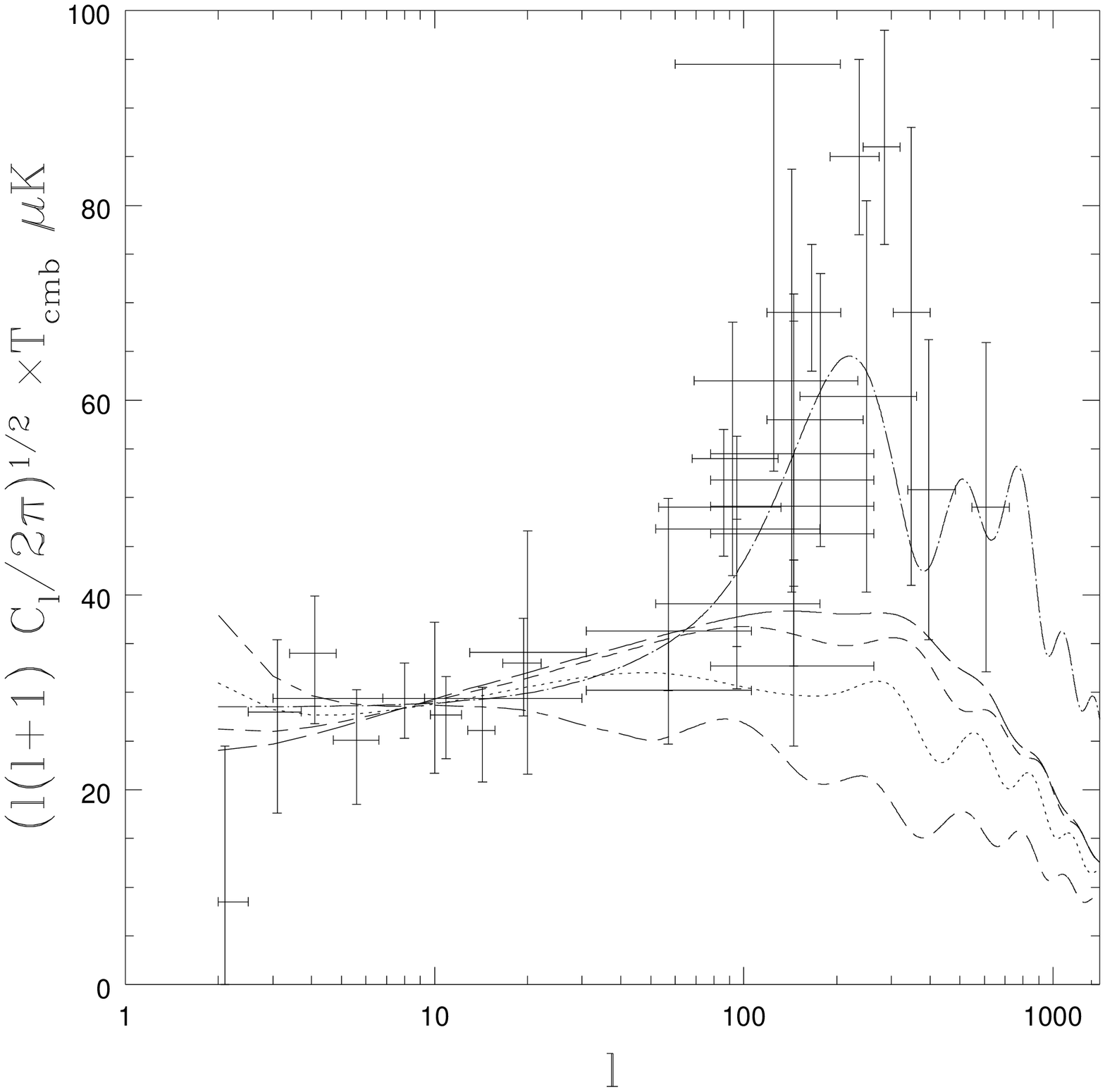,width=3.4in}}
\end{minipage}\hfill
\begin{minipage}{8.0cm}
\rightline{\psfig{file=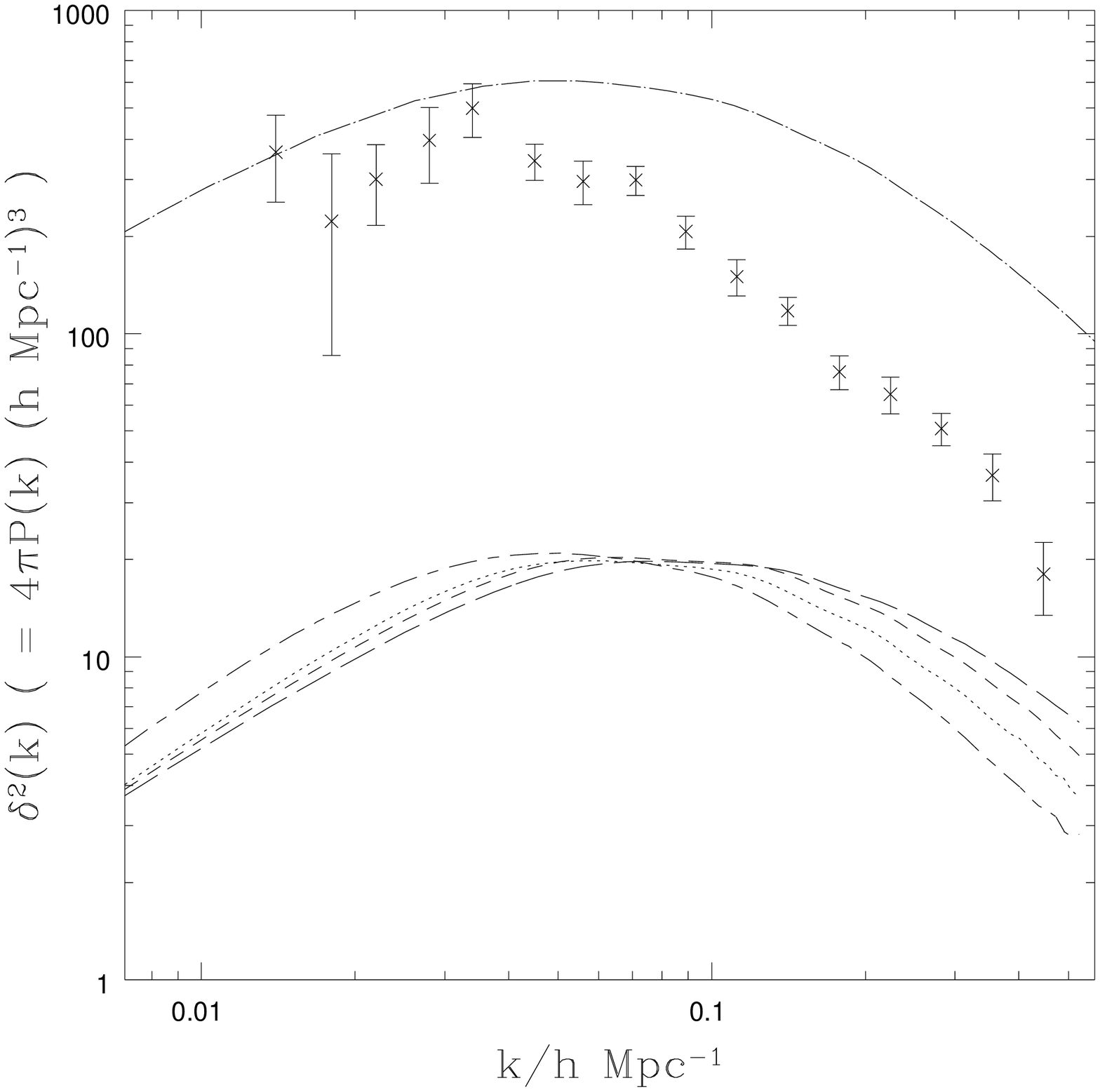,width=3.4in}}
\end{minipage}
\caption{Varying the cutoff parameter $\epsilon$:  The COBE normalized
angular power spectrum of CMB anisotropies, (left-hand graph) and the
matter power spectrum (right-hand graph) for our standard scaling
source, with various values of the parameter $\epsilon$, representing
the value of $k\tau$ at which the source is subject to a sharp
subhorizon cutoff.  
($\epsilon=2$ --
long-short-dash line, $\epsilon=5$ --
dotted line, $\epsilon=10$ --
short-dash line, $\epsilon=20$ --
long-dash line).
Observational data
and the prediction for standard CDM (dot-dash curve) are included for
comparison.}
\label{e}
\end{figure}

Another way in which we have modified our standard model is by
imposing a sharper sub-horizon cutoff in the source stress-energy. By
doing this, we hope to have covered a range of possible defect
behaviour including that of cosmic textures, whose stress-energy
tensor will exhibit a faster sub-horizon fall-off than that of strings,
reflecting the fact that they have less features on small scales. We
implement this particular modification without violating the
requirements of causality as follows. We introduce a 
parameter $\epsilon$, and specify
that the number of strings with decay times $\tau_f$ satisfying
$k\tau_f> \epsilon$ is zero, while the number of strings with decay
times satisfying $k\tau_f< \epsilon$ is unchanged. 
Results for various choices of the parameter $\epsilon$ are shown in Fig.~\ref{e}. We see that $b_{100}$ does not depend strongly on the
the value of $\epsilon$.

\begin{figure}
\setlength{\unitlength}{1cm}
\begin{minipage}{8.0cm}
\leftline{\psfig{file=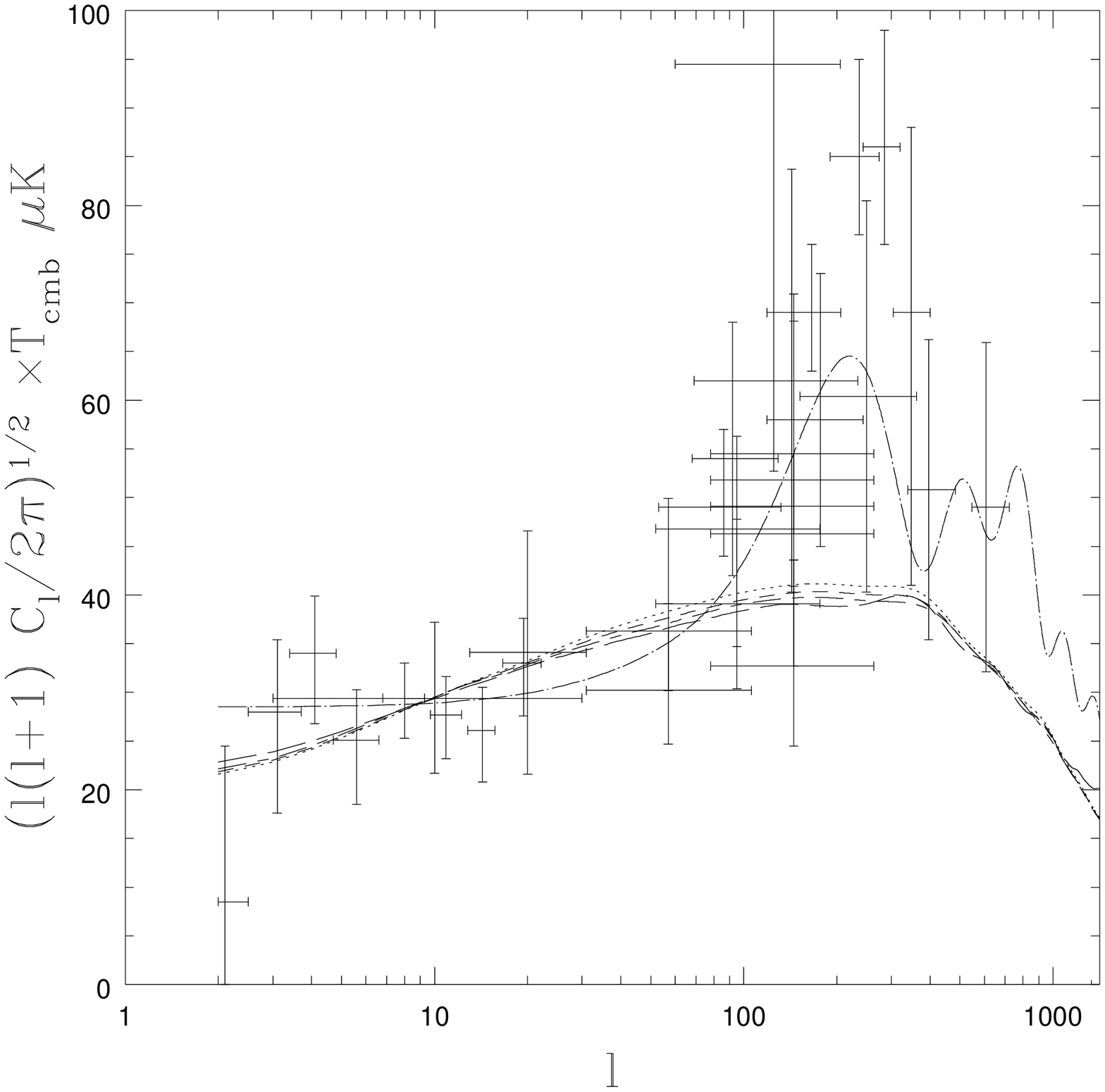,width=3.4in}}
\end{minipage}\hfill
\begin{minipage}{8.0cm}
\rightline{\psfig{file=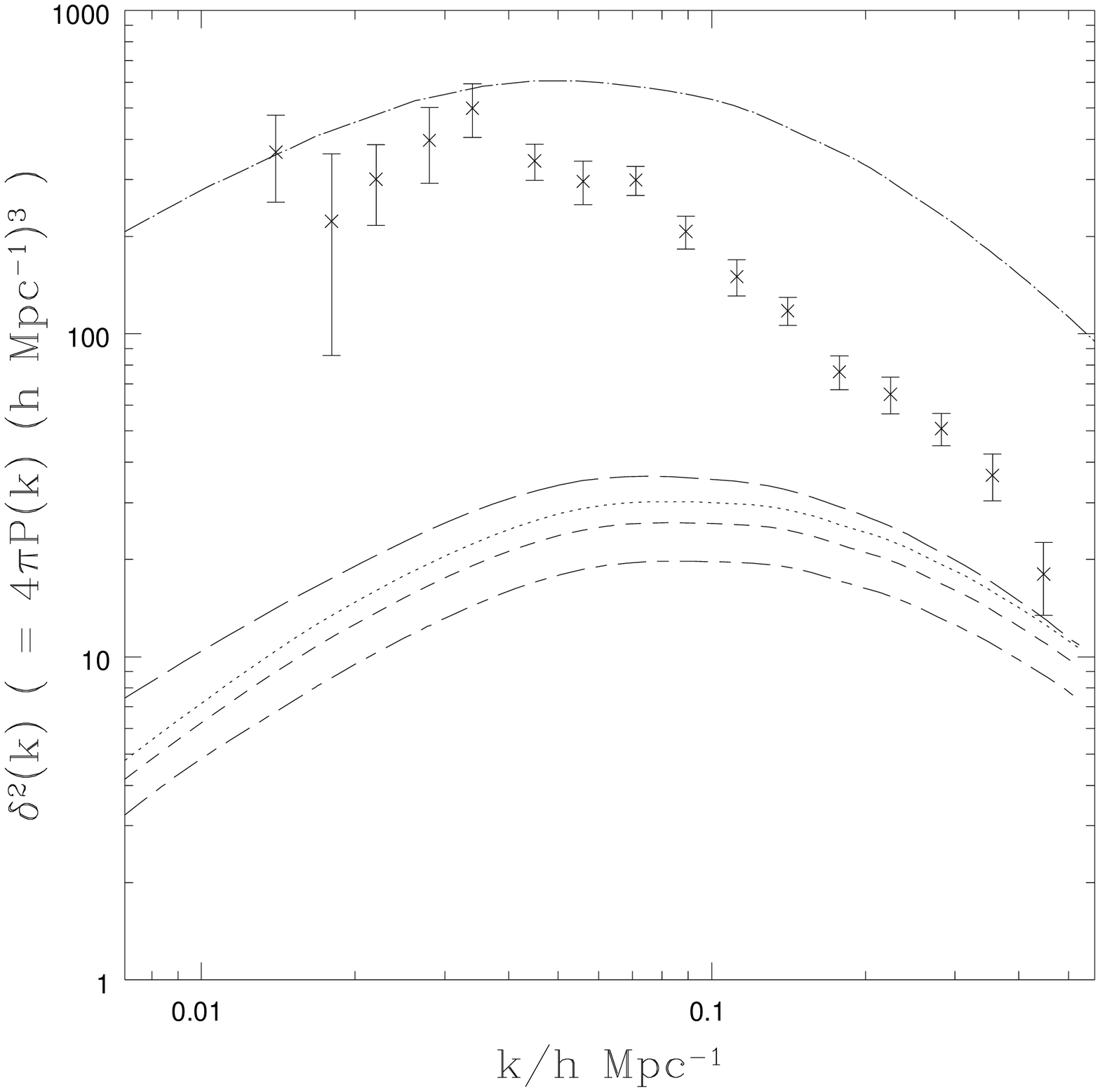,width=3.4in}}
\end{minipage}
\caption{Scaling can be enforced by varying either the string density
or the mass per unit length.  Here we plot the COBE normalized angular
power spectrum of CMB 
anisotropies (left-hand graph), and the matter power spectrum
(right-hand graph) for a scaling source, with various
values of the parameter $q$, which parameterizes the way scaling is
enforced ($q=0$ corresponds to constant mass per unit length).  We use
$v=0.65$ 
and $\xi=0.3$.  
($q=0$ --
long-short-dash line, $q=-1$ --
dotted line, $q=-1.25$ --
short-dash line, $q=-1.5$ --
long-dash line).
Observational data
and the prediction for standard CDM (dot-dash curve) are included for
comparison.}
\label{sg}
\end{figure}

Finally, we allow modifications to the way in which scaling is enforced
in our model. We note that our standard scaling model contains two
time dependent functions, one being the number of strings per unit
volume $n(\tau)$, the other $\mu(\tau)$ being the mass per unit length
of the strings. In the case of standard scaling,
$n(\tau)\sim\tau^{-3}$ while $\mu(\tau)$ is a constant. We note that
an extreme case which would also exhibit scaling behaviour would be a
model in which the number of strings $n$ was constant, but where $\mu$
evolved like $\tau^{-3/2}$. Such a case would bear some resemblance to the
inflationary mimic model of Turok \cite{Ta}, where there is a constant number of
expanding shells, but some feedback mechanism is required to modify
the shell surface density in order to introduce the correct scaling behaviour.
In fact, if we model $n(\tau)$ as $\tau^{-3+p}$ and $\mu(\tau)$ as $\tau^{q}$,
then it is easy to verify that all models satisfying $p+2q=0$ will
give rise to scaling behaviour. We modify our model by allowing
variations in the parameter $q$ from the value zero. As discussed, the
choice $q=-3/2$ represents a type of coherent limit. Results for such
variations are shown in Fig.~\ref{sg}. We see that as we approach the
coherent limit ($q=-3/2$) the value of $b_{100}$ is increased, but
still doesn't come close to the data. 

\begin{figure}
\setlength{\unitlength}{1cm}
\begin{minipage}{8.0cm}
\leftline{\psfig{file=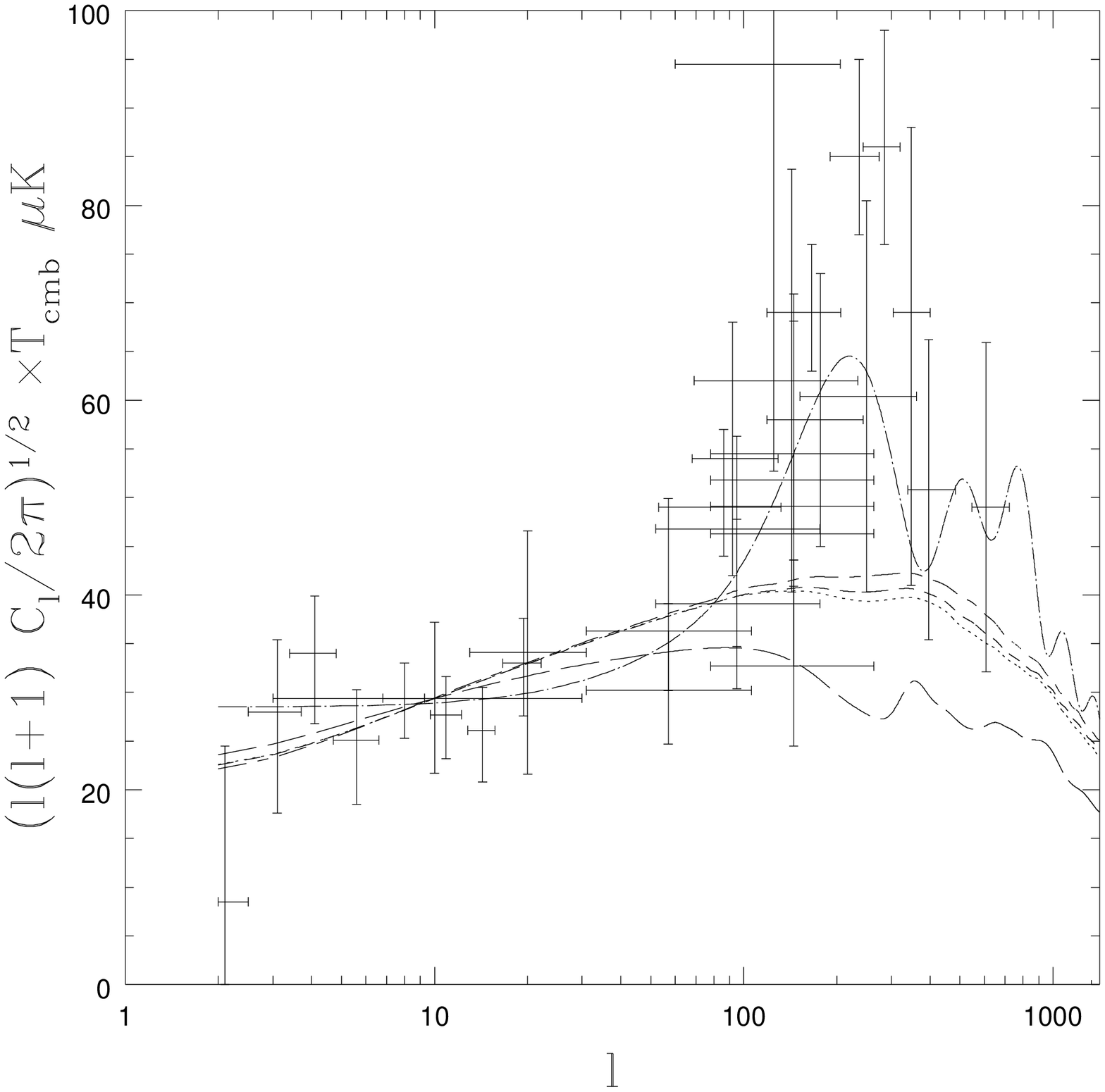,width=3.4in}}
\end{minipage}\hfill
\begin{minipage}{8.0cm}
\rightline{\psfig{file=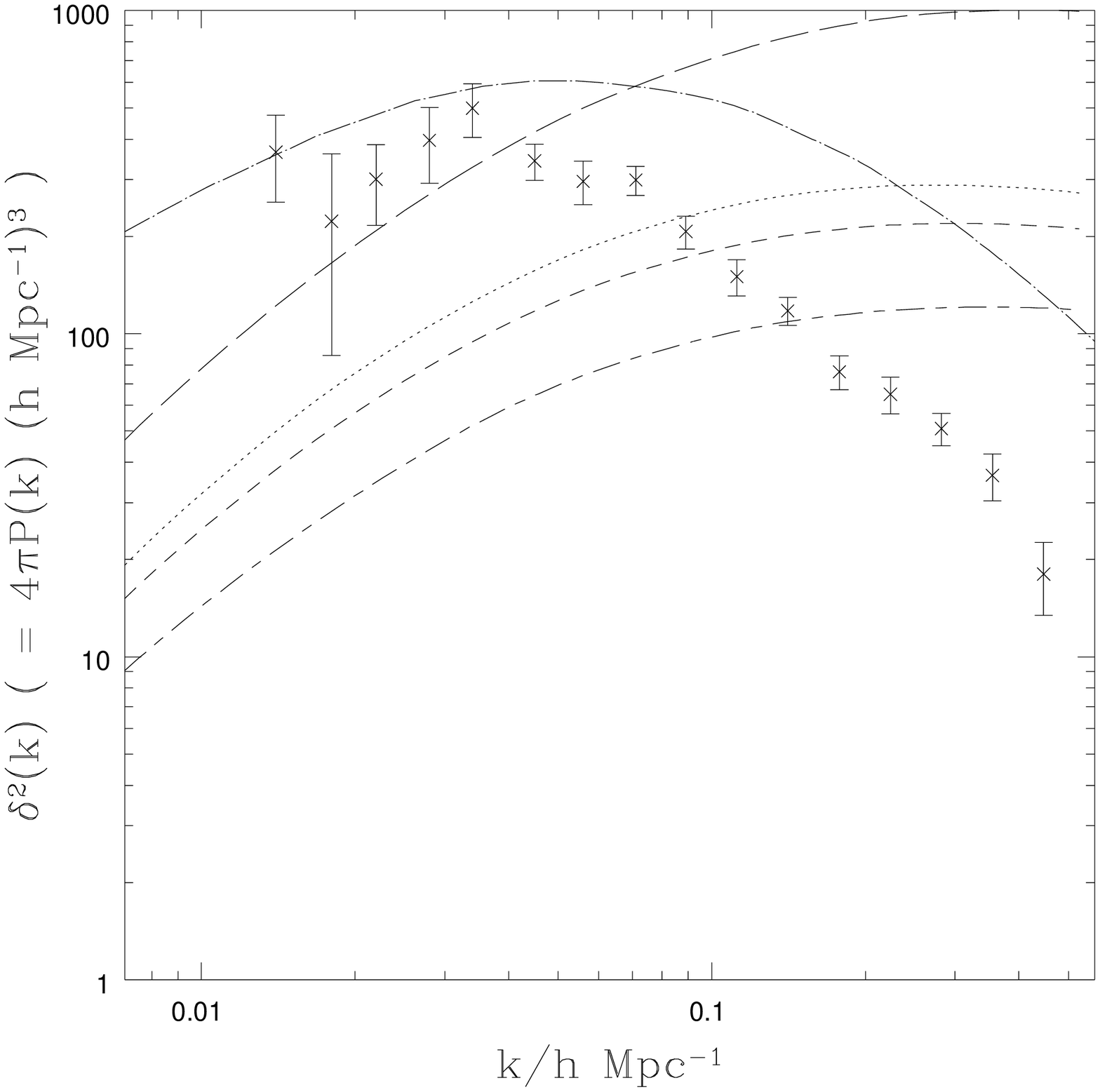,width=3.4in}}
\end{minipage}
\caption{Varying the scaling quality parameter $q$ with $v=0.0$ and
$\xi=0.001$: We plot the COBE normalized angular power spectrum of CMB
anisotropies (left-hand graph), and the matter power spectrum
(right-hand graph) for a scaling source, with 
$q=0$ --
long-short-dash line, $q=-1$ --
dotted line, $q=-1.25$ --
short-dash line, $q=-1.5$ --
long-dash line.
Observational data
and the prediction for standard CDM (dot-dash curve) are included for
comparison.}
\label{sg2}
\end{figure}

We now present one final extreme variation on our standard scaling model,
which does significantly improve the $b_{100}$ problem. Fig.~\ref{sg2}
illustrates the effect of varying the parameter $q$ in the limit that
both the string velocity and the coherence length are very small. We
find that in the coherent limit ($q=-3/2$), the matter spectrum comes
close to the data on large scales. So in the coherent limit,
the $\sigma_{100}$ normalization depends more strongly on changes in
the parameters $v$ and $\xi$ than it did in the standard, incoherent
case. Although this extreme model (like the mimic model \cite{Ta})
demonstrates that the $b_{100}$ 
problem is not a necessary consequence of the assumption of scaling,
we stress that our modifications of the degree of coherence via the
parameter $q$ as far as the coherent limit
had no physical motivation. All known defects achieve
scaling behaviour by having some kind of random decay process, such as
loop production in the case of strings, or decay into Goldstone
bosons during unwinding in the case of
textures. Consequently, scaling seems necessarily to imply incoherence,
which suggests that scaling models with constant $n(t)$ are a special,
unphysical limit. 
We see in the figure that it is only in the coherent limit that the
choice of low $\xi$ and $v$ significantly reduces the $b_{100}$ problem.

\section{Mimic inflation and variations}
 
One obvious counter example to the arguments we have put forward is
Turok's mimic inflation 
\cite{Ta}. This is a scaling active model which is quite different
from a standard defect model, which has been shown to produce the
same structure of acoustic peaks in the small scale CMB as adiabatic
theories. It is of
interest here because it also manages to overcome the difficulties of the
majority of models discussed in this paper by giving rise to a COBE normalized
matter power with no deficit of power on large scales.

Mimic inflation is also of interest because the components of the source
stress energy which are directly modelled are $\Theta_{00}+\Theta$ and
$\Theta_{D}$ (as opposed to  $\Theta_{00}$ and
$\Theta^{S}$ which we extract in our string model). The component
$\Theta_{00}+\Theta$ is particularly interesting because it accounts for essentially all of the contributions to the the scalar CMB fluctuations created at the surface of last scattering and also to the
matter power spectrum. 

In this section, we briefly explore variations on the mimic model in
which the ratio of $\Theta_{00}+\Theta$ to $\Theta_D$ is allowed
to vary freely. These variations do not respect the constraint
$\langle \Theta_{00} \Theta^S \rangle \sim k^2$ outside the
horizon. However, we believe that they are worth looking at because
they illustrate a point about the relative dependence of ISW and
matter contributions on $\Theta_{00}+\Theta$ and $\Theta_D$.

Our mimic model is based on the version presented by Hu, Spergel and
White (ref. \cite{HSW}). We take

\begin{eqnarray}
\Theta_{00}+\Theta&=& C_1 \tau^{-1/2} \frac{ \sin (Ak\tau)} {(Ak\tau)}\,, \\
\Theta_D&=&f_D C_2 \tau^{-1/2}\frac{6}{B_2^2-B_1^2} \frac{1}{k\tau}\left[
 \frac {\sin (B_1 k\tau)} {(B_1 k\tau)} -\frac {\sin (B_2 k\tau)} {(B_2 k\tau)}
\right]\,,  
\end{eqnarray} 
with 
\begin{equation}
C_1=(\tau\dot{a}/a)^{-1}\,,
\end{equation}
and
\begin{equation}
C_2=\frac{2}{3} \frac{1}{ 1 + 4 \tau \dot{a}/ {a} }\,.
\end{equation}

\begin{figure}
\setlength{\unitlength}{1cm}
\begin{minipage}{8.0cm}
\leftline{\psfig{file=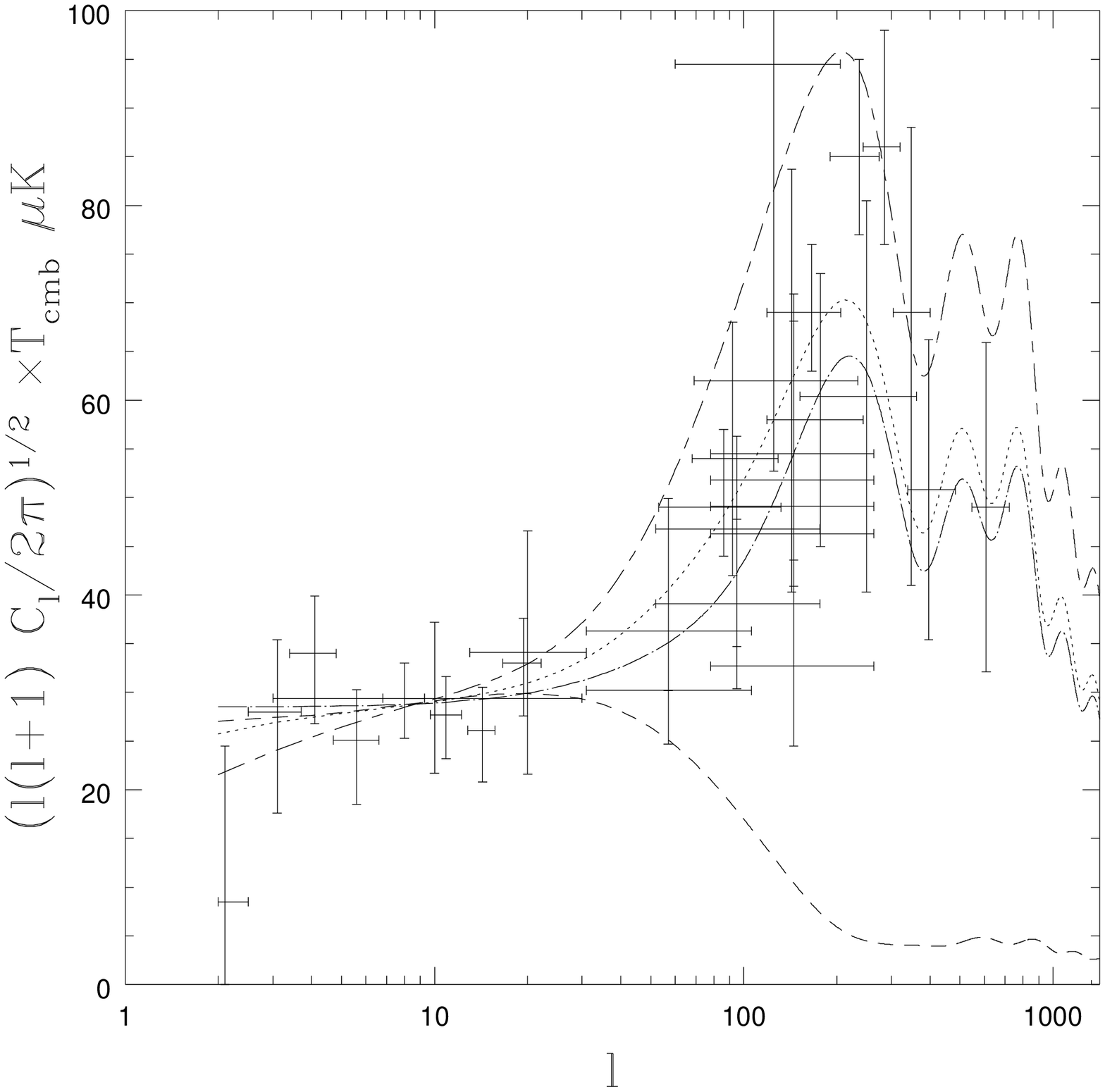,width=3.4in}}
\end{minipage}\hfill
\begin{minipage}{8.0cm}
\rightline{\psfig{file=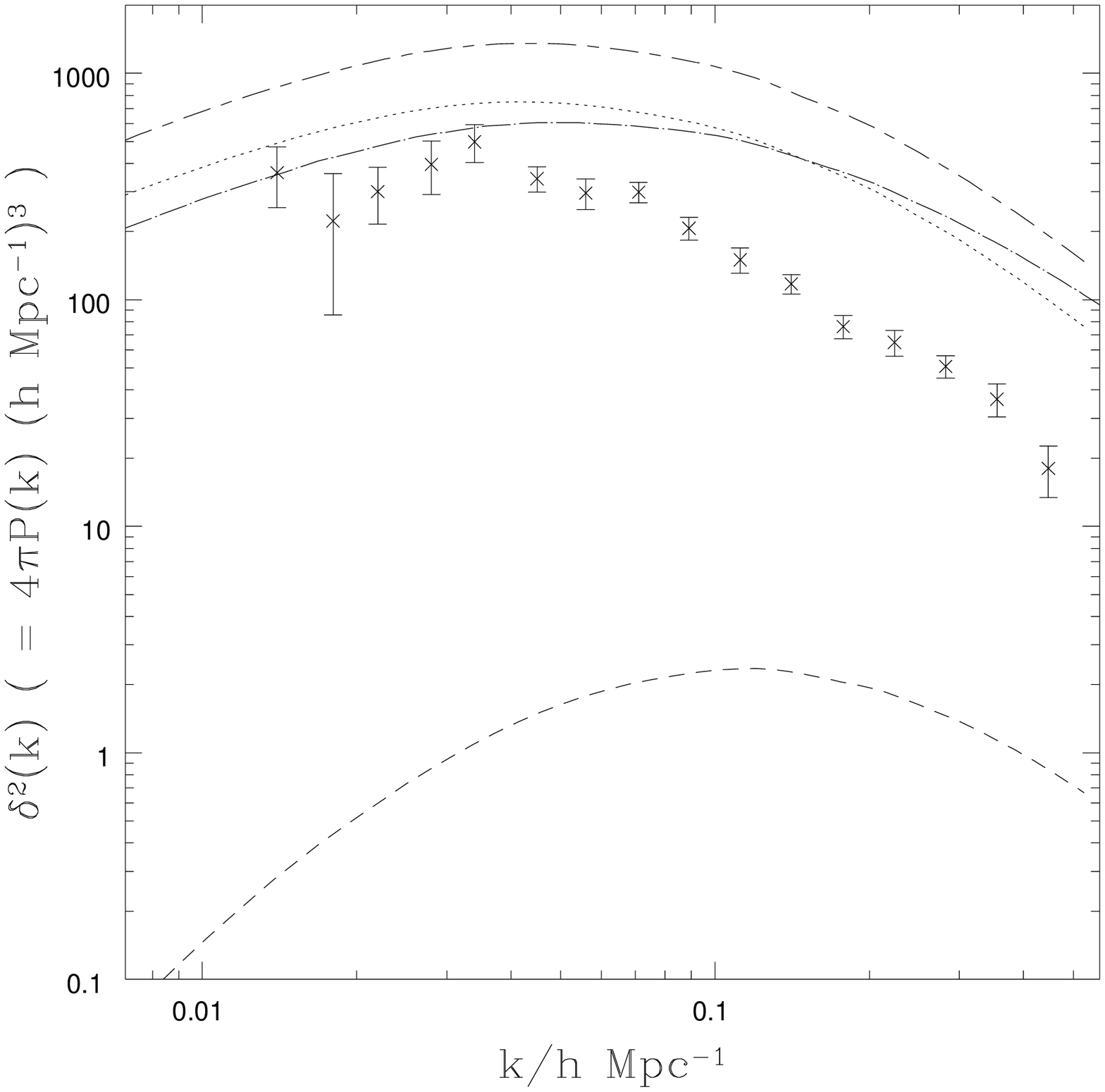,width=3.4in}}
\end{minipage}
\caption{Variations on the Mimic inflation model: 
We plot the COBE normalized angular power spectrum of CMB
anisotropies (left-hand graph), and the matter power spectrum
(right-hand graph) for variations on the mimic inflation model. The
amplitude of the $\Theta_D$ component is multiplied by a factor $f_D$,
with $f_D=1.0$ (long-short-dash line),  $f_D=0.0$ (dotted line) and
$f_D=1000$ (long-dash line). Observational data
and the prediction for standard CDM (dot-dash curve) are included for
comparison.}
\label{mimic}
\end{figure}

We have varied the model of \cite{HSW} by multiplying $\Theta_D$ by a
factor $f_D$. The 
results for such variations are shown in Fig.~\ref{mimic} (for this
figure we have made use of the values $A=1$, $B_1=0.5$, $B_2=0.5$. The
long-short-dash curves show the standard mimic model, normalized on
COBE scales. The model has an identical acoustic peak structure to
standard CDM, though the ratio of plateau to peak height is slightly
higher. The matter spectrum actually lies well above the observational
data. If the model is normalized to CMB data on scales larger than
$l=100$, then matter spectrum lies extremely close to that of standard
CDM. In other words, the mimic model matches the matter spectrum of an
adiabatic model as well as the acoustic peak structure.

The short-dash curves show the results for a run where $\Theta_D$ is
set to be 1000 times bigger than $\Theta_{00}+\Theta$. As expected,
this model gives rise to extremely small contributions for the matter
and oscillatory components, as compared to the ISW. 
The dotted curve shows the result of a run where $\Theta_D$ is
set to zero. We see that the matter spectrum in this case is of order
but
slightly lower than in the basic mimic model.  Therefore,
the relative amplitude of matter and ISW contributions seems to depend
in a non trivial way on the relative amplitude of $\Theta_D$ and
$\Theta_{00}+\Theta$, except in the limit that
$\Theta_{00}+\Theta$ is small.

\begin{figure}
\setlength{\unitlength}{1cm}
\begin{minipage}{8.0cm}
\leftline{\psfig{file=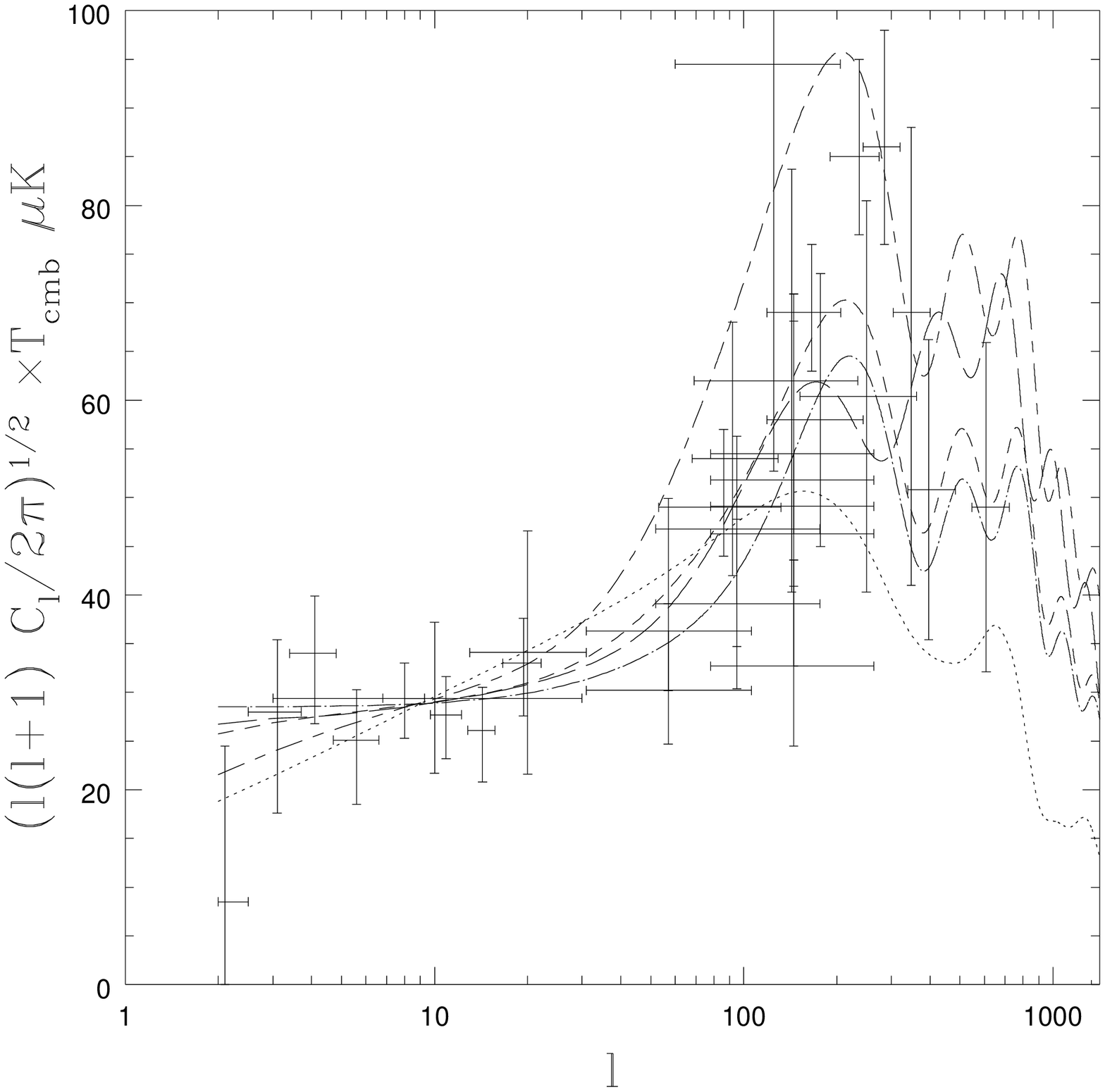,width=3.4in}}
\end{minipage}\hfill
\begin{minipage}{8.0cm}
\rightline{\psfig{file=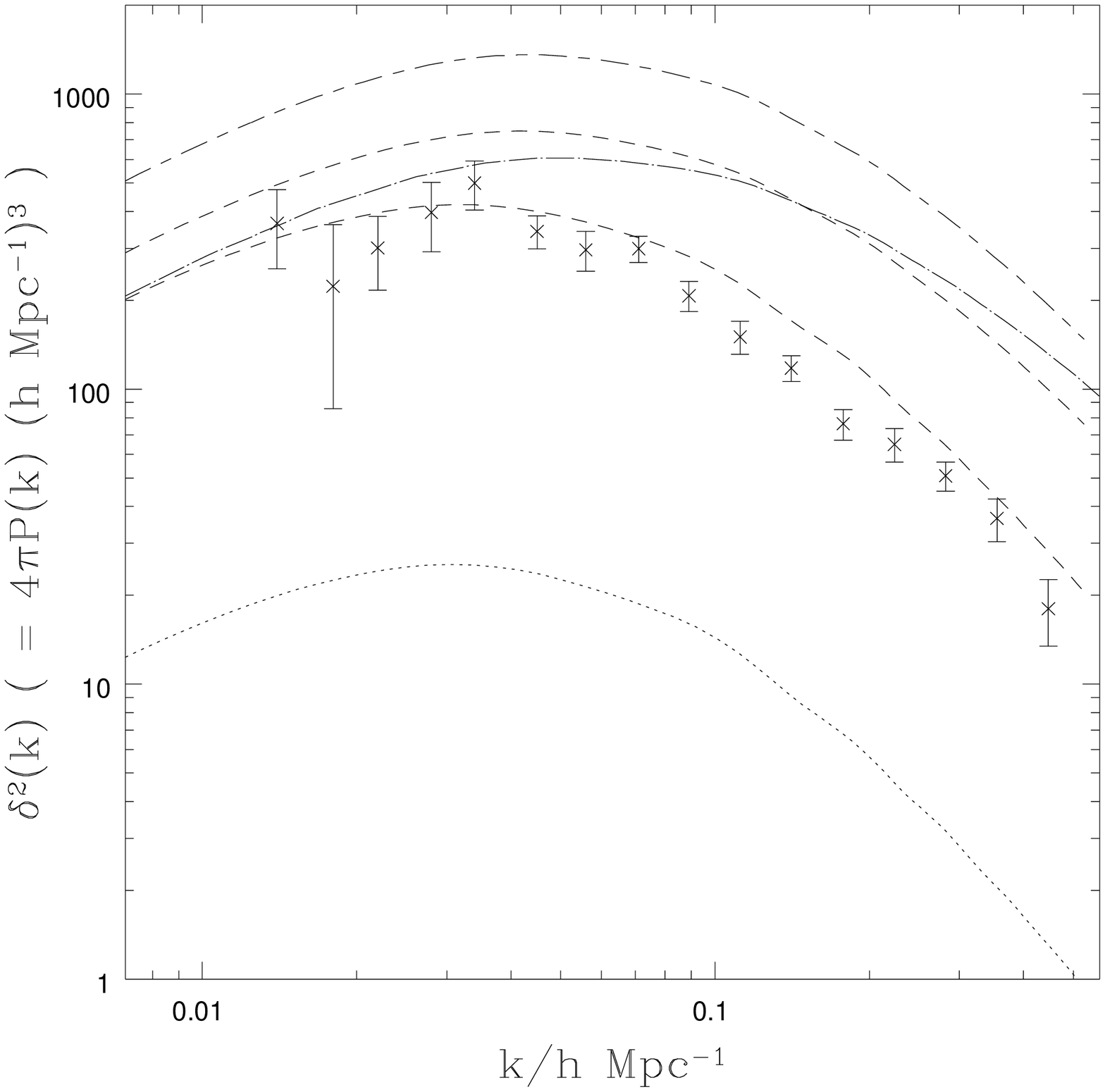,width=3.4in}}
\end{minipage}
\caption{Further variations on the Mimic inflation model: 
We plot the COBE normalized angular power spectrum of CMB
anisotropies (left-hand graph), and the matter power spectrum
(right-hand graph) for variations on the parameters $A$, $B_1$, $B_2$. 
The amplitude of these parameters is varied from the standard case by
multiplying each one by a factor $f_p$. Also the ratio
of $\Theta_{00}+\Theta$ to $\Theta_D$ is varied by multiplying
$\Theta_D$ by a factor $f_D$. We take $f_p=1.0$, $f_D=1.0$  (dot-dash
line), $f_p=0.4$, $f_D=1.0$ (dot
line), $f_p=1.0$ $f_D=0.0$ (short-dash
line), $f_p=0.4$ $f_D=0.0$ (short-dash
line). Observational data
and the prediction for standard CDM (dot-dash curve) are included for
comparison.}
\label{mimic2}
\end{figure}

In Fig.~\ref{mimic2} we show the results of varying the $f_D$ for
different choices of the parameters $A$, $B_1$ and $B_2$, which we vary
by multiplying each one by a factor $f_p$ from its original value. We
see that with $f_p=0.4$, the matter power spectrum lacks power on
large scales in the standard case of $f_D=1.0$. However, this deficit
of power can be made up for by setting $f_D=0.0$. The resulting matter
power spectrum is in strikingly good agreement with the observational
data. This is in contrast to the
case $f_P=1.0$ where the effect of reducing $f_D$ to zero is to
diminish the matter spectrum.

To summarize, the results of this section illustrate that 
models can be constructed which are capable of solving the $b_{100}$
problem. However the examples of this and previous sections have also
indicated that changes in the parameters can give contradictory
results, particularly in the coherent limit. We re-emphasize that in
the coherent limit there is no physical motivation for the concept of
scaling (which comes about as a result of random processes in all
known scaling models). Furthermore, the case which gave a good fit to
the matter spectrum did not satisfy constraints on the superhorizon
anisotropic stress that are required by isotropy. Nevertheless, these
are interesting cases which should be investigated further to see if
they could give rise to plausible models of structure formation\footnote{As we complete this work, we learned of a preprint\cite{DurKunz}, which discusses this issue.}.

\section{Discussion and conclusions}

Substantial progress has been made in understanding the predictions
from cosmic defect models of structure formation.  The modern
state-of-the-art calculations\cite{PSelTa,ABRa,ACDKSS} all represent
significant advances over previous
work\cite{AS,bsb,b&r,Peri,PST,reion,CT,DZ,ACSSV,DKLS}. While our 
`Modeling' approach is not as explicitly tied to simulations as the
work of Pen {\em et al} \cite{PSelTa} and Allen {\em et al}
\cite{ACDKSS}, its strength is that it 
allows us to probe the robustness of our key results, as a function of
variations in the defect stress-energy histories. 
 This lets us
investigate possible variations among different defect models and
possible systematic uncertainties in the simulations.

We have found a serious conflict between standard scaling defect models and the
current observational data.  This conflict can be clearly expressed
in terms of the `$b_{100}$ problem', where $b_{100}$ is the bias
on scales of $100h^{-1}$Mpc.
Current theoretical and experimental results  indicate
that the actual value of $b_{100}$ is close to unity,  but the
standard  defect models we considered, once COBE normalized,  required
$b_{100} \approx 5$ to reconcile the predictions for the density field
fluctuations with the 
observed galaxy  distribution.  The problem is robust over a wide
variations in parameters describing the defect sources and over
variations in $\Omega_b$ and $h$ as well.  It is tempting to 
exploit current uncertainties in the standard scaling picture to
resurrect the defect picture, but we have found that truly
radical deviations  from the standard scaling assumption are required
to produce a more viable model.  While one variation from the
standard model could get $b_{100} \approx 1.6$ (or even of
order 1 if a deviation from scaling is introduced as well), to get this
close is done at the expense of totally failing to fit the shape of
the matter spectrum and most of our ideas about how defects interact and evolve would also have to be wrong.

Structure formation on $100h^{-1}$Mpc scales is not strongly affected
by changes in the matter
content, and we do not expect that changes in the dark matter could save the
scaling defect models.  Probably the most interesting direction to
take the standard defect scenario is toward an open universe or one with a non vanishing cosmological constant.
There have been some suggestions \cite{VB} that
sufficient deviations from scaling might be realized in some low
$\Omega_0$ or cosmological constant models to result in a viable scenario. We
also discovered a handful of very different scaling models (see table III)
which had more reasonable values of $b_{100}$, but which have very
little to do with the standard defect scenarios.  They may however,
represent interesting directions for future model building.

Our results were anticipated in a number of ways by earlier work. Pen,
Spergel and Turok (ref.\cite{PST}) noted a serious bias problem for
specific global defect models on scales up to $20 h^{-1}$ Mpc and the
compilation of 
refs.\cite{AS} and \cite{ACSSV} in ref.\cite{WS} looks very much like
our results (and the {\em shape} of the CDM
curve in ref.\cite{AS} is  almost identical to ours).  However, in each
case it was far from clear that 
sufficient dynamic range had been achieved to predict the 
COBE normalized density field accurately.
One
of the strengths of our current method is that dynamic range is not a
problem and we feel we have covered the range of other possible
uncertainties in the `model scanning' described in sections III-V.
In cases where comparisons are appropriate, our results are 
consistent with those of Pen, Seljak and Turok \cite{PSelTa}, where
dynamic  range is also not a problem.   

Our results are extremely negative for the standard scaling defect
models and it is essential that results of this significance are
critically scrutinized.  We see four possible areas where this
scrutiny might be directed.

\begin{itemize}

\item{\em Linear Einstein-Boltzmann solver :} We have used standard
technology to solve the linearized Einstein-Boltzmann
equations. Initially, we modified a passive scalar code to include
sources and we checked this with the equivalent code used in
ref.\cite{PSelTa}. The vector and tensor codes have not at this stage
been checked. We should note, however, that an accurate
Einstein-Boltzmann solver is the ultimate quantitative authority in
this subject and that analytic fitting formulae have only limited
usage if they do not agree. Calculations based on these fitting
formulae may be wrong by up to a factor of two. 
 
\item{\em Modeling of two-point functions :} Clearly, our results
depend on our choices of defect stress-energy two-point functions,
which were designed to fit flat space string simulations. We have
extensively probed variations about these two-point functions and we
found that only much less motivated models are even marginally
consistent with the data. Nonetheless, we have illustrated that some
models, which at this stage are physically unmotivated, can fit
the data on 100$h^{-1}$Mpc. Obviously, much more work will be done on
this subject. 
 
\item{\em Bias :} It appears that $b_{100} \approx 5$ (or even $b_{100} \approx 3$) is not consistent with the data and also that defect models are
unable to generate such a large bias.  If either of these turns out
to be false the case against defects will be weakened.  

\item{\em Comparison with data :} It is essential that we have made the comparison with the data correctly. If the
current data turns out to be dramatically inaccurate, our conclusions would
of course change.  A more subtle problem might be the assumption of Gaussianity
which goes into producing the error bars used in the
comparisons between theory and data.  We believe that the dependence
of the error bars on the  sample size, be it the sky coverage in a CMB
experiment or the depth of a redshift survey, is likely to be very
model dependent. Although the central limit 
theorem seems to ensure that the comparisons we make are reasonable in
most defect models\cite{hot,r&a}, in more exotic cases\cite{joao&char}
the effect can be large. 

\end{itemize}  
 
\noindent We have carefully scrutinized all four of these issues, and, at this stage, stand by our conclusions.  However, we are happy to engage in critical dialogues on
these topics.

Much of the early calculations of defect models have
involved various heuristic arguments, particularly concentrating on
the effect of an individual defect on the microwave background or on matter accretion.
Such approaches were able to give an intuitive 
picture of the effects of defects on structure formation, but in the modern age  we feel strongly that arguments such as  `model X does not look
so bad' must be translated into expressions for the two-point functions before we are willing to take them seriously --- factors of two can be important and would obviously change the conclusions of this work.
We have also demonstrated that results from coherent models can have very
little to do with the fully active decoherent case and that it is
essential that causality be strictly enforced if physically
meaningful results are to be obtained.   We are not arguing against
the utility of heuristic discussions, but it is essential for the
heuristic pictures to be brought into contact with the modern
calculations if they are to have an impact. 

The complete failure of the bulk of the models we considered here is
in striking contrast with the (inflation motivated) nearly scale
invariant adiabatic models (which have no trouble getting
$b_{100}$ right).  These two types of model are different
in many ways and it is not necessarily easy to pin the problems on a
single physical difference.  Still, it is worth noting a number of
effects which all cause problems for the defect scenarios.

Firstly, the defect models are {\em isocurvature} models and passive
isocurvature models are know to have the CDM perturbation at a given
wavelength (relative to
temperature perturbations at the same wavelength) down by a factor of
$6$ as compared with the adiabatic case.  A similar effect is present
in the active cases\cite{JSF}, although this is not directly relevant to the 
$b_{100}$ problem, which compares temperature and CDM fluctuations on
different scales.  The scaling (or other) assumption which links these
two scales is a crucial part of the story.  This difference between
defect and adiabatic models is more directly connected with the
problem of matching the galaxy data with the smaller scale CMB
measurements, for example, within the early 
time window of Fig. \ref{standard}, a matter which is problematic for
defects, but which we have not
emphasized here.

Secondly, the growth of the CDM perturbations under gravitational
collapse is crucial to both pictures.  The fact that defect models 
lay down perturbations throughout time (as compared with the
inflationary models, for which the perturbations are `set up' early
on) means that the average defect perturbation benefits less from the
gravitational collapse.  The temperature perturbations benefit more
equally from the constant active seeding, so this is another physical
effect which suppresses the relative perturbations in matter versus radiation.

Finally, the previous effect is further enhanced by the generic
presence of vector and tensor perturbations in the active models.
These do not excite growing modes in the CDM, but make added
contributions to the temperature fluctuations, especially on COBE
scales. 

Having made these points, we must emphasize that the ability for a causal
active model to duplicate the successes of an adiabatic model are
purely a technical matter.  There is no formal `no go theorem'
against active models, it is just that the models which best match
our current ideas about defects don't work.  The best counter example
is Turok's `inflationary mimic' 
model\cite{Ta}.  We have confirmed Turok's conjecture that the mimic
model produces adiabatic-like predictions for
the CDM spectrum, as well as for the CMB anisotropies (which were the
original focus in \cite{Ta}), and we have added a few new
scaling models to the list.  We are also investigating a number of
physical processes which could produce extremely non-scaling sources.
All these models are, however, radical departures from the original
standard scaling defect picture and they require further work to determine
if they correspond to realistic physical scenarios.

The problems with the standard defect models are likely to have a significant
impact on our understanding of the origin of cosmic structure.   The
defect models were the classic examples of models of structure
formation in which Standard Big Bang (SBB) causality holds.   That is,
one started with a perfectly homogeneous universe and seeded
perturbations via causal processes in the SBB.  With the
demise of the standard defect models, the question arises whether 
{\em any} plausible SBB causal model exists.  If the answer is
negative, then this is very strong evidence for an inflationary origin
of cosmic structure, in which SBB causality is violated to produce
fluctuations of the standard `adiabatic' form.  We are currently
investigating the other possibilities of SBB causal models of
structure formation mentioned above, but it is too early to tell if a  
convincing scenario can emerge. 

\section*{Acknowledgments}

We thank U. Seljak and M. Zaldarriaga for the use of CMBFAST and in
particular Seljak for help with 
incorporating the vectors and tensors. We thank
M. Hindmarsh, L. Knox, J. Magueijo, P. Ferreira, P. Shellard,
N. Turok and G. Vincent for helpful conversations. This work was
supported by PPARC and computations
were done at the UK National Cosmology
Supercomputing Center, supported by PPARC, HEFCE and Silicon
Graphics/Cray Research. RAB was funded by PPARC grant GR/K94799 at IC and by Trinity College at DAMTP.

\def\jnl#1#2#3#4#5#6{\hang{#1, {\it #4\/} {\bf #5}, #6 (#2).}}
\def\jnltwo#1#2#3#4#5#6#7#8{\hang{#1, {\it #4\/} {\bf #5}, #6; {\it
ibid} {\bf #7} #8 (#2).}} 
\def\prep#1#2#3#4{\hang{#1, #4.}} 
\def\proc#1#2#3#4#5#6{{#1 [#2], in {\it #4\/}, #5, eds.\ (#6).}}
\def\book#1#2#3#4{\hang{#1, {\it #3\/} (#4, #2).}}
\def\jnlerr#1#2#3#4#5#6#7#8{\hang{#1 [#2], {\it #4\/} {\bf #5}, #6.
{Erratum:} {\it #4\/} {\bf #7}, #8.}}
\def\prl{Phys.\ Rev.\ Lett.}
\def\pr{Phys.\ Rev.}
\def\pl{Phys.\ Lett.}
\def\np{Nucl.\ Phys.}
\def\prp{Phys.\ Rep.}
\def\rmp{Rev.\ Mod.\ Phys.}
\def\cmp{Comm.\ Math.\ Phys.}
\def\mpl{Mod.\ Phys.\ Lett.}
\def\apj{Ap.\ J.}
\def\apjl{Ap.\ J.\ Lett.}
\def\aap{Astron.\ Ap.}
\def\cqg{Class.\ Quant.\ Grav.} 
\def\grg{Gen.\ Rel.\ Grav.}
\def\mn{MNRAS}
\def\ptp{Prog.\ Theor.\ Phys.}
\def\jetp{Sov.\ Phys.\ JETP}
\def\jetpl{JETP Lett.}
\def\jmp{J.\ Math.\ Phys.}
\def\zpc{Z.\ Phys.\ C}
\def\cupress{Cambridge University Press}
\def\pup{Princeton University Press}
\def\wss{World Scientific, Singapore}
\def\oup{Oxford University Press}

\pagebreak
\pagestyle{empty}


\begin{thebibliography}{99}

\bibitem{condensed}\jnl{V.M.H. Ruutu {\it et al}}{1996}{}{Nature}{382}{334}

\bibitem{kib1}
\jnl{T.W.B. Kibble}{1976}{}{J. Phys.}{A9}{1387}

\bibitem{VS}
\book{A. Vilenkin \& E.P.S. Shellard}{1994}{Cosmic strings and other topological defects}{\cupress}

\bibitem{HK}
\jnl{M. Hindmarsh \& T.W.B. Kibble}{1995}{}{\it Rep. Prog. Phys.}{58}{477}

\bibitem{zeld}
\jnl{Ya.B. Zel'dovich}{1980}{}{\mn}{192}{663}

\bibitem{vil}
\jnlerr{A. Vilenkin}{1981}{}{\prl}{46}{1169}{46}{1496}

\bibitem{TurSperg}
\jnl{N. Turok}{1989}{}{\prl}{63}{2625}

\bibitem{ACFM}
\jnl{A. Albrecht, D. Coulson, P.G. Ferreira and
J. Magueijo}{1996}{}{\prl}{76}{1413}

\bibitem{PSelTa}
\jnl{U.L. Pen, U. Seljak and N.Turok}{1997}{}{\prl}{79}{1615}

\bibitem{ABRa}\prep{A. Albrecht, R.A. Battye and
J. Robinson}{1997}{}{astro-ph/9707129, {\it\prl} in press}

\bibitem{ACDKSS}
\jnl{B. Allen, R.R. Caldwell, S. Dodelson, L. Knox, E.P.S. Shellard and A. Stebbins}{1997}{}{\prl}{79}{2624}

\bibitem{PST}
\jnl{U.L. Pen, D.N. Spergel and N.Turok}{1994}{}{\pr}{D49}{692}

\bibitem{CT}
\jnl{R.G. Crittenden \& N. Turok}{1995}{}{\prl}{75}{2642}

\bibitem{DGS}
\jnl{R. Durrer, A. Gangui \& M. Sakellariadou}{1996}{}{\prl}{76}{579}

\bibitem{ACSSV}
\jnl{B. Allen, R.R. Caldwell, E.P.S. Shellard, A. Stebbins and S. Veeraraghavan}{1996}{}{\prl}{77}{3061}

\bibitem{kib2}
\jnlerr{T.W.B. Kibble}{1985}{}{\np}{B252}{227}{B261}{750}

\bibitem{ACK}
\jnl{D. Austin, E.J. Copeland and
T.W.B. Kibble}{1993}{}{\pr}{D48}{5594}

\bibitem{MSa}
\jnl{C.J.A.P. Martins and E.P.S. Shellard}{1996}{}{\pr}{D54}{2535}

\bibitem{AS}
\jnltwo{A. Albrecht and A. Stebbins}{1992}{}{\prl}{68}{2121}{69}{2615}

\bibitem{AT}
\jnl{A. Albrecht and N. Turok}{1989}{}{\pr}{D40}{973}

\bibitem{BB}
\jnl{D. Bennett and F. Bouchet}{1990}{}{\pr}{D41}{2408}

\bibitem{AShe}
\jnl{B. Allen and E.P.S. Shellard}{1990}{}{\prl}{64}{119}

\bibitem{us}
\prep{R.A. Battye, J. Robinson and A. Albrecht}{1997}{}{astro-ph/9711336}

\bibitem{paul}
\prep{P.P. Avelino, E.P.S. Shellard, J.H. Wu and B. Allen}{1997}{}{In prep}

\bibitem{lineofsight}
\jnl{U. Seljak}{1994}{}{\apj}{435}{L87} \jnl{W. Hu and
N. Sugiyama}{1995}{}{\apj}{444}{489}

\bibitem{various}
\jnl{P.J.E. Pebbles and J.T. Yu}{1970}{}{\apj}{162}{815} \jnl{M.L. Wilson and J.Silk}{1981}{}{\apj}{243}{14} \jnl{P.J.E. Peebles}{1981}{}{\apj}{248}{885} \jnl{J.R Bond and A. Salzay}{1983}{}{\apj}{274}{443} \jnl{N. Vittorio and J. Silk}{1984}{}{\apj}{285}
{L39} \jnl{J.R. Bond and G. Efstathiou}{1984}{}{\apj}{285}{L45} \jnl{J. Bernstein and S. Dodelson}{1990}{}{\pr}{D41}{354} \jnl{S. Dodelson and J.M. Jubas}{1995}{}{\apj}{439}{503} \jnl{C.-P. Ma and E. Bertschinger}{1995}{}{\apj}{455}{7}

\bibitem{HW}
\prep{W. Hu and M. White}{1997}{}{astro-ph/9702170}

\bibitem{cmbfast}\jnl{U. Seljak and M. Zaldarriaga}{1996}{}{\apj}{469}{437}

\bibitem{PSelTc}\prep{N. Turok, U.L. Pen and
U. Seljak}{1997}{}{astro-ph/9706250}

\bibitem{HSW}
\jnl{W.Hu, M. White and D. Spergel}{1997}{}{\pr}{55}{3288}

\bibitem{MACFa}
\jnl{J. Magueijo, A. Albrecht, D. Coulson and P.G. Ferreira}{1996}{}{\prl}{76}{2617}

\bibitem{MACFb}
\jnl{J. Magueijo, A. Albrecht, D. Coulson and
P.G. Ferreira}{1996}{}{\pr}{D54}{3727}

\bibitem{Ta}
\jnl{N. Turok}{1996}{}{\pr}{D54}{R3686} 

\bibitem{Tb}
\jnl{N. Turok}{1996}{}{\prl}{77}{4138}

\bibitem{VHS}
\jnl{G.R. Vincent, M. Hindmarsh and M. Sakellariadou}{1997}{}{\pr}{D55}{573}

\bibitem{ABRc}
\prep{A. Albrecht, R.A. Battye and J. Robinson}{1997}{}{In prep}

\bibitem{tegmark}
The data points have been compiled by M. Tegmark at
http://www.sns.ias.edu/~max/cmb/experiments.html 

\bibitem{PD}
\jnl{J.A. Peacock and S.J. Dodds}{1994}{}{\mn}{267}{1020}

\bibitem{GF}
\jnl{E. Gaztanaga and J. Frieman}{1994}{}{\apj}{L13}{137}

\bibitem{WSDK}
\prep{J.A. Willick, M.A. Strauss, A. Dekel and
T. Kolatt}{1997}{}{astro-ph/9612240} 

\bibitem{abel} 
\prep{T. Abel {\it et al}}{1997}{}{astro-ph/9706262}

\bibitem{wandelt}
\prep{B. Wandelt}{1997}{}{In preparation} 

\bibitem{CBS}
\jnl{R.R. Caldwell, R.A. Battye and E.P.S. Shellard}{1997}{}{\pr}{D54}{7146}

\bibitem{RW}
\jnl{J. Robinson and B. Wandelt}{1996}{}{\pr}{D53}{618}

\bibitem{M}
\jnl{C.J.A.P. Martins}{1997}{}{\pr}{D55}{5208}

\bibitem{VB}
\prep{C. van de Bruck}{1997}{}{astro-ph/9705208}

\bibitem{Fer}
\jnl{P. Ferreira}{1995}{}{\prl}{74}{3522}

\bibitem{ACM}
\jnl{P.P. Avelino, R.R. Caldwell and C.J.A.P. Martins}{1997}{}{\pr}{D56}{4568}

\bibitem{DS}
\prep{R. Durrer and M. Sakellariadou}{1997}{}{astro-ph/9702028}

\bibitem{DKLS}
\prep{R. Durrer, M. Kunz, C. Lineweaver and M. Sakellariadou}{1997}{} {astro-ph/9706215}

\bibitem{DurKunz}
\prep{R. Durrer and M.Kunz}{1997}{}{astro-ph/9711133}

\bibitem{bsb} 
\jnl{D. Bennett, A. Stebbins and F. Bouchet}{1992}{}{\apj}{399}{L5} 

\bibitem{b&r}
\jnl{D. Bennett and S. Rhie}{1993}{}{\apj}{406}{L7}

\bibitem{Peri} 
\jnl{R. Moessner, L. Perivolaropoulos and R. Brandenberger}{1994}{}{\apj}{425}{365}

\bibitem{reion}
\jnl{D. Coulson, P. Ferreira, P. Graham and N. Turok}{1994}{}{Nature}{368}{27}

\bibitem{DZ}
\jnl{R. Durrer and Z.H. Zhou}{1996}{}{\pr}{D53}{5394}

\bibitem{WS}
\jnl{M. White and D. Scott}{1996}{}{Comm. on Astro.}{18}{289}

\bibitem{hot}
\jnl{N. Turok}{1996}{}{Ap. J.}{473}{L5}

\bibitem{r&a}
\jnl{J. Robinson and A. Albrecht}{1996}{}{M.N.R.A.S}{283}{733}

\bibitem{joao&char}
\prep{C. Cheung and J. Magueijo}{1997}{}{astro-ph/9707271}

\bibitem{JSF}
\jnl{A. Jaffe, A. Stebbins and J. Frieman}{1994}{}{Ap. J.}{420}{9}

\end{thebibliography}
\end{document}